\documentclass[final,3p,times]{elsarticle}
\usepackage{amssymb}
\usepackage{graphicx}% Include figure files
\usepackage{dcolumn}% Align table columns on decimal point
\usepackage{bm}% bold math
\usepackage{mathptmx}
\usepackage{array,color}
\usepackage{amsmath}
\newcommand{\ii}{\mathrm{i}}
\newcommand{\ee}{\mathrm{e}}

\newcommand{\cO}{\mathcal{O}}

\newtheorem{theorem}{Theorem}

\newtheorem{prop}{Proposition}
\begin{document}

\begin{frontmatter}



\title{Rogue wave patterns associated with Adler--Moser polynomials featuring  multiple roots in the nonlinear Schr{\"o}dinger equation}


\author[aff1]{Huian Lin}

\author[aff1]{Liming Ling$^{*}$}
\ead{linglm@scut.edu.cn}
%\affiliation[aff1]{organization={School of Mathematics, South China University of Technology},city={Guangzhou},postcode={510641}, country={China}}
\address[aff1]{School of Mathematics, South China University of Technology,Guangzhou, 510641, China}

\begin{abstract}

%This paper constructs high-order rogue wave solutions of the nonlinear Schr{\"o}dinger (NLS) equation by utilizing the Darboux transformation (DT) theory. 

In this work, we analyze the asymptotic behaviors of high-order rogue wave solutions with multiple large parameters and discover novel rogue wave patterns, including claw-like, OTR-type, TTR-type, semi-modified TTR-type, and their modified patterns. A correlation is established between these rogue wave patterns and the root structures of the Adler--Moser polynomials with multiple roots. At the positions in the $(x,t)$-plane corresponding to single roots of the Adler--Moser polynomials, these high-order rogue wave patterns asymptotically approach first-order rogue waves. At the positions in the $(x,t)$-plane corresponding to multiple roots of the Adler--Moser polynomials, these rogue wave patterns asymptotically tend toward lower-order fundamental rogue waves, dispersed first-order rogue waves, or mixed structures of these rogue waves. These structures are related to the root structures of special Adler--Moser polynomials with new free parameters, such as the Yablonskii--Vorob'ev polynomial hierarchy, among others. Notably, the positions of the fundamental lower-order rogue waves or mixed structures in these rogue wave patterns can be controlled {freely} under specific conditions.

\end{abstract}




\end{frontmatter}


\section{Introduction}

Rogue waves, also known as freak waves, extreme waves, etc., were initially used to describe abnormally large amplitude water waves that exist for a short period in localized areas of the deep ocean. In the past two decades, the concept of rogue waves has not only been applied in oceanography but also been introduced into many other fields, such as optical fibers \cite{2007Optical, 2008Optical, Kibler2010, 2014Lecaplain, akhmediev2011}, plasma physics \cite{2011plasma, tsai2016, 2017plasma, merriche2017}, Bose-Einstein condensates \cite{pethick2008, carretero2008, 2009Matter, Manikandan2014}, {finance} \cite{yan2010, yan2011vector}, superfluid helium \cite{2010Rogue}, and so on. This has sparked extensive research by many physicists and mathematicians. For instance, in 2007, Solli \textit{et al}. \cite{2007Optical} first experimentally observed anomalously large amplitude optical rogue waves in nonlinear optics. %and published related papers in the journal Nature. 
In 2010, Efimov \textit{et al}. \cite{2010Rogue} observed rogue waves in experiments with superfluid helium and discovered their occurrence shortly after the driver was turned on, in a non-equilibrium state before reaching a stable condition. Moreover, Kibler \textit{et al}. \cite{Kibler2010} observed rogue waves (called Peregrine solitons) through experiments in nonlinear fiber optics and numerical simulations. In 2011, Chabchoub \textit{et al}. \cite{Chabchoub2011} observed Peregrine solitons in a wave tank experiment. Subsequently, they \cite{Chabchoub2012} further observed a hierarchy of up to fifth-order rogue waves in wave tank experiments, consistent with the theory in $2012$. 

Given the extensive physical relevance of rogue waves, significant theoretical research has been conducted on rogue wave solutions of integrable systems, such as the AKNS integrable system \cite{Akhmediev2009, Akhmediev2010, ling2012, ling2013, wang2013, Ohta2012, yang2021, yang2024},  
the Kaup-Newell integrable system \cite{guo2013, he2014,Chan2014, 2020lin, yang2021b, lin2024chaos}, and so on, which are used to describe various physical phenomena. Moreover, due to high-order rogue wave solutions' complex and diverse dynamical behaviors, there has been a growing interest in classifying their structures. This allows for predicting later rogue wave structures from initial rogue wave waveforms. 

Since the general rogue waves was discovered, numerous scholars had explored patterns of rogue waves of NLS equation. In 2011, Akhmediev \textit{et al}. {\cite{Akhmediev2011a, Akhmediev2011b}} discovered patterns of high-order rogue wave solutions of the NLS equation, including triangular and {circular} structures. {Subsequently, they also found new rogue wave patterns for the NLS equation in Refs. \cite{Akhmediev2012, Akhmediev2013}, such as claw-like, pentagram, heptagram, and enneagram structures.} Moreover, Guo \textit{et al}. \cite{ling2012, ling2013} identified new patterns of high-order rogue wave solutions of the NLS equation: the claw-line-I, double-column, and circle-arc pattern. Meanwhile, He \textit{et al}. \cite{he2014} found new patterns in the study of high-order rogue wave solutions of the derivative nonlinear Schr{\"o}dinger (DNLS) equation, such as the modified-triangle, circle-triangle, and multi-circle pattern. In 2021, Yang \textit{et al}. \cite{yang2021, yang2021b} demonstrated that when one of the internal parameters in the rogue wave solutions of integrable equations is large, the rogue wave patterns are related to the root structures of special polynomials (Yablonskii--Vorob'ev polynomials), leading to the discovery of more intricate structural patterns.
Furthermore, Yang \textit{et al}. investigated the patterns of rogue wave solutions of the NLS equation \eqref{NLS} with multiple internal significant parameters in Ref. \cite{yang2024}. They found that these rogue wave patterns correspond to root structures of the Adler--Moser polynomials, where all roots of the polynomials are single.  However, for the cases where some roots of the Adler--Moser polynomials are not simple, {they consider this} an open question. Naturally, we ponder: do rogue wave patterns exist corresponding to the root structures of the Adler--Moser polynomials with multiple roots? If they do exist, what relationship exists between them?

Inspired by the work in Ref. \cite{yang2024}, this paper will investigate the connection between the patterns of rogue wave solutions of the NLS equation with multiple internal large parameters and root structures of the Adler--Moser polynomials with multiple roots. We can theoretically obtain more novel rogue wave patterns. Based on the locations of single roots and multiple roots of the Adler--Moser polynomials, we will divide these corresponding rogue wave patterns into two regions: the single-root region and the multiple-root region. In particular, in the multiple-root region, the high-order rogue wave patterns given in this paper asymptotically approach lower-order fundamental rogue waves, dispersed first-order rogue waves, or mixed structures of these rogue waves. The structures of these patterns in the multiple-root region are related to the root structures of new special Adler--Moser polynomials, such as the Yablonskii--Vorob'ev polynomial hierarchy, among others. The position of the multiple-root region can be regulated under specific conditions, related to the multiple roots of the Adler--Moser polynomials. However, in the rogue wave patterns discovered in previous studies, the positions of lower-order rogue waves within the patterns are relatively fixed. {Moreover, the research methods in this paper can be extended to the study of solitonic solution patterns for other integrable equations \cite{guo2013, Chan2014, ankiewicz2010rogue, yhh2024}.} Therefore, the results of this paper are significant for both experimental and theoretical studies of rogue wave patterns. 

The structure of this paper is as follows: In Sec. \ref{Sec2}, we will introduce the Darboux transformation (DT) method of the NLS equation \eqref{NLS} and then generate the formula of high-order rogue wave solutions. Moreover, we will discuss the root structures of the Adler--Moser polynomials with multiple roots. In Sec. \ref{Sec3}, we will analyze asymptotic behaviors and pattern structures of high-order rogue wave solutions for the NLS equation \eqref{NLS}. Sec. \ref{Sec4} will present some cases of rogue wave patterns with multiple internal large parameters. Sec. \ref{Sec5} will provide the proofs of the main results of this paper. Finally, we will summarize and discuss the results of this paper in Sec. \ref{Sec6}.

\section{Preliminaries}\label{Sec2}

In this section, we will introduce the DT method of the NLS equation and then present the determinant formula of the high-order rogue wave solutions. Furthermore, for the convenience of subsequent studies on rogue wave patterns, we will also introduce the Adler--Moser polynomials and discuss their root structures with multiple roots.

\subsection{DT of the NLS equation}

The NLS equation {reads}
\begin{equation}\label{NLS}
	\begin{aligned}
		&\ii {q}_{t}+{q}_{xx}+2 | {q}|^{2}{q} ={0}, 
	\end{aligned}
\end{equation}
where $q=q(x,t) $ is the complex function, and the subscripts $x$ and $t$ denote the partial derivative. This equation is completely integrable and serves as one of the fundamental research models in many fields, such as water waves, optical fibers, biology, plasma physics, Bose-Einstein condensates, and so on \cite{2007Optical, Hasegawa1995, agrawal2000, 2000Nonlinear, Pitaevskii2003, yan2010}.
The Lax pair of NLS equation  \eqref{NLS} are given in Ref. {\cite{akns1974, ling2012}}, as follows:
\begin{equation}\label{laxp}
\begin{array}{ll}
	\Phi_{x}=\mathbf{U}\Phi, & \Phi_{t}=\mathbf{V}\Phi,\\
	 \mathbf{U} =\ii (\lambda\sigma_{3}+ \mathbf{Q}), & \mathbf{V} = \ii\left( 2\lambda^{2}\sigma_{3} + 2\lambda \mathbf{Q} -\sigma_{3}(\mathbf{Q}^{2}+\ii\mathbf{Q}_{x}) \right),\\
	\sigma_{3}=\begin{bmatrix}
		1 & 0\\
		0 & -1
	\end{bmatrix}, &
	\mathbf{Q}=\begin{bmatrix}
		0 & q^{*}\\
		q & 0
	\end{bmatrix},
\end{array}
\end{equation}
where $\Phi=\Phi(\lambda;x,t)$ is the complex matrix spectral function and $\lambda \in \mathbb{C}$ is the spectral parameter. For the above Lax pair, the zero curvature equation $ \mathbf{U}_{t}-\mathbf{V}_{x} +[ \mathbf{U}, \mathbf{V} ]=0$ can derive the NLS equation \eqref{NLS}, where $[ \mathbf{U}, \mathbf{V} ] \equiv \mathbf{U}\mathbf{V} - \mathbf{V}\mathbf{U}$ is a commutator. 

Next, we recall the DT of the AKNS system presented in Refs. \cite{1991Matveev, terng2000, 2005gu, ling2021, dong2022}. The $N$-fold DT,
\begin{equation}\label{ndt1}
	\Phi^{[N]}= \mathbf{T}_{N}(\lambda;x,t)\Phi,
\end{equation} 
can convert the system \eqref{laxp} into a new one
\begin{equation}\label{laxp2}
	\Phi^{[N]}_{x}=\mathbf{U}^{[N]}\Phi^{[N]}, \quad \Phi^{[N]}_{t}=\mathbf{V^{[N]}}\Phi^{[N]}, \quad 	\mathbf{U}^{[N]} =\mathbf{U}|_{q=q^{[N]}}, \quad \mathbf{V}^{[N]} =\mathbf{V}|_{q=q^{[N]}}.
\end{equation}
The $N$-fold Darboux matrix $\mathbf{T}_{N}(\lambda;x,t)$ is presented in the following theorem. 

\begin{theorem}\label{th:dt}
	If ${q}(x,t) \in \mathbf{L}^{\infty}(\mathbb{R}^{2})\cap \mathbf{C}^{\infty}(\mathbb{R}^{2})$ and the matrix function $\Phi(\lambda;x,t)$ is the solution of Lax pair \eqref{laxp}, then the $N$-fold Darboux matrix is expressed as
	\begin{equation}\label{ndt2}
		\mathbf{T}_{N}=\mathbb{I} +\mathbf{Y}_{N}\mathbf{M}^{-1}\mathbf{D}^{-1}\mathbf{Y}_{N}^{\dagger},
	\end{equation}
	where $\mathbb{I}$ is a second-order identity matrix, $\dagger$ denotes the conjugate transpose,
	\begin{equation}\label{yn}
	\begin{aligned}
		&\mathbf{Y}_{N}=\left( \phi_{1}, \phi_{2}, \cdots, \phi_{N} \right), \quad \mathbf{D}=\mathrm{diag}\left( \lambda-\lambda_{1}^{*}, \lambda-\lambda_{2}^{*}, \cdots, \lambda-\lambda_{N}^{*} \right), \quad
		\mathbf{M}=\left( \frac{\phi_{i}^{\dagger}\phi_{j}}{\lambda_{i}^{*}-\lambda_{j}}\right)_{1\leq i,j \leq N},\\
  		&\phi_{i}=\left( \phi_{i,1}, \phi_{i,2} \right)^T= f_{i}\, \Phi(\lambda_{i};x,t)(\Phi(\lambda_{i};0,0))^{-1}(1,-\ii)^{T},
	\end{aligned}
	\end{equation}
 and $f_{i}=f_{i}(x,t)$ $(1\leq i\leq N)$ are arbitrary functions regarding $x$ and $t$.
	Additionally, a new potential function can be obtained by the following B{\"a}cklund transformation, as follows:
	\begin{equation}\label{qnfor1}
		q^{[N]}(x,t)=q(x,t) + 2\mathbf{Y}_{N,2}\mathbf{M}^{-1}\mathbf{Y}_{N,1}^{\dagger},
	\end{equation}
	where the vector $\mathbf{Y}_{N,i}$ denotes the $i$-row vector of $\mathbf{Y}_{N}$ with $i=1,2$.
\end{theorem}
For more details on the $N$-fold DT, refer to Refs. \cite{terng2000, ling2021,dong2022}.

\subsection{Rogue wave solutions of the NLS equation}\label{subsec2.2}

Here, we apply the above $N$-fold DT to construct the determinant formula of the $ N $th-order rogue wave solutions for the NLS equation \eqref{NLS}. First, we set a nonzero seed solution (i.e. a plane wave solution) for the Lax pair \eqref{laxp}, as follows:
\begin{equation}\label{seed}
	q(x,t)=b\ee^{\ii\zeta}, \quad \zeta = cx+(2b^{2}-c^{2})t,
\end{equation}
where $b$ and $c$ are arbitrary real constants. Then, we can calculate the fundamental solution
\begin{equation}\label{eigenf}
\begin{aligned}
	& \Phi=\ee^{-\frac{\ii}{2}\zeta \sigma_{3}}\mathbf{E}(\lambda)\ee^{\ii\xi\omega\sigma_{3}}, 
\end{aligned}
\end{equation}
by substituting the seed solution \eqref{seed} into the system \eqref{laxp}, where
\begin{equation}\label{xiom}  	
	\mathbf{E}(\lambda)=\frac{1}{2\xi}\begin{pmatrix}
		1 & 1\\
		\frac{2b}{c+2\lambda +2\xi}  & \frac{2b}{c+2\lambda -2\xi}
	\end{pmatrix}, \quad\xi=\sqrt{(\lambda +\frac{c}{2})^{2}+b^{2}}, \quad \omega=x+(2\lambda -c)t +a(\lambda),
\end{equation}
and the arbitrary parameter $a(\lambda)$ is independent of $x$ and $t$. Note that the arbitrary parameter $a(\lambda)$ in the function \eqref{eigenf} is a key factor influencing the dynamic structures of rogue wave solutions for the NLS equation \eqref{NLS}, which will be discussed in detail later. 

Since NLS equation \eqref{NLS} admits Galilean and scaling invariance, we take $b=1$ and $c=0$ without loss of generality. Then, based on the seed solution $q=\ee^{2\ii t}$, we can rewrite the solution formula \eqref{qnfor1} as
\begin{equation}\label{qnfor2}
	q^{[N]}(x,t)= \frac{\det(\mathbf{M}^{(1)})}{\det(\mathbf{M}^{(0)})} \ee^{2\ii t},  
\end{equation}
with
\begin{equation}
\begin{aligned}
	&\mathbf{M}^{(n)} = \left( M_{i,j}^{(n)}\right)_{1\leq i,j \leq N}, \quad	n=0,1,\\
	&M_{i,j}^{(0)}= \frac{\phi_{i}^{\dagger}\phi_{j}}{-2\ii (\lambda_{i}^{*}-\lambda_{j})}, \quad 
	M_{i,j}^{(1)}= M_{i,j}^{(0)} + \ii \ee^{-2\ii t} \phi_{j,2}\phi_{i,1}^{*}.
\end{aligned}
\end{equation}
Furthermore, according to the definition \eqref{yn} of $\phi_{i}$ with $\Phi(\lambda_{i};x,t)$ given in \eqref{eigenf}, the elements $M_{i,j}^{(n)} $ in the matrix $\mathbf{M}^{(n)}$ ($n=0,1$) are simplified as the following quadric form
\begin{equation}\label{mij1}
\begin{aligned}
&M_{i,j}^{(0)}=\frac{f_{i}^{*}f_{j}}{4\xi_{i}^{*}\xi_{j}} {\left( \lambda_{i}^{*}+\xi_{i}^{*}+\ii, \lambda_{i}^{*}-\xi_{i}^{*}-\ii \right)  \ee^{-\ii\xi_{i}^{*}\omega_{i}^{*}\sigma_{3}}}
	\begin{pmatrix}
		\frac{1}{-\ii(\lambda_{i}^{*}+\xi_{i}^{*}) +\ii(\lambda_{j} +\xi_{j})} & \frac{1}{-\ii(\lambda_{i}^{*}+\xi_{i}^{*}) +\ii(\lambda_{j} -\xi_{j})} \\
		\frac{1}{-\ii(\lambda_{i}^{*}-\xi_{i}^{*}) +\ii(\lambda_{j} +\xi_{j})} & \frac{1}{-\ii(\lambda_{i}^{*}-\xi_{i}^{*}) +\ii(\lambda_{j} -\xi_{j})}
	\end{pmatrix}
	{\begin{pmatrix}
		\lambda_{j}+\xi_{j}-\ii \\ \xi_{j}-\lambda_{j} +\ii
	\end{pmatrix}\ee^{\ii\xi_{j}\omega_{j}\sigma_{3}}},\\
	&M_{i,j}^{(1)}=\frac{f_{i}^{*}f_{j}}{4\xi_{i}^{*}\xi_{j}}{\left( \lambda_{i}^{*}+\xi_{i}^{*}+\ii, \lambda_{i}^{*}-\xi_{i}^{*}-\ii \right)  \ee^{-\ii\xi_{i}^{*}\omega_{i}^{*}\sigma_{3}}}
	\begin{pmatrix}
		\frac{\frac{\lambda_{i}^{*}+\xi_{i}^{*}}{\lambda_{j}+\xi_{j}}}{-\ii(\lambda_{i}^{*}+\xi_{i}^{*}) +\ii(\lambda_{j} +\xi_{j})} & \frac{\frac{\lambda_{i}^{*}+\xi_{i}^{*}}{\lambda_{j}-\xi_{j}}}{-\ii(\lambda_{i}^{*}+\xi_{i}^{*}) +\ii(\lambda_{j} -\xi_{j})} \\
		\frac{\frac{\lambda_{i}^{*}-\xi_{i}^{*}}{\lambda_{j}+\xi_{j}}}{-\ii(\lambda_{i}^{*}-\xi_{i}^{*}) +\ii(\lambda_{j} +\xi_{j})} & \frac{\frac{\lambda_{i}^{*}-\xi_{i}^{*}}{\lambda_{j}-\xi_{j}}}{-\ii(\lambda_{i}^{*}-\xi_{i}^{*}) +\ii(\lambda_{j} -\xi_{j})}
	\end{pmatrix}
	{\begin{pmatrix}
		\lambda_{j}+\xi_{j}-\ii \\ \xi_{j}-\lambda_{j} +\ii
	\end{pmatrix}\ee^{\ii\xi_{j}\omega_{j}\sigma_{3}}}.
\end{aligned}
\end{equation}

Then, we consider further simplifying the elements $M_{i,j}^{(n)}$ $(n=0,1)$ in Eq. \eqref{mij1}. For convenience, we {introduce} some notations:
\begin{equation}\label{bij1}
	\begin{aligned}
	&B_{i,j}^{(0)}(h_{i}^{*}, \hat{h}_{j})= \frac{f_{i}^{*}f_{j}}{4\xi_{i}^{*}\xi_{j}} \, \frac{{(\lambda_{i}^{*}+\xi_{i}^{*}+\ii)(\lambda_{j}+\xi_{j}-\ii ) }}{-\ii(\lambda_{i}^{*}+\xi_{i}^{*}) +\ii(\lambda_{j} +\xi_{j})} \,\ee^{\ii \xi_{j}\omega_{j}-\ii \xi_{i}^{*}\omega_{i}^{*}}, \\
	&B_{i,j}^{(1)}(h_{i}^{*}, \hat{h}_{j})= \frac{f_{i}^{*}f_{j}}{4\xi_{i}^{*}\xi_{j}}\,\dfrac{{(\lambda_{i}^{*}+\xi_{i}^{*}+\ii)(	\lambda_{j}+\xi_{j}-\ii )} }{-\ii(\lambda_{i}^{*}+\xi_{i}^{*}) +\ii(\lambda_{j} +\xi_{j})} \,\frac{\lambda_{i}^{*}+\xi_{i}^{*}}{\lambda_{j}+\xi_{j}}\, \ee^{\ii \xi_{j}\omega_{j}-\ii \xi_{i}^{*}\omega_{i}^{*}}, \quad h_{j}=\hat{h}_{j}=\ii (\lambda_{j}+\xi_{j}).
	\end{aligned}
\end{equation}
By choosing the suitable parameter
\begin{equation}\label{cj}
	f_{j}=2\xi_{j}\ee^{\ii(\lambda_{j}x +\ii(2\lambda_{j}^{2}+1)t) +\ii \lambda_{j}\, a(\lambda_{j})},
\end{equation}
we can derive
\begin{equation}\label{bij2}
	\begin{aligned}
	&B_{i,j}^{(0)}(h_{i}^{*}, \hat{h}_{j})= \frac{(h_{i}^{*}+1)(\hat{h}_{j}+1)}{h_{i}^{*}+\hat{h}_{j}} \ee^{\hat{h}_{j}x -\hat{h}_{j}^{2}\ii t +a(\lambda_{j})\hat{h}_{j} +h_{i}^{*}x+(\hat{h}_{i}^{*})^{2}\ii t +(a(\lambda_{i})h_{i})^{*}}, \\
	&B_{i,j}^{(1)}(h_{i}^{*}, \hat{h}_{j})= \frac{(h_{i}^{*}+1)(\hat{h}_{j}+1)}{h_{i}^{*}+\hat{h}_{j}} \left( -\frac{h_{i}^{*}}{\hat{h}_{j}}\right)  \ee^{\hat{h}_{j}x -\hat{h}_{j}^{2}\ii t +a(\lambda_{j})\hat{h}_{j} +h_{i}^{*}x+(\hat{h}_{i}^{*})^{2}\ii t +(a(\lambda_{i})h_{i})^{*}}.
\end{aligned}	
\end{equation}

Next, let all spectral parameters $\lambda_{i}, \lambda_{j} \rightarrow -\ii$, then we have $ \xi_{i}^{*}, \xi_{j}\rightarrow 0$ and $h_{i}^{*}, \hat{h}_{j} \rightarrow 1$. Then, by applying the expansion
\begin{equation}\label{ak1}
\begin{aligned}
&\hat{h}_{j}x -\hat{h}_{j}^{2}\ii t +a(\lambda_{j})\hat{h}_{j} +h_{i}^{*}x+({h}_{i}^{*})^{2}\ii t +(a(\lambda_{i})h_{i})^{*} =\sum_{k=0}^{\infty}\left( A_{k}^{+}(\ln h_{i}^{*})^{k} + A_{k}^{-}(\ln \hat{h}_{j})^{k}\right),\\
 & A_{k}^{+}=\frac{x+2^{k}\ii t}{k!}+ \hat{a}_{k}^{*}, \quad A_{k}^{-}=\left(A_{k}^{+} \right)^{*}, 
\end{aligned} 
\end{equation}
with $
(a(\lambda_{i})h_{i})^{*}=\sum_{k=0}^{\infty} \hat{a}_{k}^{*}(\ln {h}_{i}^{*})^{k}$ and $ a(\lambda_{j})\hat{h}_{j}=\sum_{k=0}^{\infty} \hat{a}_{k}(\ln \hat{h}_{j})^{k} $,
the expression \eqref{bij2} of $ B_{i,j}^{(n)}(h_{i}^{*}, \hat{h}_{j})$ $(n=0,1)$ can be rewritten as
\begin{equation}\label{bij3}
\begin{aligned}
	&B_{i,j}^{(n)}(h_{i}^{*}, \hat{h}_{j}) =  \dfrac{(-1)^{n}2}{1-\frac{(h_{i}^{*}-1)(\hat{h}_{j}-1)}{(h_{i}^{*}+1)(\hat{h}_{j}+1)}}  \exp\left( \sum_{k=0}^{\infty}\left( A_{k}^{+}(\ln h_{i}^{*})^{k} + A_{k}^{-}(\ln \hat{h}_{j})^{k} + n(\ln h_{i}^{*}-\ln \hat{h}_{j})\right) \right) \\
	&= {(-1)^{n}}{2} \sum_{v=0}^{\infty} \left( \frac{\ln h_{i}^{*} \, \ln \hat{h}_{j}}{4} \right)^{v} \exp \left( v\sum_{l=1}^{\infty}s_{l}\left((\ln h_{i}^{*})^{l}+ (\ln \hat{h}_{j})^{l} \right) +\sum_{k=0}^{\infty}\left( A_{k}^{+}(\ln h_{i}^{*})^{k} + A_{k}^{-}(\ln \hat{h}_{j})^{k} + n(\ln h_{i}^{*}-\ln \hat{h}_{j})\right)  \right),
\end{aligned}
\end{equation}
where $s_{l}$ is defined by
\begin{equation}\label{sj1}
	\sum_{l=1}^{\infty} s_{l} \epsilon^{l} =\ln \left( \frac{2}{\epsilon} \tanh\frac{\epsilon}{2} \right), 
\end{equation}
with $s_{2l+1}=0$ for $l\geq 0$. 

In addition, if we multiply both the numerator and denominator by a common factor for the determinant formula \eqref{qnfor2}, the solution $q^{[N]}(x,t)$ remains invariant. Here, we take the following notations:
\begin{equation}\label{bij4}
\begin{aligned}
	\hat{B}_{i,j}^{(n)}(h_{i}^{*}, \hat{h}_{j}) =&\dfrac{{B}_{i,j}^{(n)}(h_{i}^{*}, \hat{h}_{j})}{2\exp\left( \sum_{k=0}^{\infty}\left( A_{2k}^{+}(\ln h_{i}^{*})^{2k} + A_{2k}^{-}(\ln \hat{h}_{j})^{2k}\right)  \right)}\\
	=& (-1)^{n} \sum_{v=0}^{\infty} \left( \frac{\ln h_{i}^{*} \, \ln \hat{h}_{j}}{4} \right)^{v} \exp \left( v\sum_{l=1}^{\infty}s_{2l}\left((\ln h_{i}^{*})^{2l}+ (\ln \hat{h}_{j})^{2l} \right) \right.\\
	& \left. +\sum_{k=0}^{\infty}\left( A_{2k+1}^{+}(\ln h_{i}^{*})^{2k+1} + A_{2k+1}^{-}(\ln \hat{h}_{j})^{2k+1} + n(\ln h_{i}^{*}-\ln \hat{h}_{j})\right)  \right)\\
 =& \sum_{l,k=0}^{\infty} \tau^{(n)}_{k,l} (\ln h_{i}^{*})^{k} (\ln \hat{h}_{j})^{l}, \quad n=0,1,
\end{aligned}
\end{equation}
with
\begin{equation}\label{tau0}
	\tau_{k,l}^{(n)} =\frac{1}{k!}(\partial_{\ln h_{i}^{*}})^{k} \frac{1}{l!}(\partial_{\ln \hat{h}_{j}})^{l} \hat{B}_{i,j}^{(n)}(h_{i}^{*}, \hat{h}_{j}) |_{h_{i}^{*}=\hat{h}_{j}=1}.
\end{equation}
Thus, for the the expression \eqref{mij1} of $M_{i,j}^{(n)}$ $(n=0,1)$, we can calculate
\begin{equation}\label{mij2}
\begin{aligned}
	\dfrac{ M_{i,j}^{(n)}}{2\exp\left( \sum_{k=0}^{\infty}\left( A_{2k}^{+}(\ln h_{i}^{*})^{2k} + A_{2k}^{-}(\ln \hat{h}_{j})^{2k}\right)  \right) } = &\hat{B}_{i,j}^{(n)}(h_{i}^{*}, \hat{h}_{j}) -\hat{B}_{i,j}^{(n)}\left( h_{i}^{*}, \frac{1}{\hat{h}_{j}}\right)  -\hat{B}_{i,j}^{(n)}\left( \frac{1}{h_{i}^{*}}, \hat{h}_{j}\right)  +\hat{B}_{i,j}^{(n)}\left( \frac{1}{h_{i}^{*}}, \frac{1}{\hat{h}_{j}}\right) \\
	=&4\, \sum_{l,k=0}^{\infty} \tau^{(n)}_{2k+1,2l+1} (\ln h_{i}^{*})^{2k+1} (\ln \hat{h}_{j})^{2l+1}.
\end{aligned}	
\end{equation}

For simplification, we define some new notations:
\begin{equation}\label{xpm}
\begin{aligned}
	&\mathbf{x}^{\pm}(n)=\left( x_{1}^{\pm}(n), 0, x_{3}^{\pm}, 0, \ldots, 0, x_{2k+1}^{\pm}, 0, \ldots \right), \quad \mathbf{s}=(0, s_{2}, 0, s_{4}, 0,\ldots, 0, s_{2k}, 0,\ldots),\\
	&x_{1}^{+}(n)=A_{1}^{+} + n = x+2\ii t+ n + a_{1}, \quad x_{1}^{-}(n)=A_{1}^{-} - n = x-2\ii t- n + a_{1}^{*},\\
	&x_{2k+1}^{+}=A_{2k+1}^{+}=\frac{x+2^{2k+1}\ii t}{(2k+1)!}+a_{2k+1}, \quad x_{2k+1}^{-}=A_{2k+1}^{-}=(x_{2k+1}^{+})^{*}, \quad k\geq 1,
\end{aligned}
\end{equation}
where $a_{2k+1}=\hat{a}_{2k+1}^{*}$ $(k\geq 0)$ are given in Eq. \eqref{ak1}. By applying a coordinate transformation $x\rightarrow x+\Re{(a_{1})}$, $ t\rightarrow t+\frac{\Im{(a_{1})}}{2}$, we can eliminate the terms of $a_{1}$ and $a_{1}^{*}$ in $x_{1}^{\pm}(n)$, where $\Re{(a_{1})} $ and $\Im{(a_{1})}$ represent the real and imaginary parts of $a_{1}$, respectively. Thus, we can assume $a_{1}=0$ without loss of generality.

Now, combining the above formulas \eqref{qnfor2}-\eqref{mij1} and \eqref{bij4}-\eqref{mij2}, we can generate the formula of $ N $th-order rogue wave solution for the NLS equation \eqref{NLS}, as follows:
\begin{equation}\label{qn1}
\begin{aligned}
	&q^{[N]}(x,t)=\frac{\tau^{(1)}}{\tau^{(0)}}e^{2\ii t}, \quad \tau^{(n)}=\det_{1\leq i,j\leq N}\left( \tau_{2i-1,2j-1}^{(n)}\right), \\
	& \tau_{i,j}^{(n)} =(-1)^{n} \sum_{v=0}^{\min(i,j)} \frac{1}{4^{v}}S_{i-v}(\mathbf{x}^{+}(n)+v\mathbf{s}) S_{j-v}(\mathbf{x}^{-}(n)+v\mathbf{s}),
\end{aligned}
\end{equation}
where the Schur polynomial $ S_{k}(\mathbf{x}) $ with $ \mathbf{x}=(x_{1},x_{2},\ldots) $ is defined by
\begin{equation}\label{schur}
	\sum_{k=0}^{\infty}S_{k}(\mathbf{x})\epsilon^{k}=\exp\left( \sum_{k=1}^{\infty} x_{k}\epsilon^{k}\right). 
\end{equation}

Note that if $q^{[N]}(x,t)$ is the solution of NLS equation \eqref{NLS}, so is $-q^{[N]}(x,t)$. Thus, for the convenience of analysis later, we ignored the term $(-1)^{n}$ of $\tau_{i,j}^{(n)}$ for the rogue wave solution formula \eqref{qn1} in the later text. Moreover, {when internal parameters $a_{2k+1}=0 $ $(k\geq 0)$, the amplitude of the $N$th-order fundamental rogue wave solution $q^{[N]}(x,t)$ \eqref{qn1} reaches its maximum value at the origin $ (x,t)=(0,0) $.} For $ N=1 $, we obtain the first-order rogue wave solution $q^{[1]}(x,t)=\hat{q}^{[1]}(x,t)\ee^{2\ii t}$ of NLS equation \eqref{NLS} with
\begin{equation}\label{q1}
\hat{q}^{[1]}(x,t)= 1-\frac{4(4\ii t +1)}{4x^{2}+16t^{2}+1}.
\end{equation} 

\subsection{Adler--Moser polynomials and their root structures}\label{subsec22}

In 1978, Adler and Moser \cite{am1978} constructed the special polynomials (i.e., Adler--Moser polynomials) to generate rational solutions of the Korteweg-de Vries (KdV) equation. The Adler--Moser polynomials $ \Theta_{N}(z) $ can be expressed as the following determinant \cite{am2009}:
\begin{equation}\label{amp}
\Theta_{N}(z)= c_{N} \det_{1\leq i,j\leq N}\left( \theta_{2i-j}(z)\right),
\end{equation}
where $ c_{N}=\prod_{k=1}^{N}(2k-1)!! $, all $ \theta_{k}(z) $'s are the special Schur polynomials defined by
\begin{equation}\label{schur2}
\sum_{k=0}^{\infty}\theta_{k}(z)\epsilon^{k}=\exp\left( z\epsilon+ \sum_{j=1}^{\infty} \kappa_{j}\epsilon^{2j+1}\right),
\end{equation}
$ \theta_{k}(z)=0 $ for $ k<0 $, $ \theta_{k+1}'(z)=\theta_{k}(z) $, and $ \kappa_{j} $ $ (j\geq1) $ are arbitrary complex constants. Since $  \theta_{k}(z) $ is the $ k $-order polynomial, we can determine that the degree of the polynomial $ \Theta_{N}(z) $ is $ \frac{N(N+1)}{2} $. Especially, if there is only one nonzero complex parameter $ \kappa_{m}=-\frac{2^{2m}}{2m+1} $, $ \Theta_{N}(z) $ can be reduced to the following Yablonskii--Vorob'ev polynomial hierarchy \cite{yv2003}:
\begin{equation}\label{yvpm}
Q_{N}^{[m]}(z)=c_{N} \det_{1\leq i,j\leq N}\left( p_{2i-j}^{[m]}(z) \right), \quad  p_{k}^{[m]}(z)= \theta_{k}(z;\kappa_{m}),
\end{equation}
where $ \theta_{k}(z;\kappa_{m}) $ represents $ \theta_{k}(z) $ with only one parameter $ \kappa_{m} $. 

Root structures of the Yablonskii--Vorob'ev polynomial hierarchy $Q_{N}^{[m]}(z)$ have been studied in Refs. \cite{yang2021, Taneda2000, Fukutani2000, yv2003, Buckingham2014, Balogh2016}. From these literature, it is found that $Q_{N}^{[m]}(z)$ has $\Gamma_{0}$-multiple zero root and $\Gamma_{N}$ nonzero roots, where
\begin{equation}\label{gamma}
     \begin{aligned}     
     &\Gamma_{0}=\frac{N_{0}(N_{0}+1)}{2}, \quad \Gamma_{N}=\frac{1}{2}\left(N(N+1)-N_{0}(N_{0}+1)\right),\\
     &N_{0}= \left\{
	\begin{array}{ll}
		N\mod{(2m+1)}, & 0\leq N \mod (2m+1) \leq m,\\ 
            2m- (N\mod (2m+1)), & N \mod (2m+1) > m,
	\end{array}
	\right.
     \end{aligned}
\end{equation}
and the notation $ (N\mod (2m+1)) $ represents the remainder of $N$ divided by $2m+1$. For the Yablonskii--Vorob'ev polynomial $Q_{N}^{[1]}(z)$, all nonzero roots of are simple, as shown in Ref. \cite{Fukutani2000}. Additionally, Clarkson \textit{et al}. \cite{yv2003} proposed a conjecture that all nonzero roots of the Yablonskii--Vorob'ev polynomial hierarchy $Q_{N}^{[m]}(z)$ $(m> 1)$ are simple. As this conjecture holds in all examples presented in this paper, we assume its validity here. 

Next, we will discuss the root structures of Adler--Moser polynomial $ \Theta_{N}(z) $ with multiple nonzero parameters, which is crucial for our subsequent investigation of rogue wave patterns. For the NLS equation \eqref{NLS}, Yang \textit{et al}. \cite{yang2021} studied {rogue} wave patterns corresponding to root structures of the polynomials $ Q_{N}^{[m]}(z) $ in detail, including triangle pattern, pentagon pattern, heptagon pattern, and so on. Furthermore, they have recently explored rogue wave patterns associated with the root structures of the polynomials $ \Theta_{N}(z) $ \cite{yang2024}, where $ \Theta_{N}(z) $ involve multiple nonzero complex parameters $ \kappa_{j} $. However, they only explored the specific scenario where all roots of $ \Theta_{N}(z) $ are simple. Now, we will discuss the cases of $ \Theta_{N}(z) $ existing multiple roots. Note that the polynomials $ \Theta_{N}(z) $ discussed in this article all possess at least $ 2 $ nonzero complex parameters $ \kappa_{j}$.

\begin{figure*}[!htbp]
\centering
\includegraphics[width=\textwidth]{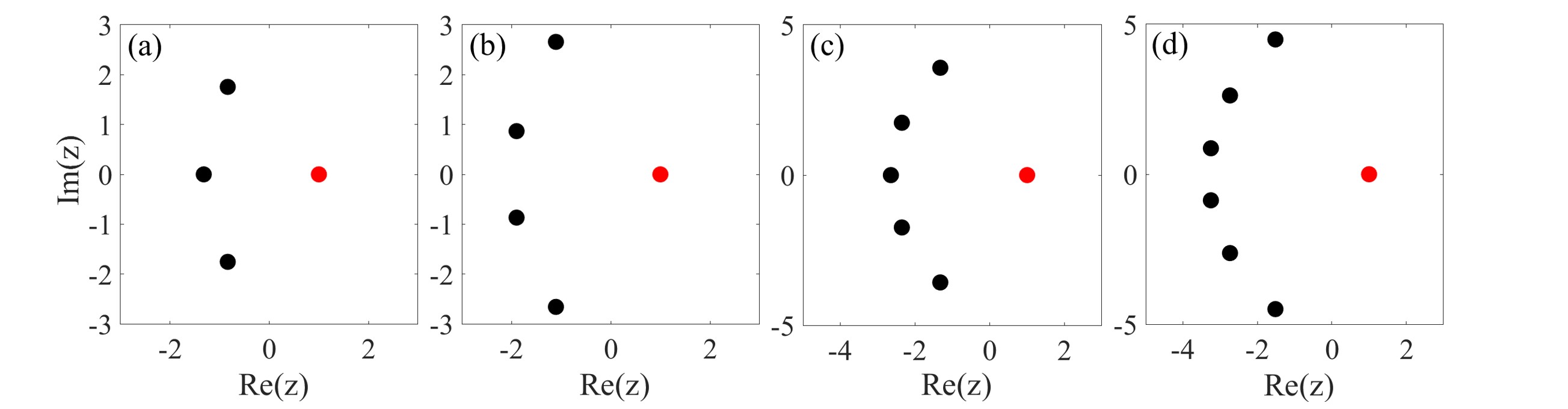}
\caption{The claw-like root structures of the Adler--Moser polynomials $ \Theta_{N}(z) $ with $ N=3, 4, 5 ,6$, $ \kappa_{j}=\frac{z_{0}^{2j+1}}{2j+1} $ $ (1\leq j \leq N-1) $, and $  z_{0}=1$.  (a) $ \Theta_{3}(z) $ has a nonzero triple root $ z_{0}=1 $. (b) $ \Theta_{4}(z) $ has a nonzero sixfold root $ z_{0}=1 $. (c) $ \Theta_{5}(z) $ has a nonzero decuple root $ z_{0}=1 $. (d) $ \Theta_{6}(z) $ has a nonzero fifteen-multiple root $ z_{0}=1 $. Black points represent single roots, while red points represent multiple roots.}
\label{Fig1}
\end{figure*}

Here, we present the first few Adler--Moser polynomials
\begin{equation}\label{ampf}
\begin{aligned}
	\Theta_{1}(z)=& z, \quad \Theta_{2}(z)=z^{3}-3\kappa_{1}, \quad \Theta_{3}(z)= z^{6} -15\kappa_{1}z^{3} +45\kappa_{2}z-45\kappa_{1}^{2},\\
	\Theta_{4}(z)=& z^{10}-45\kappa_{1}z^{7} +315\kappa_{2}z^{5} -1575\kappa_{3}z^{3} +4725\kappa_{1}\kappa_{2}z^{2} -4725\kappa_{1}^{3}z -4725\kappa_{2}^{2} +4725\kappa_{1}\kappa_{3}.
\end{aligned}
\end{equation}
Then, we numerically study the polynomial $ \Theta_{N}(z) $ with parameters $ (\kappa_{1}, \kappa_{2}, \ldots, \kappa_{N-1}) $ defined by
\begin{equation}\label{kap0}
	\kappa_{j}=\frac{z_{0}^{2j+1}}{2j+1}, \quad z_{0}\in \mathbb{C}\setminus \{0\}, \quad 1\leq j\leq N-1, 
\end{equation} 
and find that this polynomial has only one nonzero $ \frac{N(N-1)}{2} $-multiple root $ z_{0} $ and $ N $ nonzero single roots, as shown in Fig. \ref {Fig1}. Based on this numerical evidence, we suppose in this paper that when the parameters $ (\kappa_{1}, \kappa_{2}, \ldots, \kappa_{N-1}) $ are defined by Eq. \eqref{kap0}, the polynomial $ \Theta_{N}(z) $ $ (N\geq 3) $ has one nonzero $ \frac{N(N-1)}{2} $-multiple root and $ N $ nonzero single roots. We refer to this type of root structure of $ \Theta_{N}(z) $ as the claw-like structure. 

\begin{figure*}[!htbp]
\centering
\includegraphics[width=0.8\textwidth]{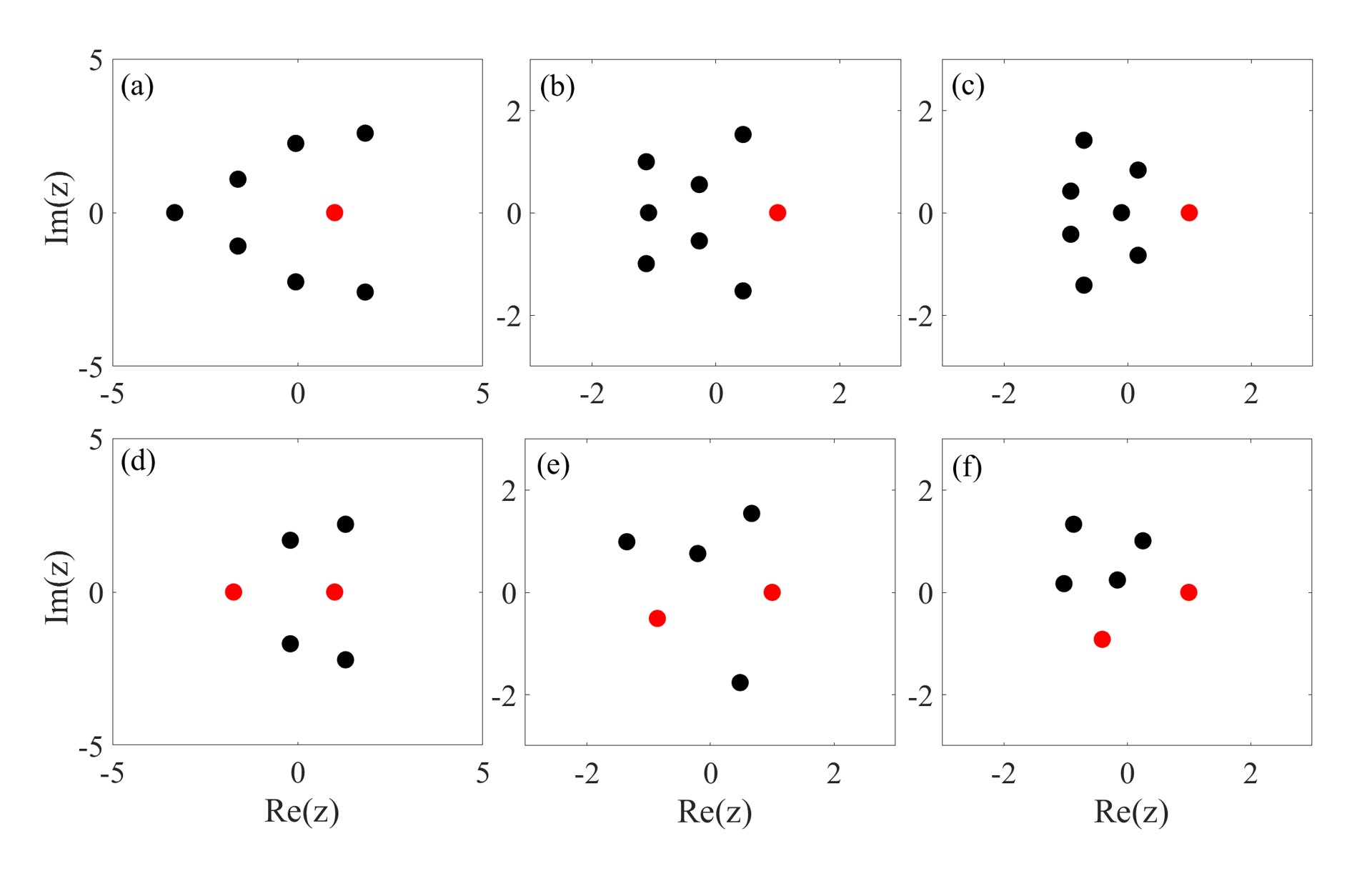}
\caption{In the first row, these are the OTR root structures of the Adler--Moser polynomial $ \Theta_{4}(z) $ with only one triple root $ z_{0}=1 $. From $ (a) $ to $ (d) $: $(\kappa_{1}, \kappa_{2}, \kappa_{3})$ $ = $ $(-\frac{2}{3}, \frac{1}{5}, \frac{1}{7})$, $(0, -\frac{1}{45}, -\frac{1}{189})$, $(\frac{3\sqrt{5}-5}{30}, 0, \frac{-7\sqrt{5}+15}{350}) $. In the second row, these are the TTR root structures of the Adler--Moser polynomial $ \Theta_{4}(z) $ with two distinct nonzero triple roots $(z_{0,1},z_{0,2})$. From $(d)$ to $(f)$: $(z_{0,1},z_{0,2})$ $ \approx $ $ (-1.73,1) $, $ (-0.86-0.51\ii,1) $, $ (-0.40-0.92\ii,1) $, and $(\kappa_{1}, \kappa_{2}, \kappa_{3}) $ $ \approx $ $ (-0.32,-0.28,0.063)$, $  (-0.029+0.028\ii, -0.032+0.0077\ii, -0.0052-0.0016\ii) $, $ (-0.060+0.098\ii,-0.0011+0.0045\ii,0.0016+0.0013\ii) $. In all panels, the black points represent single roots, while red points represent multiple roots.}
\label{Fig2}
\end{figure*}

Furthermore, when appropriate parameters $ (\kappa_{1}, \kappa_{2}, \ldots, \kappa_{N-1}) $ are chosen, the polynomial $ \Theta_{N}(z) $ $ (N\geq 4) $ exhibits other structures containing multiple roots. For example, when
\begin{equation}\label{kap1}
	\kappa_{1}\ne \frac{z_{0}^{3}}{3}, \quad \kappa_{2}=-\frac{z_{0}^{6}-15\kappa_{1}z_{0}^{3}-45\kappa_{1}^{2}}{45z_{0}}, \quad \kappa_{3}=-\frac{z_{0}^{9} -189\kappa_{1}^{2}z_{0}^{3} -189\kappa_{1}^{3}}{189z_{0}^{2}},
\end{equation}
and  $ z_{0}\in \mathbb{C}\setminus \{0\} $, the polynomial $ \Theta_{4}(z) $ has at least one nonzero triple root $ z=z_{0} $, as show in Fig. \ref {Fig2}. In particular, when $ \Theta_{4}(z) $ has only one nonzero triple root $ z=z_{0} $, we refer to this type of root structure of $ \Theta_{4}(z) $ as the OTR (one triple root) structure. By selecting different parameters $(\kappa_{1},\kappa_{2},\kappa_{3})$ satisfying the condition \eqref{kap1}, we find that the OTR root structure of $ \Theta_{4}(z) $ exhibits only three distinct distribution models, along with varying degrees of translation and rotation. We illustrate these three models of the OTR root structure in Figs. \ref {Fig2} $(a)$-$(c)$ with $ z_{0}=1 $. 
Additionally, if 
\begin{equation}\label{kap2}
	\kappa_{1}=\frac{{z}_{0,2}^{3}}{105}\left( 49\left( \frac{z_{0,1}}{{z}_{0,2}}\right) ^{5} +188\left( \frac{z_{0,1}}{{z}_{0,2}}\right)^{4} +313\left( \frac{z_{0,1}}{{z}_{0,2}}\right)^{3} +308\left( \frac{z_{0,1}}{{z}_{0,2}}\right)^{2} +178\left( \frac{z_{0,1}}{{z}_{0,2}}\right) \right), 
\end{equation}
$ \kappa_{2} $ and  $ \kappa_{3} $ satisfy Eq. \eqref{kap1} with $z_{0}=z_{0,1}$, and $ z_{0,i}\ne0 $ $( i=1,2)$ satisfy
\begin{equation}\label{kap3}
	49\left( \frac{z_{0,1}}{{z}_{0,2}}\right)^{6} +237\left( \frac{z_{0,1}}{{z}_{0,2}}\right)^{5} +501\left( \frac{z_{0,1}}{{z}_{0,2}}\right)^{4} +631\left( \frac{z_{0,1}}{{z}_{0,2}}\right)^{3} +501\left( \frac{z_{0,1}}{{z}_{0,2}}\right)^{2} +237\left( \frac{z_{0,1}}{{z}_{0,2}}\right) +49=0,
\end{equation}
then $ \Theta_{4}(z) $ has two distinct nonzero triple roots that are $ z=z_{0,1} $ and $ z={z}_{0,2} $. Here, we call this type of root structure the TTR (two triple roots) structure for the polynomial $ \Theta_{4}(z) $. Similarly, by taking different parameters $(\kappa_{1},\kappa_{2},\kappa_{3})$ satisfying Eqs. \eqref{kap2} and \eqref{kap3}, we can find that the TTR root structure of $ \Theta_{4}(z) $ possesses only three distinct distribution models, as well as varying degrees of translation and rotation. We plot these three distinct models for the TTR root structure of $ \Theta_{4}(z) $ in Figs. \ref {Fig2} $(d)$-$(f)$.

Next, we will present two propositions of $ \theta_{k}(z) $ for the later proof of rogue wave patterns. For convenience, we denote $ \theta_{k}(z_{0}) $ as $ \theta_{k} $ in the later text and introduce some notations:
\begin{equation}\label{nthe}
	\begin{array}{ll}
		\Theta_{3,1}(z_{0})=\begin{vmatrix}
			\theta_{1} & \theta_{0} & 0 \\
			\theta_{3} & \theta_{2} & \theta_{1} \\
			\theta_{7} & \theta_{6} & \theta_{5}
		\end{vmatrix}, &
		\Theta_{3,2}(z_{0})=\begin{vmatrix}
			\theta_{1} & \theta_{0} & 0 \\
			\theta_{5} & \theta_{4} & \theta_{3} \\
			\theta_{7} & \theta_{6} & \theta_{5}
		\end{vmatrix}, \\
		\Theta_{4,1}(z_{0})=\begin{vmatrix}
			\theta_{1} & \theta_{0} & 0 & 0 \\
			\theta_{3} & \theta_{2} & \theta_{0} & 0 \\
			\theta_{5} & \theta_{4} & \theta_{2} & \theta_{1} \\
			\theta_{7} & \theta_{6} & \theta_{4} & \theta_{3}
		\end{vmatrix}, &
		\Theta_{4,2}(z_{0})=\begin{vmatrix}
			\theta_{1} & \theta_{0} & 0 & 0 \\
			\theta_{3} & \theta_{2} & \theta_{0} & 0 \\
			\theta_{5} & \theta_{4} & \theta_{2} & \theta_{0} \\
			\theta_{7} & \theta_{6} & \theta_{4} & \theta_{2}
		\end{vmatrix}.
	\end{array}
\end{equation}

\begin{prop}\label{pro1a}
	If $ \kappa_{j}=\frac{z_{0}^{2j+1}}{2j+1}$ $ (j\geq 1) $ and $ z_{0}\in \mathbb{C}\setminus \{0\} $, then the equations 
	\begin{equation}\label{the2j}
		\begin{vmatrix}
			\theta_{1} & \theta_{0} \\
			\theta_{2j+1} & \theta_{2j}
		\end{vmatrix}= 0, \quad j\geq 1,
	\end{equation}
	are satisfied.
\end{prop}
This Proposition will be proved in Sec. \ref{ssec1a}.

\begin{prop}\label{pro1}
If $ \kappa_{1}\ne \frac{z_{0}^{3}}{3} $, $ z_{0}\in \mathbb{C}\setminus \{0\} $, and $ (\kappa_{2}, \kappa_{3}) $ are given in Eq. \eqref{kap1}, then the determinants $ \Theta_{3}(z_{0}) $, $ \Theta_{3,i}(z_{0}) $ $(i=1,2)$, and $ \Theta_{4,1}(z_{0}) $ are all equal to zero, and 
\begin{equation}\label{the2}
	\begin{aligned}
		(\Theta_{3}(z_{0}))'\ne0, \quad \Theta_{4,2}(z_{0})\ne 0, \quad
		\begin{vmatrix}
			\theta_{1} & \theta_{0} \\
			\theta_{2j+1} & \theta_{2j}
		\end{vmatrix}\ne 0, \quad 1\leq j\leq 3,
	\end{aligned}
\end{equation}
where $\Theta_{3,i}(z_{0}), \Theta_{4,1}(z_{0})$ and $\Theta_{4,2}(z_{0})$ are defined by Eq. \eqref{nthe}.
\end{prop}
The proof of this Proposition will be presented in Sec. \ref{ssec1}.

\section{Asymptotics of rogue wave solutions with multiple internal large parameters}\label{Sec3}

It is evident that rogue wave solution $ q^{[N]}(x,t) $ \eqref{qn1} of NLS equation \eqref{NLS} contain $ N-1 $ free internal complex parameters $ a_{2j+1} $ $ (1\leq j\leq N-1) $. When only one of these parameters $ a_{2j+1} $ in the rogue wave solution $ q^{[N]}(x,t) $ of NLS equation \eqref{NLS} is large, the rogue wave patterns correspond to root structures of the Yablonskii--Vorob’ev polynomials, as detailed in \cite{yang2021}. Furthermore, when multiple parameters among $ (a_{3}, a_{5}, \ldots, a_{2N-1}) $ are large, the corresponding rogue wave patterns are related to the root structures of the Adler--Moser polynomials. The scenario where the Adler--Moser polynomials exclusively possess single roots has been studied in Ref. \cite{yang2024}. This section will explore the relationship between patterns of the rogue wave solution $ q^{[N]}(x,t) $ \eqref{qn1} and the root structures of the Adler--Moser polynomial $ \Theta_{N}(z) $ with multiple roots.

\subsection{Asymptotics of the claw-like and modified claw-like rogue wave patterns}\label{subsec3.1}

For the rogue wave solution $ q^{[N]}(x,t) $ $ (N\geq 3) $, when taking the internal large parameters
\begin{equation}\label{para0}
	a_{2j+1}=\kappa_{j,2j+1}A^{2j+1}+\kappa_{j,1}A, \quad 1\leq j\leq N-1 , \quad |A|\gg 1,
\end{equation}
with large complex constant $ A $, free complex constant $ \kappa_{j,1} $, and $\kappa_{j,2j+1}=\kappa_{j}$ given in Eq. \eqref{kap0}, we can obtain the rogue wave patterns: claw-like pattern and modified claw-like pattern, by choosing appropriate values of $ \kappa_{j,1} $. Based on the proof of Theorem 3 below, we can conclude that the parameter $ \kappa_{N-1,1} $ does not affect the structure of the rogue wave patterns. Thus, we set $ \kappa_{N-1,1}=0 $ without loss of generality. 

In addition, the structures of the claw-like and modified claw-like patterns for rogue wave solution $ q^{[N]}(x,t) $ \eqref{qn1} are related to the root structures of the Adler--Moser polynomial $ \Theta_{N}(z) $, where the parameters $ (\kappa_{1}, \kappa_{2}, \ldots, \kappa_{N-1}) $ of $ \Theta_{N}(z) $ are given in \eqref{kap0}, and $ \Theta_{N}(z) $ has a nonzero $ \frac{N(N-1)}{2} $-multiple root and $ N $ single roots. According to the locations of single roots and multiple roots of $ \Theta_{N}(z) $, we categorize the rogue wave patterns into two regions: the single-root region and the multiple-root region. Then, we will analyze asymptotic behaviors of such two types of patterns in the theorems below. In particular, these patterns exhibit the same asymptotic behavior in the single-root region. However, their asymptotic behaviors and structures differ in the multiple-root region and are influenced by the parameters $ \kappa_{j,1} $ $(1\leq j\leq N-2)$ in Eq. \eqref{para0}. 

\begin{theorem}[Single-root region \cite{yang2024}]\label{Theo1}
	For the high-order rogue wave solution $ q^{[N]}(x,t) $ \eqref{qn1} with $ N\geq3 $, let the internal large parameters $ (a_{3}, a_{5}, \ldots, a_{2N-1} ) $ be defined by Eq. \eqref{para0}, and $ (x_{0}, t_{0}) $ and $ (\check{x}_{0}, \check{t}_{0}) $ be the $ \frac{N(N-1)}{2} $-multiple root and the single root of the Adler--Moser polynomial $ \Theta_{N}(A^{-1}(x+ 2\ii t)) $ with the parameters \eqref{kap0}, respectively. When $ \sqrt{(x-x_{0})^{2}+(t-t_{0})^{2}}> \cO(|A|) $ and $ |A|\gg 1 $, the rogue wave solution $ q^{[N]}(x,t) $ \eqref{qn1} with arbitrary complex parameters $ \kappa_{j,1} $ $ (1\leq j\leq N-2) $ in the internal large parameters \eqref{para0} asymptotically separate into $ N $ first-order rogue waves $ \hat{q}^{[1]}(x-\check{x}_{0}, t-\check{t}_{0})\ee^{2\ii t} $, where $ \hat{q}^{[1]}(x,t) $ is given in Eq. \eqref{q1}. 
 
 Likewise, when $ \sqrt{(x-\check{x}_{0})^{2}+(t-\check{t}_{0})^{2}}= \cO(1) $ and $ |A|\gg 1 $, the solution $ q^{[N]}(x,t) $ \eqref{qn1} has the asymptotics:
	\begin{equation}\label{asym0}
		q^{[N]}(x,t) = \hat{q}^{[1]}(x-\check{x}_{0}, t-\check{t}_{0})\ee^{2\ii t} +\cO(|A|^{-1}).
	\end{equation}
\end{theorem}
The detailed proof of Theorem \ref{Theo1} refers to Ref. \cite{yang2024}.

\begin{theorem}[Multiple-root region]\label{Theo2}

For the high-order rogue wave solution $ q^{[N]}(x,t) $ \eqref{qn1} with $ N\geq3 $, let the internal large parameters $ (a_{3}, a_{5}, \ldots, a_{2N-1} ) $ be defined by Eq. \eqref{para0}, and $z_{0}=A^{-1}(x_{0} +2\ii t_{0}) $ be the $ \frac{N(N-1)}{2} $-multiple root of the Adler--Moser polynomial $ \Theta_{N}(z) $ with the parameters \eqref{kap0}. In the multiple-root region with $ \sqrt{(x-x_{0})^{2}+(t-t_{0})^{2}}\leq \cO(|A|) $, based on different values of $ (\kappa_{1,1}, \kappa_{2,1}, \ldots, \kappa_{N-2,1}) $ in the internal large parameters, the solution $ q^{[N]}(x,t) $ \eqref{qn1} admits the following asymptotics:
\begin{itemize}
    \item[(1).] If taking the parameters
    \begin{equation}\label{para01}
		\kappa_{j,1}=-\frac{\Re(z_{0}\ee^{\ii \arg A}) +2^{2j}\ii \Im(z_{0}\ee^{\ii \arg A})}{(2j+1)!}\ee^{-\ii \arg A},  \quad 1\leq j \leq N-2,
	\end{equation}
    with $\arg A$ being the principal value of the argument of $A$, then when $ \sqrt{(x-x_{0})^{2}+(t-t_{0})^{2}}= \cO(1) $ and $ |A|\gg 1 $, the claw-like pattern of the solution $ q^{[N]}(x,t) $ \eqref{qn1} asymptotically approaches to an $ (N-1) $th-order fundamental rogue wave $ \hat{q}^{[N-1]}(x-x_{0}, t-t_{0})\ee^{2\ii t} $, and exists the asymptotic expression:
	\begin{equation}\label{asym1}
		q^{[N]}(x,t) = \hat{q}^{[N-1]}(x-{x}_{0}, t-{t}_{0})\ee^{2\ii t} +\cO(|A|^{-1}),
	\end{equation}
    where $\hat{q}^{[N-1]}(x, t)={q}^{[N-1]}(x, t) \ee^{-2\ii t} $, and ${q}^{[N-1]}(x, t)$ is given by the formula \eqref{qn1} with all internal large parameters being zero.
	
	\item[(2).] If the parameters $ \kappa_{j,1} $ $ (1\leq j \leq m-1, 1\leq m \leq N-2) $ satisfy Eq. \eqref{para01} but $\kappa_{m,1}$ does not, then when $ |A|\gg 1 $ and $ \sqrt{(x-x_{0})^{2}+(t-t_{0})^{2}}\leq \cO(|A|) $, the modified claw-like pattern of the solution $ q^{[N]}(x,t) $ \eqref{qn1} asymptotically split into $ \Gamma_{N-1} $ first-order rogue waves $ \hat{q}^{[1]}(x-\tilde{x}_{0},t-\tilde{t}_{0})\ee^{2\ii t} $ and an $N_{0}$-order fundamental rogue wave $ \hat{q}^{[N_{0}]}(x-x_{0}, t-t_{0})\ee^{2\ii t} $, where ${q}^{[N_{0}]}(x, t)$ is given by the formula \eqref{qn1} with all internal large parameters being zero, $ \hat{q}^{[1]}(x,t) $ is given in Eq. \eqref{q1}, $  (\tilde{x}_{0},\tilde{t}_{0}) $ are determined by
	\begin{equation}\label{xt2}
		\tilde{x}_{0}+2\ii \,\tilde{t}_{0} =\bar{z}_{0}\left[ -\frac{2m+1}{2^{2m}}\left(\frac{\Re(z_{0}A)+2^{2m}\ii \Im(z_{0}A)}{(2m+1)!} +\kappa_{m,1} A \right) \right]^{1/(2m+1))} +z_{0}A,
	\end{equation}
    $ \bar{z}_{0} $ is the root of the Yablonskii--Vorob'ev polynomial $ Q_{N-1}^{[m]}(z) $, and $\Gamma_{N-1}$ and $N_{0}$ are defined by Eq. \eqref{gamma} with $N$ replaced by $N-1$. 

    Likewise, when $ |A|\gg 1 $ and $ \sqrt{(x-\tilde{x}_{0})^{2}+(t-\tilde{t}_{0})^{2}}= \cO(1) $, the solution $ q^{[N]}(x,t) $ \eqref{qn1} has the asymptotics:
	\begin{equation}\label{asym2}
		q^{[N]}(x,t) = \hat{q}^{[1]}(x-\tilde{x}_{0},t-\tilde{t}_{0})\ee^{2\ii t} +\cO(|A|^{-1/(2m+1)}).
	\end{equation}
    When $ |A|\gg 1 $ and $ \sqrt{(x-x_{0})^{2}+(t-t_{0})^{2}}= \cO(1) $, the solution $ q^{[N]}(x,t) $ admits the asymptotics:
    \begin{equation}\label{asym21}
		q^{[N]}(x,t) = \hat{q}^{[N_{0}]}(x-{x}_{0},t-{t}_{0})\ee^{2\ii t} +\cO(|A|^{-1}).
	\end{equation}

\end{itemize}
\end{theorem}
We will prove this Theorem in Sec. \ref{proof.Theo2}.

From the corresponding results of Ref. \cite{yang2024} and the proof of Theorem \ref{Theo2}, we can easily find that when $ |A|\rightarrow \infty $ and $ (x,t) $ is not near the locations of the first-order rogue waves and the lower-order rogue wave in the above claw-like and modified claw-like patterns, the rogue wave solution $ q^{[N]}(x,t) $ \eqref{qn1} asymptotically approaches to the plane wave background $ \ee^{2\ii t} $. Particularly, if $N_{0}=0$ in case $(2)$ of Theorem \ref{Theo2}, then when $ |A|\rightarrow \infty $, the rogue wave solution $ q^{[N]}(x,t) $ \eqref{qn1} also asymptotically approaches to the plane wave background $ \ee^{2\ii t} $ at the position $(x_{0}, t_{0})$ of the $ (x,t) $-plane.

Furthermore, based on the asymptotics of the claw-like and modified claw-like patterns given in Theorems \ref{Theo1}-\ref{Theo2} for the rogue wave solution $ q^{[N]}(x,t) $ \eqref{qn1}, where the internal large parameters $ (a_{3}, a_{5},\ldots, a_{2N-1}) $ are defined by Eq. \eqref{para0}, we can provide more concise asymptotic expression as follows:
\begin{itemize}
	\item[(1).] If the complex parameters $ (\kappa_{1,1}, \kappa_{2,1}, \ldots, \kappa_{N-2,1}) $ are given in Eq. \eqref{para01} in the internal large parameters \eqref{para0}, then when $ |A|\gg 1 $, the solution $ q^{[N]}(x,t) $ \eqref{qn1} has the following asymptotic expression:
	\begin{equation}\label{asymc1}
		\left|q^{[N]}(x,t)\right| = \left|q^{[N-1]}(x-{x}_{0}, t-{t}_{0})\right| +\sum_{(\check{x}_{0}, \check{t}_{0})} \left(\left|{q}^{[1]}(x-\check{x}_{0}, t-\check{t}_{0})\right|-1 \right) +\cO(\left|A\right|^{-1}),
	\end{equation}
	where $({x}_{0}, {t}_{0})$ is the $ \frac{N(N-1)}{2} $-multiple root of the Adler--Moser polynomial $ \Theta_{N}(A^{-1}(x+2\ii t)) $ with free parameters \eqref{kap0}, and $(\check{x}_{0}, \check{t}_{0})$ traverses $ N $ single roots of $ \Theta_{N}(A^{-1}(x+2\ii t)) $.
	
	\item[(2).] If the parameters $ \kappa_{l,1} $ $ (1\leq l \leq m-1, 1\leq m \leq N-2) $ satisfy Eq. \eqref{para01} but $\kappa_{m,1}$ does not in the internal large parameters \eqref{para0}, then when $ |A|\gg 1 $, we obtain the following asymptotic expression:
    \begin{equation}\label{asymc2}
	\begin{aligned}
	    \left|q^{[N]}(x,t)\right| =& \, \left|q^{[N_{0}]}(x-{x}_{0}, t-{t}_{0})\right|+\sum_{(\tilde{x}_{0}, \tilde{t}_{0})}  \left(\left|q^{[1]}(x-\tilde{x}_{0}, t-\tilde{t}_{0})\right|-1 \right) \\
            & + \sum_{(\check{x}_{0}, \check{t}_{0})} \left(\left|q^{[1]}(x-\check{x}_{0}, t-\check{t}_{0})\right|-1 \right) +\cO(|A|^{-1/3}) ,
	\end{aligned}	
	\end{equation}
	where {$ |q^{[0]}(x-{x}_{0}, t-{t}_{0})|=1 $},  $(\check{x}_{0}, \check{t}_{0})$ traverses $ N $ single roots of the Adler--Moser polynomial $ \Theta_{N}(A^{-1}(x+2\ii t)) $ with free parameters \eqref{kap0}, $(\tilde{x}_{0}, \tilde{t}_{0})$ is defined by Eq. \eqref{xt2} with $\bar{z}_{0}$ traversing $ \Gamma_{N-1} $ nonzero single roots of the Yablonskii--Vorob'ev polynomial $ Q_{N-1}^{[m]}(z) $, and $N_{0}$ and $ \Gamma_{N-1} $ is defined by Eq. \eqref{gamma} with $N$ replaced by $N-1$.
\end{itemize}

It is worth noting that by modifying the form \eqref{para0} of the large internal parameters $ (a_{3}, a_{5}, \ldots, a_{2N-1} ) $, we can obtain new modified claw-like patterns of the rogue wave solution $ q^{[N]}(x,t) $, whose asymptotic behaviors are consistent with that of Theorem \ref{Theo1} in the single-root region, but differ that of Theorem \ref{Theo2} in the multiple-root region. For instance, we set the internal large parameters $ (a_{3}, a_{5}, \ldots, a_{2N-1} ) $ of the rogue wave solution $ q^{[N]}(x,t) $ $ (N\geq 3) $ to
\begin{equation}\label{parab1}
	a_{2j+1}=\kappa_{j,2j+1}A^{2j+1} + \hat{\kappa}_{j,1} A^{(2j+1)/3}, \quad a_{2N+1}=\kappa_{N-1,2N-1}A^{2N-1}, \quad 1\leq j\leq N-2,
\end{equation}
or,
\begin{equation}\label{parab2}
	a_{2j+1}=\kappa_{j,2j+1}A^{2j+1}+ \kappa_{j,1}A + \hat{\kappa}_{j,2} A^{(2j+1)/(2N+1)}, \quad a_{2N+1}=\kappa_{N-1,2N-1}A^{2N-1}, \quad 1\leq j\leq N-2, 
\end{equation}
where $ A $ is the large complex constant, $ \hat{\kappa}_{j,i} $ $ (i=1,2) $ are arbitrary free complex constants, and $ \kappa_{j,2j+1}=\kappa_{j} $ and $ \kappa_{j,1} $ are given in Eq. \eqref{kap0} and \eqref{para01}, respectively. Then, we can generate new modified claw-like rogue wave patterns whose structures in the multiple-root region correspond to the root structure of the Adler–Moser polynomial $ \Theta_{N-1}(z) $ with the free parameters $ (\hat{\kappa}_{1,i}, \hat{\kappa}_{2,i}, \ldots, \hat{\kappa}_{N-2,i}) $ $ (i=1,2) $. Moreover, based on the proof of Theorem \ref{Theo2}, we add an extra constant term $ \hat{\kappa}_{j,3}A^{(2j+1)/3(2m+1)} $ $ (\hat{\kappa}_{j,3}\in \mathbb{C}) $ to $ a_{2j+1} $ $ (1\leq j\leq N_{0}-1) $ in the large internal parameters \eqref{para0} for the modified claw-like rogue wave patterns presented in Theorem \ref{Theo2} that contain the $ N_{0} $th-order fundamental rogue wave in the multiple-root region. This allows us to alter the structure of this $ N_{0} $th-order fundamental rogue wave to correspond to the root structure of the Adler–Moser polynomial $ \Theta_{N_{0}}(z) $.

We can analyze asymptotic behaviors of these new modified claw-like patterns similarly to Theorems \ref{Theo1}-\ref{Theo2}. Thus, we will not elaborate further here. But, we will provide several examples in Sec. \ref{Sec4} to shown these modified claw-like rogue wave patterns.

\subsection{Asymptotics of other rogue wave patterns with multiple internal large parameters}\label{subsec3.2}

Next, we will investigate whether there exist other patterns corresponding to other multiple root structures of the Adler--Moser polynomial $ \Theta_{N}(z) $ for the high-order rogue wave solution $ q^{[N]}(x,t) $ \eqref{qn1}. 
Here, we focus on a concrete case of $N=4$. Then, other patterns of arbitrary $ N $th-order rogue wave solutions $ q^{[N]}(x,t) $ \eqref{qn1} can be similarly generated.

According to the discussion in Sec. \ref{subsec22}, the cases of multiple roots in the Adler--Moser polynomial $ \Theta_{4}(z) $ can be categorized into three types: only one nonzero sextuple root, only one nonzero triple root, and two distinct nonzero triple roots. The root structure of only one sextuple root corresponds to the claw-like pattern and modified claw-like pattern of the rogue wave solution $ q^{[4]}(x,t) $. Its asymptotic behaviors are shown in Theorems \ref{Theo1} and \ref{Theo2} above. Now, we provide the asymptotics analysis of rogue wave patterns of $ q^{[4]}(x,t) $ corresponding to two other root structures of $ \Theta_{4}(z) $: OTR structure (only one nonzero triple root) and TTR structure (two distinct nonzero triple roots).

When $ \Theta_{4}(z) $ has only one nonzero triple root $ z_{0} $, we set the internal large parameters
\begin{equation}\label{para2}
	a_{3}=\kappa_{1,3}A^{3} +\kappa_{1,1}A, \quad a_{5}=\kappa_{2,5}A^{5}+\kappa_{2,3}A^{3}, \quad a_{7}=\kappa_{3,7}A^{7}+\kappa_{3,5}A^{5},   \quad |A| \gg 1,
\end{equation}
for the fourth-order rogue wave solution $ q^{[4]}(x,t) $ of the NLS equation \eqref{NLS}, where $ A $ is a large constant, the parameters $ \kappa_{j,2j+1}=\kappa_{j} $ $ (j=1,2,3) $ are defined by Eq. \eqref{kap1}, and $ \kappa_{1} $ does not satisfy Eq. \eqref{kap2}. Then, by utilizing the formula \eqref{qn1} of rogue wave solution and choosing the appropriate complex parameters $ (\kappa_{1,1}, \kappa_{2,3}, \kappa_{3,5}) $, we can obtain the rogue wave patterns of the solution $ q^{[4]}(x,t) $ corresponding to the OTR structure of $ \Theta_{4}(z) $. If $ (\kappa_{1,1},\kappa_{2,3},\kappa_{3,5}) $ satisfy the parameter equation $ \rho(z_{0},\kappa_{1,1},\kappa_{1,3},\kappa_{2,3},\kappa_{3,5})=0 $ with
\begin{equation}\label{para4}
	\rho(z_{0},\kappa_{1,1},\kappa_{1,3},\kappa_{2,3},\kappa_{3,5})=\frac{\Re(z_{0}\ee^{\ii \arg A}) +4\ii \Im(z_{0}\ee^{\ii \arg A})}{6}\ee^{-\ii \arg A}+\kappa_{1,1}  -{\kappa_{2,3}}  \dfrac{3z_{0}({z_{0}}^{3}+6\kappa_{1,3})}{({z_{0}}^{3}-3\kappa_{1,3})^{2}}+\kappa_{3,5} \dfrac{9{z_{0}}^{2}}{({z_{0}}^{3}-3\kappa_{1,3})^{2}},  
\end{equation} 
and $ z_{0} $ is the nonzero triple root of $ \Theta_{4}(z) $, then we can obtain the OTR-type patterns of rogue wave solution $ q^{[4]}(x,t) $. If the parameters $ (\kappa_{1,1},\kappa_{2,3},\kappa_{3,5})$ do not satisfy the parameter equation $ \rho(z_{0},\kappa_{1,1},\kappa_{1,3},\kappa_{2,3},\kappa_{3,5})=0 $, we can generate the modified OTR-type patterns of rogue wave solution $ q^{[4]}(x,t) $. Their asymptotic behaviors are illustrated by the following proposition.

\begin{prop}\label{prop3}
	Let $ z_{0}=A^{-1}(x_{0}+2\ii t_{0}) $ be the only nonzero triple root of the Adler--Moser polynomial $ \Theta_{4}(z) $ with free parameters $ (\kappa_{1}, \kappa_{2}, \kappa_{3}) $ given in Eq. \eqref{kap1} and $ \kappa_{1} $ not satisfying Eq. \eqref{kap2}. When the internal large parameters $ (a_{3}, a_{5}, a_{7}) $ are defined by Eq. \eqref{para2} with $ \kappa_{j,2j+1}=\kappa_{j} $ $ (1\leq j\leq 3) $, the rogue wave solution $ q^{[4]}(x,t) $ of the NLS equation \eqref{NLS} admits the following asymptotics: 
	\begin{enumerate}
	   
		\item[(1).] In the multiple-root region with $ \sqrt{(x-x_{0})^{2}+(t-t_{0})^{2}} \leq \cO(|A|) $. 

  \begin{itemize}
      \item[(a).] If $ (\kappa_{1,1},\kappa_{2,3},\kappa_{3,5}) $ satisfy the parameter equation $ \rho(z_{0},\kappa_{1,1},\kappa_{1,3},\kappa_{2,3},\kappa_{3,5})=0 $ in Eq. \eqref{para4}, then when $ |A|\gg 1 $, the OTR-type patterns of the rogue wave solution $ q^{[4]}(x,t) $ asymptotically approach a second-order fundamental rogue wave $ \hat{q}^{[2]}(x-x_{0}, t-t_{0})\ee^{2\ii t} $, where $\hat{q}^{[2]}(x, t)= {q}^{[2]}(x, t)\ee^{-2\ii t}$ and does not contain large parameters. 
      
      Meanwhile, when $ \sqrt{(x-{x}_{0})^{2}+(t-{t}_{0})^{2}}= \cO(1) $ and $ |A|\gg 1 $, the solution $ q^{[4]}(x,t) $ has the following asymptotics:
		\begin{equation}\label{asym3}
			q^{[4]}(x,t) = \hat{q}^{[2]}(x-x_{0}, t-t_{0})\ee^{2\ii t} +\cO(|A|^{-1}).
		\end{equation}

      \item[(b).] If the parameters $ (\kappa_{1,1},\kappa_{2,3},\kappa_{3,5})$ do not satisfy the parameter equation $ \rho(z_{0},\kappa_{1,1},\kappa_{1,3},\kappa_{2,3},\kappa_{3,5})=0 $ in Eq. \eqref{para4}, the modified OTR-type patterns of the rogue wave solution $ q^{[4]}(x,t) $ asymptotically split into three first-order rogue waves $ \hat{q}^{[1]}(x-\tilde{x}_{0},t-\tilde{t}_{0})\ee^{2\ii t} $, where  $ \hat{q}^{[1]}(x,t) $ is given in Eq. \eqref{q1}, and $  (\tilde{x}_{0},\tilde{t}_{0}) $ are determined by 
      \begin{equation}\label{tixta0}
      	\tilde{x}_{0}+2\ii \tilde{t}_{0} = \bar{z}_{0}A^{-1/3}+z_{0}A,
      \end{equation}
      with the only triple root $z_{0}$ of $ \Theta_{4}(z) $ and the single root $ \bar{z}_{0} $ of the polynomial $ 	Q_{2}(\bar{z}) $. Here, the polynomial $Q_{2}(\bar{z})$ represents the special Adler--Moser polynomial $\Theta_{2}(\bar{z})$ defined by Eq. \eqref{amp} with only one free parameter $ \kappa_{1}=\rho(z_{0},\kappa_{1,1},\kappa_{1,3},\kappa_{2,3},\kappa_{3,5}) $ given in Eq. \eqref{para4}.
      
      Likewise, when $ \sqrt{(x-\tilde{x}_{0})^{2}+(t-\tilde{t}_{0})^{2}}= \cO(1) $ and $ |A|\gg 1 $, the solution $ q^{[4]}(x,t) $ satisfies the following asymptotic expression:
		\begin{equation}\label{asym4}
			q^{[4]}(x,t) = \hat{q}^{[1]}(x-\tilde{x}_{0},t-\tilde{t}_{0})\ee^{2\ii t} +\cO(|A|^{-1/3}).
		\end{equation}
  \end{itemize}

		\item[(2).] In the single-root region with $ \sqrt{(x-x_{0})^{2}+(t-t_{0})^{2}}> \cO(|A|) $, when $ |A|\gg1 $, the OTR-type and modified OTR-type patterns of the solution $ q^{[4]}(x,t) $ all possess seven first-order rogue waves $ \hat{q}^{[1]}(x-\check{x}_{0}, t-\check{t}_{0})\ee^{2\ii t} $, where $ (\check{x}_{0}, \check{t}_{0}) $ is the single root of $ \Theta_{4}(A^{-1}(x+2\ii t)) $. Their asymptotic expressions are consistent with that in Eq. \eqref{asym0}. 
		
		\item[(3).] When $ |A|\rightarrow \infty $, and $ (x,t) $ is not near the locations of the first-order rogue waves and lower-order rogue wave in the OTR-type and the modified OTR-type rogue wave patterns, the solution $ q^{[4]}(x,t) $ asymptotically approaches to the plane wave background $ \ee^{2\ii t} $.
	\end{enumerate}
	
\end{prop}
The proof process is outlined in Sec. \ref{prop3.proof}

On the other hand, when the Adler--Moser polynomial $ \Theta_{4}(z) $ has two distinct nonzero triple roots, we assume that the parameters $ (\kappa_{1}, \kappa_{2}, \kappa_{3}) $ and two nonzero triple roots $ z_{0, i} $ $ (i=1, 2) $ of $ \Theta_{4}(z) $ are determined by Eqs. \eqref{kap1}-\eqref{kap3}, and the internal large parameters $ (a_{3}, a_{5}, a_{7}) $ of the rogue wave solution $ q^{[4]}(x,t) $ for NLS equation \eqref{NLS} are defined by Eq. \eqref{para2}. 
Then, by choosing appropriate parameters $ (\kappa_{1,1},\kappa_{2,3},\kappa_{3,5}) $, we can obtain the patterns of the rogue wave solution $ q^{[4]}(x,t) $ for NLS equation \eqref{NLS} corresponding to TTR structures of $ \Theta_{4}(z) $. If $ (\kappa_{1,1},\kappa_{2,3},\kappa_{3,5}) $ simultaneously satisfy the parameter equations $ \rho(z_{0,i},\kappa_{1,1},\kappa_{1,3},\kappa_{2,3},\kappa_{3,5})=0 $ $ (i=1,2) $ in Eq. \eqref{para4}, then we can generate the TTR-type patterns of the rogue wave solution $ q^{[4]}(x,t) $, {which exist two discrete second-order fundamental rogue waves in the multiple-root region}. If $ (\kappa_{1,1},\kappa_{2,3},\kappa_{3,5}) $ only satisfy one of the parameter equations $ \rho(z_{0,i},\kappa_{1,1},\kappa_{1,3},\kappa_{2,3},\kappa_{3,5})=0 $ $ (i=1,2) $ in Eq. \eqref{para4}, then we can obtain the semi-modified TTR-type patterns of the rogue wave solution $ q^{[4]}(x,t) $, {which features a second-order fundamental rogue wave and a triangle formed by three first-order rogue waves in the multiple-root region}. If $ (\kappa_{1,1},\kappa_{2,3},\kappa_{3,5}) $ do not satisfy the parameter equations $ \rho(z_{0,i},\kappa_{1,1},\kappa_{1,3},\kappa_{2,3},\kappa_{3,5})=0 $ $ (i=1,2) $ in Eq. \eqref{para4}, we can yield the modified TTR-type rogue wave patterns, which features two separate triangles in the multiple-root region, each formed by three first-order rogue waves. Their asymptotic behaviors are presented in the following proposition.

\begin{prop}\label{prop4}
	Let $ z_{0,i}=A^{-1}(x_{0}^{(i)}+2\ii t_{0}^{(i)}) $ $ (i=1,2) $ be two distinct nonzero triple roots of the Adler--Moser polynomial $ \Theta_{4}(z) $ with free parameters $ (\kappa_{1}, \kappa_{2}, \kappa_{3}) $ defined by Eqs. \eqref{kap1}-\eqref{kap3}. When the internal large parameters $ (a_{3}, a_{5}, a_{7}) $are defined by \eqref{para2} with $ \kappa_{j,2j+1}=\kappa_{j} $ $ (j=1,2,3) $, the rogue wave solution $ q^{[4]}(x,t) $ for NLS equation \eqref{NLS} admits the following asympototics: 
	\begin{itemize}
		\item[(1).] In the multiple-root region with $ \sqrt{(x-x_{0}^{(1)})^{2}+(t-t_{0}^{(1)})^{2}}\leq \cO(|A|) $ or  $ \sqrt{(x-x_{0}^{(2)})^{2}+(t-t_{0}^{(2)})^{2}}\leq \cO(|A|) $.
        \begin{enumerate}[(a).]
            \item  If $ (\kappa_{1,1},\kappa_{2,3},\kappa_{3,5}) $ simultaneously satisfy the parameter equations $ \rho(z_{0,i}, \kappa_{1,1}, \kappa_{1,3}, \kappa_{2,3}, \kappa_{3,5})=0 $ $ (i=1,2) $ in Eq. \eqref{para4}, then when $ |A|\gg 1 $, the TTR-type patterns of the rogue wave solution $ q^{[4]}(x,t) $ asymptotically approach two second-order fundamental rogue wave $ \hat{q}^{[2]}(x-x_{0}^{(i)}, t-t_{0}^{(i)}) \ee^{2\ii t} $ with $\hat{q}^{[2]}(x,t)={q}^{[2]}(x,t)\ee^{-2\ii t}$ and not containing large parameters. When $ \sqrt{(x-x_{0}^{(i)})^{2}+(t-t_{0}^{(i)})^{2}}= \cO(1) $ and $ |A|\gg 1 $, the solution $ q^{[4]}(x,t) $ has the asymptotic expressions:
		\begin{equation}\label{asym5}
			q^{[4]}(x,t) = \hat{q}^{[2]}(x-x_{0}^{(i)}, t-t_{0}^{(i)}) \ee^{2\ii t} +\cO(|A|^{-1}), \quad i=1,2.
		\end{equation}
            
            \item If $ (\kappa_{1,1},\kappa_{2,3},\kappa_{3,5}) $ satisfy the parameter equation $ \rho(z_{0}, \kappa_{1,1}, \kappa_{1,3}, \kappa_{2,3}, \kappa_{3,5})=0 $ with $ z_{0}=z_{0,i_{1}} $ but not with $ z_{0}=z_{0,i_{2}} $ in Eq. \eqref{para4}, then when $ |A|\gg 1 $, the semi-modified TTR-type patterns of the solution $ q^{[4]}(x,t) $ asymptotically approach a second-order fundamental rogue wave $ \hat{q}^{[2]}(x-x_{0}^{(i_{1})}, t-t_{0}^{(i_{1})})\ee^{2\ii t} $, and three discrete first-order rogue waves $ \hat{q}^{[1]}(x-\tilde{x}_{0}^{(i_{2})},t-\tilde{t}_{0}^{(i_{2})}) \ee^{2\ii t} $, where $ \hat{q}^{[1]}(x,t) $ is given in Eq. \eqref{q1}, $(\tilde{x}_{0}^{(i_{2})}, \tilde{t}_{0}^{(i_{2})})$ are defined by
            \begin{equation}\label{xt30}
			\tilde{x}_{0}^{(i_{2})}+2\ii \,\tilde{t}_{0}^{(i_{2})} =\bar{z}_{0,i_{2}}A^{1/3} +z_{0,i_{2}}A, 
		\end{equation}
            $ \bar{z}_{0,i_{2}} $ is the root of the polynomial $ {Q}_{2,i_{2}}(\bar{z}) $, $i_{1}\ne i_{2}$, and $i_{1}, i_{2}=1,2$. Here, the polynomial ${Q}_{2,i_{2}}(\bar{z})$ represents the special Adler--Moser polynomial $\Theta_{2}(\bar{z})$ defined by Eq. \eqref{amp} with only one free parameter $ \kappa_{1}=\rho(z_{0,i_{2}},\kappa_{1,1},\kappa_{1,3},\kappa_{2,3},\kappa_{3,5}) $ given in Eq. \eqref{para4}.
            
            Moreover, when $ \sqrt{(x-x_{0}^{(i_{1})})^{2}+(t-t_{0}^{(i_{1})})^{2}}= \cO(1) $ and $ |A|\gg 1 $, the solution $ q^{[4]}(x,t) $ has the asymptotic expression:
            \begin{equation}\label{asym5b}
			q^{[4]}(x,t) = \hat{q}^{[2]}(x-x_{0}^{(i_{1})}, t-t_{0}^{(i_{1})})\ee^{2\ii t} +\cO(|A|^{-1}). 
		\end{equation}
            When $ \sqrt{(x-\tilde{x}_{0}^{(i_{2})})^{2}+(t-\tilde{t}_{0}^{(i_{2})})^{2}}= \cO(1) $ and $ |A|\gg 1 $, the solution $ q^{[4]}(x,t) $ has the asymptotics:
            \begin{equation}\label{asym5c}
			q^{[4]}(x,t) = \hat{q}^{[1]}(x-\tilde{x}_{0}^{(i_{2})},t-\tilde{t}_{0}^{(i_{2})}) \ee^{2\ii t} +\cO(|A|^{-1/3}).
		\end{equation}
            
            \item If $ (\kappa_{1,1},\kappa_{2,3},\kappa_{3,5})$ do not satisfy the parameter equations $ \rho(z_{0,i}, \kappa_{1,1}, \kappa_{1,3}, \kappa_{2,3}, \kappa_{3,5})=0 $ $ (i=1,2) $ in Eq. \eqref{para4}, then when $ |A|\gg 1 $, the modified TTR-type patterns of the solution $ q^{[4]}(x,t) $ asymptotically split into three first-order rogue waves $ \hat{q}^{[1]}(x-\tilde{x}_{0}^{(i)},t-\tilde{t}_{0}^{(i)})\ee^{2\ii t} $ near the point $(x_{0}^{(i)},t_{0}^{(i)})$ of the $(x,t)$-plane, where 
		\begin{equation}\label{xt3}
			\tilde{x}_{0}^{(i)}+2\ii \,\tilde{t}_{0}^{(i)} =\bar{z}_{0,i}A^{1/3} +z_{0,i}A, \quad i=1,2,
		\end{equation}
	    with $ \bar{z}_{0,i} $ defined by Eq. \eqref{xt30}. 
	    
	    Meanwhile, when $ \sqrt{(x-\tilde{x}_{0}^{(i)})^{2}+(t-\tilde{t}_{0}^{(i)})^{2}} = \cO(1) $ and $ |A|\gg 1 $, the solution $ q^{[4]}(x,t) $ admits the following asymptotics:
		\begin{equation}\label{asym6}
			q^{[4]}(x,t) = \hat{q}^{[1]}(x-\tilde{x}_{0}^{(i)},t-\tilde{t}_{0}^{(i)}) \ee^{2\ii t} +\cO(|A|^{-1/3}), \quad i=1,2.
		\end{equation}
  
        \end{enumerate}

	\item[(2).] In the single-root region with $ \sqrt{(x-x_{0}^{(i)})^{2}+(t-t_{0}^{(i)})^{2}}> \cO(|A|) $ $(i=1,2)$, when $ |A|\gg 1 $, the TTR-type, semi-modified TTR-type, and modified TTR-type patterns of the rogue wave solution $ q^{[4]}(x,t) $ all possess four first-order rogue waves $ \hat{q}^{[1]}(x-\check{x}_{0},  {t}-\check{t}_{0})\ee^{2\ii t} $ with the single root $ (\check{x}_{0}, \check{t}_{0}) $ of $ \Theta_{4}(A^{-1}(x+2\ii t)) $. Their asymptotic expressions are consistent with that in Eq. \eqref{asym0}. 
		
	\item[(3).] When $ |A|\rightarrow \infty $, and $ (x,t) $ is not near the locations of the above first-order rogue waves and lower-order rogue waves in the TTR-type, semi-modified TTR-type and modified TTR-type rogue wave patterns, the solution $ q^{[4]}(x,t) $ asymptotically approaches to the plane wave background $ \ee^{2\ii t} $.
	\end{itemize}
\end{prop}

The proof of Proposition \ref{prop4} is similar to that of Proposition \ref{prop3}. Here, we only require simultaneous consideration of the parameters $ (\kappa_{1,1},\kappa_{2,3},\kappa_{3,5}) $ in the internal large parameters $ (a_{3}, a_{5}, a_{7}) $ \eqref{para2} for the solution $ q^{[4]}(x,t) $ near two points $(x_{0}^{(i)}, t_{0}^{(i)}) $ $ (i=1,2) $ defined by Proposition \ref{prop4} on the $(x,t)$-plane. Therefore, we omit details of the proof here.

Now, we can provide more concise asymptotic expressions for the OTR-type, TTR-type, semi-modified TTR-type and their modified patterns of the rogue wave solution $ q^{[4]}(x,t) $, as follows:

\begin{itemize}

\item[(1).] When $z_{0}$ is the only nonzero triple root of the Adler--Moser polynomial $ \Theta_{4}(z) $ with free parameters $ (\kappa_{1},\kappa_{2},\kappa_{3}) $ given in Eq. \eqref{kap1}, and the internal large parameters $ (a_{3}, a_{5}, a_{7}) $ and $ (\kappa_{1,3}, \kappa_{2,5}, \kappa_{3,7}) $ of the rogue wave solution $ q^{[4]}(x,t) $ are defined by Proposition \ref{prop3}, we obtain the following conclusions.
	\begin{enumerate}
	 	\item[(a).] If $ (\kappa_{1,1}, \kappa_{2,3}, \kappa_{3,5}) $ satisfy the parameter equation $ \rho(z_{0},\kappa_{1,1},\kappa_{1,3},\kappa_{2,3},\kappa_{3,5})=0 $ in Eq. \eqref{para4}, then when $ |A|\gg 1 $, the modulus of the solution $ q^{[4]}(x,t) $ can be approximately expressed as
	 	\begin{equation}\label{asymc3}
	 		\left|q^{[4]}(x,t)\right| = \left|q^{[2]}(x-{x}_{0}, t-{t}_{0})\right| +\sum_{(\check{x}_{0}, \check{t}_{0})} \left(\left|q^{[1]}(x-\check{x}_{0}, t-\check{t}_{0})\right|-1\right) +\cO(|A|^{-1}),
	 	\end{equation}
	 	where $z_{0}=A^{-1}({x}_{0}+2\ii {t}_{0})$ is the only nonzero triple root of the Adler--Moser polynomial $ \Theta_{4}(z) $, and $(\check{x}_{0}, \check{t}_{0})$ traverses seven single roots of $ \Theta_{4}(A^{-1}(x+2\ii t)) $.
	 	
	 	\item[(b).] If $ (\kappa_{1,1}, \kappa_{2,3}, \kappa_{3,5}) $ do not satisfy the parameter equation $ \rho(z_{0},\kappa_{1,1},\kappa_{1,3},\kappa_{2,3},\kappa_{3,5})=0 $ in Eq. \eqref{para4}, then when $ |A|\gg 1 $, we have
	 	\begin{equation}\label{asymc4}
	 		\left|q^{[4]}(x,t)\right| = 1+\sum_{(\check{x}_{0}, \check{t}_{0})} \left(\left|q^{[1]}(x-\check{x}_{0}, t-\check{t}_{0})\right|-1\right) +\sum_{(\tilde{x}_{0}, \tilde{t}_{0})} \left(\left|q^{[1]}(x-\tilde{x}_{0}, t-\tilde{t}_{0})\right|-1\right) +\cO(|A|^{-1/3}) ,
	 	\end{equation}
	 	where $(\check{x}_{0}, \check{t}_{0})$ traverses seven single roots of the Adler--Moser polynomial $ \Theta_{4}(A^{-1}(x+2\ii t)) $, and $(\tilde{x}_{0}, \tilde{t}_{0})$ is given in Eq. \eqref{tixta0} with $\bar{z}_{0}$ traversing three single roots of the polynomial $ Q_{2}(\bar{z}) $ defined by Proposition \ref{prop3}.
	\end{enumerate}

\item[(2).]  When $ z_{0,i} $ $ (i=0,1) $ are two distinct nonzero triple roots of the Adler--Moser polynomial $ \Theta_{4}(z) $ with free parameters $ (\kappa_{1}, \kappa_{2}, \kappa_{3}) $ defined by Eq. \eqref{kap1}-\eqref{kap3}, and the internal large parameters $ (a_{3}, a_{5}, a_{7}) $ and  $ (\kappa_{1,3}, \kappa_{2,5}, \kappa_{3,7}) $ of the rogue wave solution $ q^{[4]}(x,t) $ are defined by Proposition \ref{prop4}, we have the following conclusions.
	\begin{enumerate}
		\item[(a).] If$ (\kappa_{1,1},\kappa_{2,3},\kappa_{3,5}) $ simultaneously satisfy the parameter equations $ \rho(z_{0,i}, \kappa_{1,1}, \kappa_{1,3}, \kappa_{2,3}, \kappa_{3,5})=0 $ $ (i=1,2) $ in Eq. \eqref{para4}, then when $ |A|\gg 1 $, the solution $ q^{[4]}(x,t) $ admits the following asymptotic expression:
		\begin{equation}\label{asymc5}
			\left|q^{[4]}(x,t)\right| = 1+ \sum_{i=0}^{1} \left(\left|q^{[2]}(x-{x}_{0}^{(i)}, t-{t}_{0}^{(i)}) \right|-1\right)+\sum_{(\check{x}_{0}, \check{t}_{0})} \left(\left|q^{[1]}(x-\check{x}_{0}, t-\check{t}_{0})\right|-1\right) +\cO(|A|^{-1}),
		\end{equation}
		where $({x}_{0}^{(i)}, {t}_{0}^{(i)})$  $(i=1,2)$ are two triple roots of the Adler--Moser polynomial $ \Theta_{4}(A^{-1}(x+2\ii t)) $, and $(\check{x}_{0}, \check{t}_{0})$ traverses four single roots of the Adler--Moser polynomial $ \Theta_{4}(A^{-1}(x+2\ii t)) $.

            \item[(b).] If $ (\kappa_{1,1},\kappa_{2,3},\kappa_{3,5}) $ satisfy the parameter equation $ \rho(z_{0}, \kappa_{1,1}, \kappa_{1,3}, \kappa_{2,3}, \kappa_{3,5})=0 $ with $ z_{0}=z_{0,i_{1}} $ but not with $ z_{0}=z_{0,i_{2}} $ $(i_{1}\ne i_{2}, i_{1},i_{2}=1,2)$ in Eq. \eqref{para4}, then when $ |A|\gg 1 $, we obtain
            \begin{equation}\label{asymc7}
			\begin{aligned}
			    \left|q^{[4]}(x,t)\right| =  &\sum_{(\tilde{x}_{0}^{(i_{2})},\, \tilde{t}_{0}^{(i_{2})})} \left(\left|q^{[1]}(x-\tilde{x}_{0}^{(i_{2})}, t-\tilde{t}_{0}^{(i_{2})})\right|-1\right) +\sum_{(\check{x}_{0}, \check{t}_{0})} \left(\left|q^{[1]}(x-\check{x}_{0}, t-\check{t}_{0})\right|-1\right) \\
             &+\left|q^{[2]}(x-{x}_{0}^{(i_{1})}, t-{t}_{0}^{(i_{1})})\right| + \cO(|A|^{-1/3}),
			\end{aligned}
		\end{equation}
            where $({x}_{0}^{(i)}, {t}_{0}^{(i)})$ $(i=1,2)$ are two distinct triple roots of the polynomial $ \Theta_{4}(A^{-1}(x+2\ii t)) $, $(\tilde{x}_{0}^{(i)}, \tilde{t}_{0}^{(i)})$ are given in Eq. \eqref{xt30} with $\bar{z}_{0,i}$ traversing three single roots of the polynomial $ Q_{2,i}(\bar{z}) $ defined by Proposition \ref{prop4}, and $(\check{x}_{0}, \check{t}_{0})$ traverses four single roots of the Adler--Moser polynomial $ \Theta_{4}(A^{-1}(x+2\ii t)) $. 
            
		\item[(c).] If $ (\kappa_{1,1}, \kappa_{2,3}, \kappa_{3,5})$ do not satisfy the parameter equations $ \rho(z_{0,i}, \kappa_{1,1}, \kappa_{1,3}, \kappa_{2,3}, \kappa_{3,5})=0 $ $ (i=1,2) $ in Eq. \eqref{para4}, then when $ |A|\gg 1 $, we yield
		\begin{equation}\label{asymc6}
			\left|q^{[4]}(x,t)\right| = 1+ \sum_{i=0}^{1}\left( \sum_{(\tilde{x}_{0}^{(i)}, \tilde{t}_{0}^{(i)})} \left(\left|q^{[1]}(x-\tilde{x}_{0}^{(i)}, t-\tilde{t}_{0}^{(i)})\right|-1\right) \right)  +\sum_{(\check{x}_{0}, \check{t}_{0})} \left(\left|q^{[1]}(x-\check{x}_{0}, t-\check{t}_{0})\right|-1\right) + \cO(|A|^{-\frac{1}{3}}),
		\end{equation}
		where $(\tilde{x}_{0}^{(i)}, \tilde{t}_{0}^{(i)})$ is given in Eq. \eqref{xt3} with $\bar{z}_{0,i}$ traversing three single roots of the polynomial $ Q_{2,i}(z) $ defined by Proposition \ref{prop4}, and $(\check{x}_{0}, \check{t}_{0})$ traverses four single roots of the Adler--Moser polynomial $ \Theta_{4}(A^{-1}(x+2\ii t)) $.
	\end{enumerate}

\end{itemize}

The results of Proposition \ref{prop3} and \ref{prop4} indicate that when selecting specific internal large parameters $ (a_{3}, a_{5}, a_{7}) $, the fourth-order rogue wave solution $ q^{[4]}(x,t) $ exhibits patterns corresponding to all root structures of the Adler--Moser polynomial $ \Theta_{4}(z) $ with multiple roots.

Our method can be employed to investigate the correspondence between the patterns of the $ N $-order rogue wave solution $ q^{[N]}(x,t) $ \eqref{qn1} and the root structures of the Adler--Moser polynomial $ \Theta_{N}(z) $ with multiple roots. We assume that some roots of $ \Theta_{N}(z) $ with free complex parameters $ (\kappa_{1}, \kappa_{2}, \ldots, \kappa_{N-1}) $ are multiple roots, and set the internal large parameters $ (a_{3}, a_{5}, \ldots, a_{2N-1}) $ of the rogue wave solution $ q^{[N]}(x,t) $ \eqref{qn1} for the NLS equation \eqref{NLS} to be
\begin{equation}\label{para7}
	 a_{2j+1}=\kappa_{j,2+1}A^{2j+1}+\cO{(A^{2j-1})}, \quad \kappa_{j,2j+1}=\kappa_{j}, \quad 1\leq j\leq N-1, \quad |A|\gg 1.
\end{equation} 
Then, by selecting appropriate free parameter terms of $\cO{(A^{2j-1})}$, we can obtain many rogue wave patterns associated with the root structures of the Adler--Moser polynomial $ \Theta_{N}(z) $, similar to those presented in this paper, such as the claw-like, OTR-type, and TTR-type patterns, as well as their corresponding semi-modified and modified rogue wave patterns. These rogue wave patterns are composed of first-order rogue waves and lower-order fundamental rogue waves. Similarly, we can categorize these rogue wave patterns into multiple-root and single-root regions based on the positions of the multiple roots and single roots of the Adler--Moser polynomial $ \Theta_{N}(z) $. Then, it is found that these rogue wave patterns all asymptotically approach distributed first-order rogue waves in the single-root region. However, in the multiple-root region, these patterns asymptotically approach lower-order fundamental rogue waves or mixed structures of lower-order fundamental rogue waves and discrete first-order rogue waves, which are related to the parameter terms of $\cO{(A^{2j-1})}$. For the lower-order rogue waves in the multiple-root region, their positions correspond to the corresponding multiple roots of $ \Theta_{N}(z) $. For the mixed structures of lower-order fundamental rogue waves and discrete first-order rogue waves in the multiple-root region, the distribution of these wave peaks is not only related to multiple roots of the general Adler--Moser polynomial $ \Theta_{N}(z) $ but also the root structure of new special Adler--Moser polynomials, such as the Yablonskii--Vorob'ev polynomial hierarchy, among others.

Combining with the results presented by Yang \textit{et al}. in Ref. \cite{yang2024}, we can affirm that $ N $th-order rogue wave solutions $ q^{[N]}(x,t) $ of NLS equation \eqref{NLS} certainly have patterns corresponding to the entire root structures of the Adler--Moser polynomial $ \Theta_{N}(z) $.

\section{Examples of rogue wave patterns with multiple internal large parameters}\label{Sec4}

This section will provide specific examples of the patterns with multiple internal large parameters for the high-order rogue wave solution of NLS equation \eqref{NLS}, such as the claw-like patterns of $q^{[N]}(x,t)$ $(N=3, 4, 5, 6)$, the modified claw-like patterns of $ q^{[6]}(x,t) $ and $ q^{[4]}(x,t) $, and the OTR-type, TTR-type, semi-modified TTR-type, and their modified patterns of $q^{[4]}(x,t)$, based on the asymptotic analysis of the rogue wave solutions in Sec \ref{Sec3}. Additionally, we will present the corresponding dynamic figures for these rogue wave patterns in Figs. \ref{Fig3a}-\ref{Fig5}, as well as the predicted locations of all wave peaks. This further illustrates and substantiates our analytical findings in this paper.

\begin{enumerate}[(1).]
	
\item Cases of the claw-like patterns for the rogue wave solution $q^{[N]}(x,t)$.
 
When taking the internal large parameters 
\begin{equation}\label{pa01}
	a_{2j+1}=\frac{1}{2j+1}A^{2j+1} -\frac{1}{(2j+1)!} A, \quad 1\leq j\leq N-1,
\end{equation}
we can generate the claw-like rogue wave patterns of $ q^{[N]}(x,t) $ of the $ N $th-order rogue wave solution $ q^{[N]}(x,t) $ \eqref{qn1}. Here, we provide the cases of $N=3, 4, 5, 6$ and plot them in Fig. \ref{Fig3a}. It is found that these patterns of $q^{[N]}(x,t)$ all have $N$ first-order rogue waves in the single-root region and an $ (N-1)$th-order fundamental rogue wave in the multiple-root region. The quantity and locations of the wave peaks all correspond to the root structures of the Adler--Moser polynomial $ \Theta_{N}(z) $ with the parameters $\kappa_{j}=\frac{1}{2j+1}$ $ (1\leq j \leq N-1) $ and a $ \frac{N(N-1)}{2} $-multiple zero root $ z_{0}=1 $. Their approximation error is all $ \cO(|A|^{-1}) $. 

\begin{figure*}[!htbp]
	\centering
	\includegraphics[width=\textwidth]{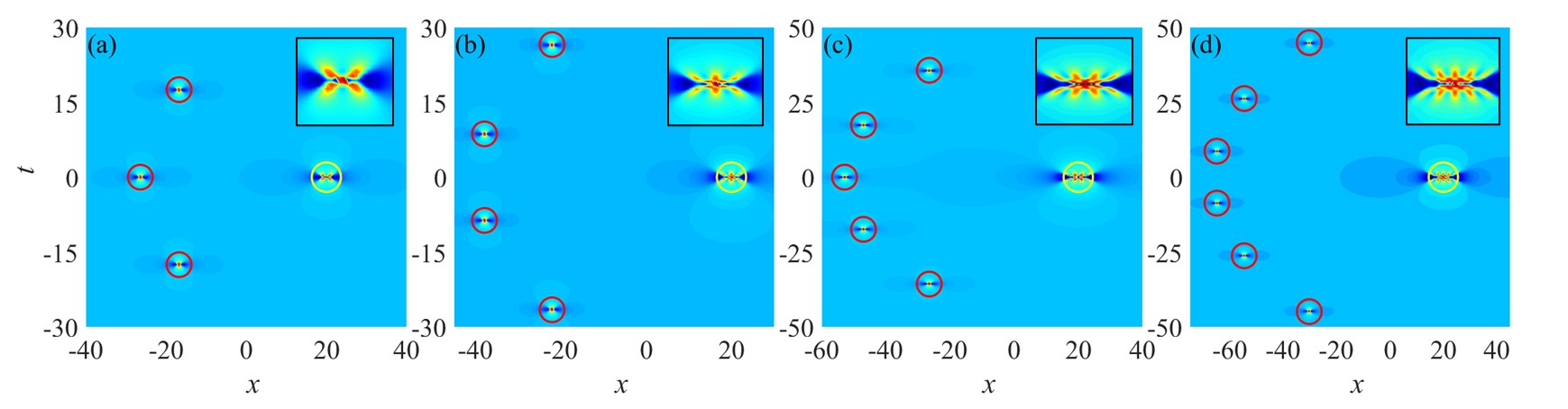}
	\caption{The claw-like patterns of the $ N $th-order rogue wave solution $ q^{[N]}(x,t) $ for NLS equation \eqref{NLS}, where $ N=3,4,5,6 $ from $ (a) $ to (d), and the internal large parameters $ a_{2j+1} $ $ (1\leq j\leq N-1) $ are given in Eq. \eqref{para0} with $\kappa_{j,2j+1}=\frac{1}{2j+1}$, $\kappa_{j,1} =-\frac{1}{(2j+1)!}$, and $ A=20 $. The top right corner of each figure displays a detailed image of the multiple-root region, which is approximately an $ (N-1) $th-order fundamental rogue wave. These red and yellow circles represent the predicted positions of first-order rogue waves in the single-root region and lower-order rogue waves in the multiple-root region, respectively.}
	\label{Fig3a}
\end{figure*}

\item Cases of the modified claw-like patterns for the rogue wave solution $ q^{[6]}(x,t) $ and $ q^{[4]}(x,t) $.

When the internal large parameters $a_{2j+1} $ $ (1\leq j\leq N-1) $ of the rogue wave solution $ q^{[N]}(x,t) $ are defined by Eq. \eqref{para0} with 
\begin{equation}\label{pa02}
\begin{aligned}
	&\kappa_{j,2j+1}=\frac{1}{2j+1}, \quad 1\leq j\leq N-1,\\ 
	&\kappa_{l,1} =-\frac{1}{(2l+1)!},  \quad \kappa_{m,1}\ne -\frac{1}{(2m+1)!}, \quad 1\leq l \leq m-1, 1\leq m \leq 4,
\end{aligned}
\end{equation}
we can obtain the modified claw-like rogue wave patterns. Here, we provide the cases of the modified claw-like patterns for the rogue wave solution $ q^{[6]}(x,t) $, as shown in Fig. \ref{Fig3b}. 

It is evident that the dynamic behaviors of these modified claw-like patterns are consistent with those of the corresponding claw-like patterns in the single-root region, as shown in Figs. \ref{Fig3b} $ (a)$-$(d) $. They exhibit six first-order rogue waves in the single-root region, which corresponds to all single roots of the Adler--Moser polynomial $ \Theta_{6}(z) $ with the parameters $\kappa_{j}=\frac{1}{2j+1}$ $ (1\leq j \leq 5) $. And their approximation error in the single-root region is $ \cO(|A|^{-1}) $. However, in the multiple-root region, these modified claw-like rogue wave patterns of $q^{[6]}(x,t)$ have the mixed structures of first-order rogue waves and lower-order fundamental rogue wave, where the locations and quantity of the wave peaks are both associated with the multiple roots of $ \Theta_{N}(z) $ and the root structures of the Yablonskii--Vorob'ev polynomial $Q_{N-1}^{[m]}(z)$. Since their approximation error near the positions corresponding to the single roots of $Q_{N-1}^{[m]}(z)$ is $ \cO(|A|^{-1/(2m+1)}) $, we need to select sufficiently large value of $ A $. As shown in Figs. \ref{Fig3b} $ (e)$-$(h) $, we focus on illustrating the triangular, pentagonal, heptagonal, and ring structures formed by first-order rogue waves and lower-order fundamental rogue waves in the modified claw-like rogue wave patterns of $q^{[6]}(x,t)$ in the multiple-root region. These structures correspond to the root structures of the Yablonskii--Vorob'ev polynomials $Q_{N-1}^{[m]}(z)$ $ (m=1,2,3,4) $, respectively. Specifically, in Fig. \ref{Fig3b} $ (h) $, a third-order fundamental rogue wave is in the center of the ring structure and its asymptotic errors is $ \cO(|A|^{-1}) $, corresponding to a sextuple zero root of the polynomial $Q_{N-1}^{[4]}(z)$.

\begin{figure*}[!htbp]
	\centering
	\includegraphics[width=\textwidth]{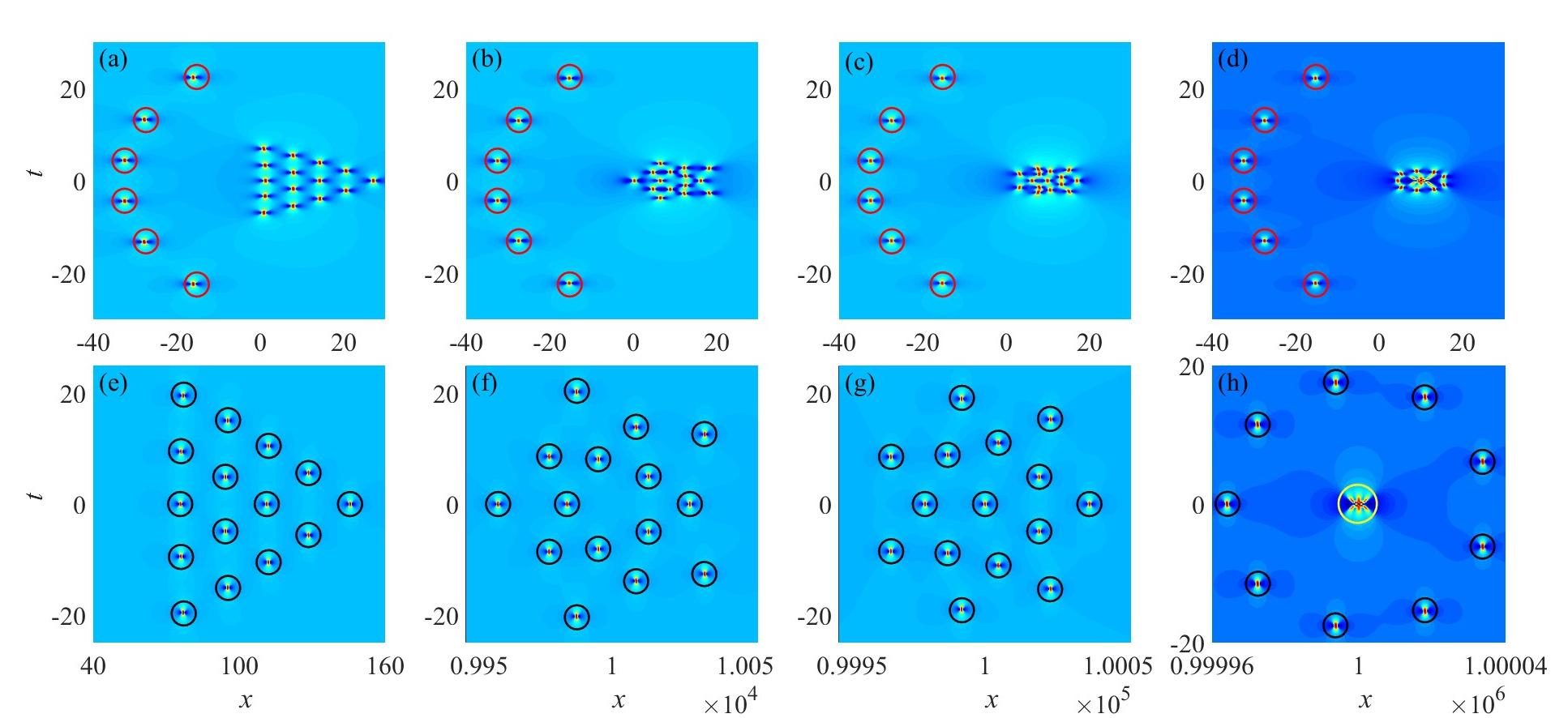}
	\caption{The modified claw-like patterns of the rogue wave solution $ q^{[6]}(x,t) $ for NLS equation \eqref{NLS}, whose the internal large parameters $ a_{2j+1} $ $ (1\leq j\leq 5) $ are given in \eqref{para0} with $\kappa_{j,2j+1}=\frac{1}{2j+1}$. From $ (a) $ to $ (d) $: $ A=10 $, and $ (\kappa_{1,1}, \kappa_{2,1}, \kappa_{3,1}, \kappa_{4,1}, \kappa_{5,1})= $ $ (5,0,0,0,0) $, $ (-\frac{1}{3!}, 5,0,0,0) $, $ (-\frac{1}{3!}, -\frac{1}{5!},5,0,0) $, $ (-\frac{1}{3!}, -\frac{1}{5!}, -\frac{1}{7!},5,0) $. 
	From $ (e) $ to $ (h) $: $ A=10^{2}, 10^{4}, 10^{5}, 10^{6} $, and $ (\kappa_{1,1}, \kappa_{2,1}, \kappa_{3,1}, \kappa_{4,1}, \kappa_{5,1})= $ $ (10,0,0,0,0) $, $ (-\frac{1}{3!}, 10,0,0,0) $, $ (-\frac{1}{3!}, -\frac{1}{5!}, 10^{2},0,0) $, $ (-\frac{1}{3!}, -\frac{1}{5!}, -\frac{1}{7!}, 10^{3},0) $. The first row displays the complete dynamic behaviors of the modified claw-like rogue wave patterns, while the second row only illustrates the dynamic behaviors of these patterns in the multiple-root region. These red, yellow, and black circles represent the predicted positions of first-order rogue waves in the single-root region, lower-order rogue wave in the multiple-root region, and first-order rogue waves in the multiple-root region, respectively.}
	\label{Fig3b}
\end{figure*}

Moreover, we add extral special constant to $ a_{2j+1} $ in the initial large parameters for the patterns in Figs. \ref{Fig3b} $ (d) $ and $ (h) $, as follows:
\begin{equation}\label{pa03}
\begin{aligned}
	&a_{2j+1}=\frac{1}{2j+1}A^{2j+1}+\kappa_{j,1}A+\hat{\kappa}_{j,3}A^{(2j+1)/27},  \quad A=10, \quad 1\leq j\leq 5,\\
	&(\kappa_{1,1}, \kappa_{2,1}, \kappa_{3,1}, \kappa_{4,1}, \kappa_{5,1})= (-\frac{1}{3!}, -\frac{1}{5!}, -\frac{1}{7!}, 10,0),\quad \hat{\kappa}_{l,3}=0, \quad 3\leq l\leq 5.
\end{aligned}
\end{equation}
Then, by choosing the free parameters $ (\hat{\kappa}_{1,3},\hat{\kappa}_{2,3}) $ as one of the values below
\begin{equation}\label{pa04}
	(1,1), \quad (-2\ii, 2\ii),
\end{equation}
we obtain new modified claw-like rogue wave patterns of $q^{[6]}(x,t)$, as shown in Fig. \ref{Fig3d} $ (a) $ and $ (c) $. These patterns differ from the ones shown in Fig. \ref{Fig3b} $ (d) $ only near the center of the multiple-root region. Because these patterns have large approximation errors in the multiple-root region, we need to choose a sufficiently large value for the large parameter $ A $ to more clearly demonstrate the asymptotic behaviors of these patterns in the multiple-root region. As shown in Figs. \ref{Fig3d} $ (b) $ and $ (d) $, we have detailed the multiple-root regions of these new modified claw-like rogue wave patterns, whose internal large parameters $ a_{2j+1} $ $ (1\leq j\leq 5) $ are defined by Eq. \eqref{pa03} with the different values of the free parameters 
\begin{equation}\label{pa05}
\begin{aligned}
	&A=10^{6}, \quad (\kappa_{1,1}, \kappa_{2,1}, \kappa_{3,1}, \kappa_{4,1}, \kappa_{5,1})= (-\frac{1}{3!}, -\frac{1}{5!}, -\frac{1}{7!}, 10^{3},0), \quad \hat{\kappa}_{l,3}=0, \quad 3\leq l\leq 5,
\end{aligned}
\end{equation}
and $ (\hat{\kappa}_{1,3},\hat{\kappa}_{2,3}) $ equal to one of  
\begin{equation}\label{pa06}
	(10,10), \quad (-5\ii, 10\ii).
\end{equation}
Compared to those in Fig. \ref{Fig3b} $ (h) $, the asymptotic behaviors of the nine first-order rogue waves forming the circle is consistent, but the asymptotic behavior near the center of the circle is entirely different. Near the center of the multiple-root region, these models are composed of a triangular formation of six first-order rogue waves, related to the root structure of the new Adler--Moser polynomial with free parameters $ (\hat{\kappa}_{1,3},\hat{\kappa}_{2,3}) $ given in Eq. \eqref{pa06}, and the asymptotic error is $ \cO(|A|^{-1/27}) $.

\begin{figure*}[!htbp]
	\centering
	\includegraphics[width=\textwidth]{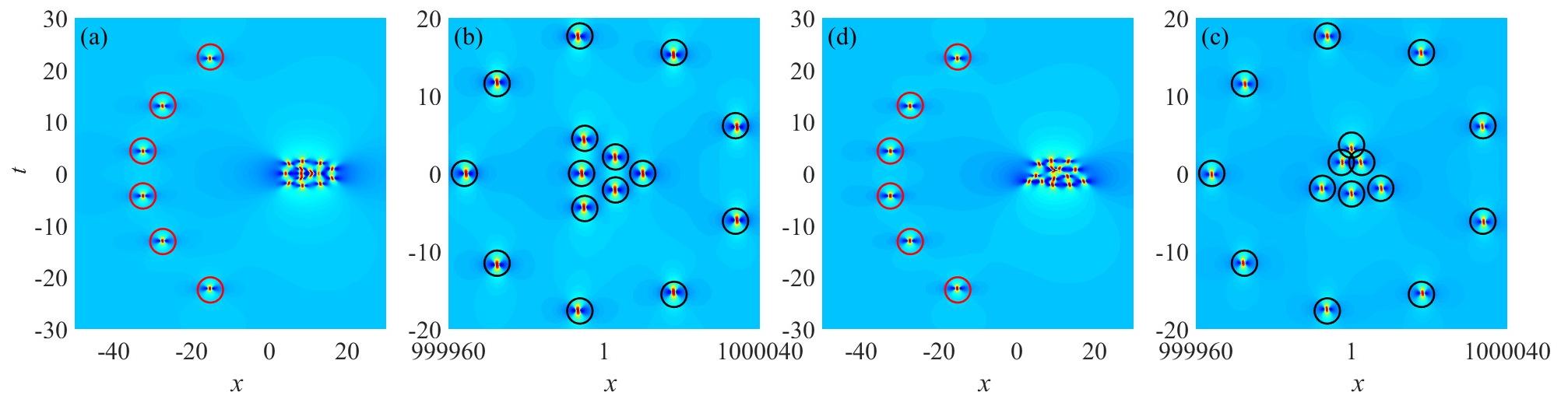}
	\caption{The modified claw-like patterns for the rogue wave solution $ q^{[6]}(x,t) $ of NLS equation \eqref{NLS} with the internal large parameters $a_{2j+1} $ $ (1\leq j\leq 5) $ defined by Eq. \eqref{pa03}. In $ (a) $ and $ (c) $, the complete dynamic behaviors are presented, and the free parameters in the large parameters $a_{2j+1} $ $ (1\leq j\leq 5) $ are given by Eq. \eqref{pa03}-\eqref{pa04}. While in $ (b) $ and $ (d) $, the dynamic behaviors in the multiple-root region are only presented, and the free parameters in the large parameters $a_{2j+1} $ $ (1\leq j\leq 5) $ are given by Eq. \eqref{pa05}-\eqref{pa06}. These red and black circles represent the predicted positions of first-order rogue waves in the single-root region and the multiple-root region, respectively.}
	\label{Fig3d}
\end{figure*}

Furthermore, for the rogue wave solution $ q^{[4]}(x,t) $ of NLS equation \eqref{NLS}, we can also choose the internal large parameters $(a_{3},a_{5},a_{7}) $ in the form defined by Eq. \eqref{parab1} or \eqref{parab2} with the free parameters
\begin{equation}\label{khk1}
\begin{aligned}
	&(\kappa_{1,1}, \kappa_{2,5}, \kappa_{3,7})=(\frac{1}{3}, \frac{1}{5}, \frac{1}{7}), \quad	(\kappa_{1,1}, \kappa_{2,1})=(-\frac{1}{3!}, -\frac{1}{5!} ),\\
	& (\hat{\kappa}_{1,1}, \hat{\kappa}_{2,1})=(2, 5), \quad (\hat{\kappa}_{1,2}, \hat{\kappa}_{2,2})=(5\ii, 5\ii).
\end{aligned}
\end{equation}
Then, we can obtain new modified claw-like rogue wave patterns of $ q^{[4]}(x,t) $ that are different from those in Fig. \ref{Fig3b}, as shown in Fig. \ref{Fig3c}. These patterns in the multiple-root region all have six first-order rogue waves, which corresponds to the root structure of the Adler--Moser polynomials $ \Theta_{4}(z) $ with parameters $ (\hat{\kappa}_{1,i}, \hat{\kappa}_{2,i}) $ $ (i=1,2) $. And their approximation errors in the multiple-root region are $ \cO(|A|^{-1/3}) $ for the pattern in Fig. \ref{Fig3c} $ (a) $ and  $ \cO(|A|^{-1/9}) $ for the pattern in Fig. \ref{Fig3c} $ (b) $, respectively. However, in the single-root region, these patterns exhibit the same asymptotic behaviors as in Theorem \ref{Theo1}, asymptotically separating into four first-order rogue waves with an approximation error $ \cO(|A|^{-1}) $.

\begin{figure*}[!htbp]
	\centering
	\includegraphics[width=0.7\textwidth]{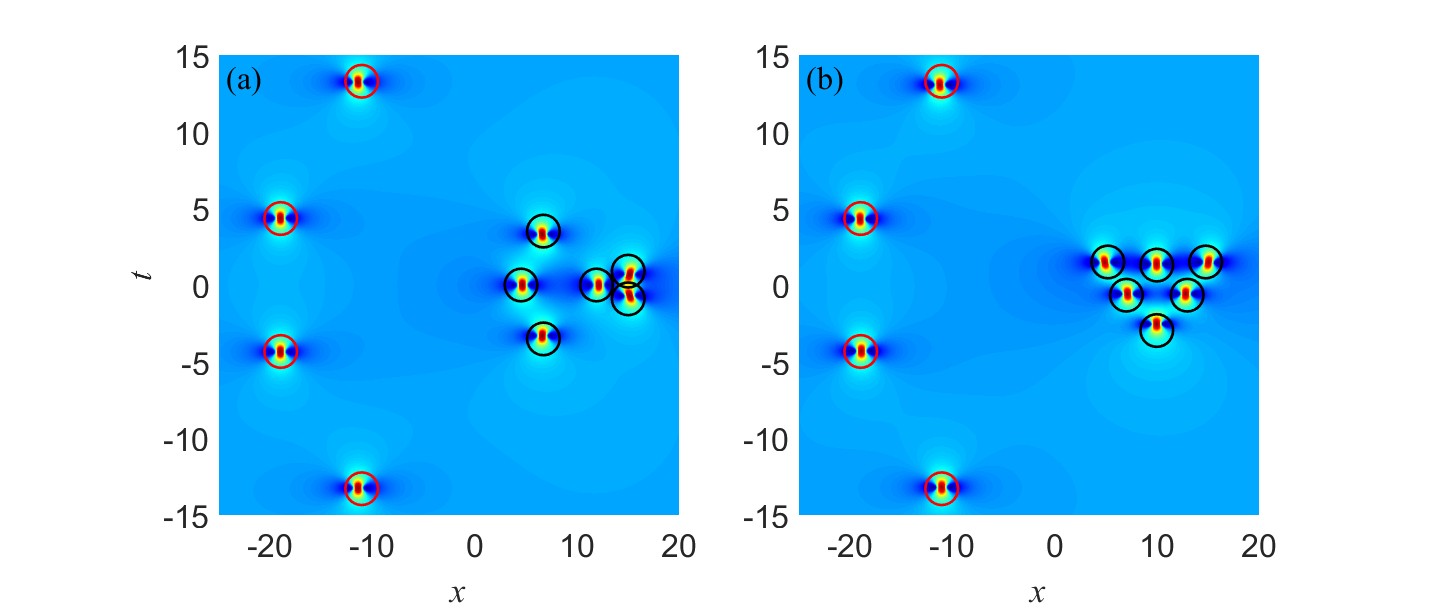}
	\caption{The modified claw-like patterns of the rogue wave solution $ q^{[4]}(x,t) $ for NLS equation \eqref{NLS}. $ (a) $ The internal large parameters $(a_{3},a_{5},a_{7}) $ of $ q^{[4]}(x,t) $ are defined by \eqref{parab1} with $ A=10 $, $ (\kappa_{1,1}, \kappa_{2,5}, \kappa_{3,7})=(\frac{1}{3}, \frac{1}{5}, \frac{1}{7}) $, and $ (\hat{\kappa}_{1,1}, \hat{\kappa}_{2,1})=(2, 5) $. 
	 $ (b) $ The internal large parameters $(a_{3},a_{5},a_{7}) $ of $ q^{[4]}(x,t) $ are defined by \eqref{parab2} with $ A=10 $, $ (\kappa_{1,1}, \kappa_{2,5}, \kappa_{3,7})=(\frac{1}{3}, \frac{1}{5}, \frac{1}{7}) $, $ (\kappa_{1,1}, \kappa_{2,1})=(-\frac{1}{3!}, -\frac{1}{5!} ) $, and $ (\hat{\kappa}_{1,2}, \hat{\kappa}_{2,2})=(5\ii, 5\ii) $. These red and black circles represent the predicted positions of first-order rogue waves in the single-root region and the multiple-root region, respectively.}
	\label{Fig3c}
\end{figure*}

\item Cases of the OTR-type and modified OTR-type patterns of the rogue wave solution $q^{[4]}(x,t)$.

When the internal large parameters $ (a_{3}, a_{5}, a_{7}) $ of the solution $ q^{[4]}(x,t) $ are defined by Eq. \eqref{para2}, we choose three sets of values of $ (\kappa_{1,3}, \kappa_{2,5}, \kappa_{3,7})  $:
	\begin{equation}\label{kapa1}
		\begin{aligned}
			&\left( -\frac{2}{3}, \frac{1}{5}, \frac{1}{7}\right), \quad \left( \frac{\sqrt{5}}{10}- \frac{1}{6}, 0, \frac{\sqrt{5}}{50}- \frac{3}{70}\right), \\ 
			&\left(-\frac{1}{3}-\frac{h_{1}(h_{1}(3\sqrt{3}\ii +13)-196)}{8232}, \frac{1}{5}-\frac{(\sqrt{3}\ii +9)h_{1}(h_{1}-14)}{8232}, 0 \right),
		\end{aligned}
	\end{equation}
with $ \quad h_{1}=\left(-2548+588 \ii \sqrt{3}\right)^{\frac{1}{3}} $. This results in the Adler--Moser polynomial $ \Theta_{4}(z) $ with the parameters $ (\kappa_{1}, \kappa_{2}, \kappa_{3}) =(\kappa_{1,3}, \kappa_{2,5}, \kappa_{3,7})$ having only one triple root $ z_{0}=1 $. Further, if taking $ (\kappa_{1,1}, \kappa_{2,3}, \kappa_{3,5}) =(0,0,0) $ for the internal large parameters $ (a_{3}, a_{5}, a_{7}) $ \eqref{para2}, we can obtain the modified OTR-type patterns for the rogue wave solution $ q^{[4]}(x,t) $. If $ (\kappa_{1,1}, \kappa_{2,3}, \kappa_{3,5}) =(-\frac{1}{6},0,0 ) $, we can generate the OTR-type patterns for the rogue wave solution $ q^{[4]}(x,t) $. Their dynamical diagrams are plotted in Fig. \ref{Fig3}. From the figures of the modified OTR-type patterns, it is observed that seven first-order rogue waves exist in the single-root region, whose positions are related to all single roots of the corresponding Adler--Moser polynomial $ \Theta_{4}(z) $. Meanwhile, there are three first-order rogue waves in the multiple-root region of the modified OTR-type patterns of $ q^{[4]}(x,t) $, whose positions are both correlated with the triple root $ z_{0}=1 $ of $ \Theta_{4}(z) $ and the root structure of the polynomial $ Q_{2}(\bar{z}) $ defined by Proposition \ref{prop3}. Furthermore, for the OTR-type rogue wave patterns, we find that the quantity and positions of the first-order rogue waves in the single-root region are consistent with the above modified OTR-type patterns. However, in the multiple-root region, the structures of the OTR-type patterns are different, featuring a second-order fundamental rogue wave.

\item Cases of the TTR-type, semi-modified TTR-type, and modified TTR-type patterns of the rogue wave solution $q^{[4]}(x,t)$.
 
When the internal large parameters $ (a_{3}, a_{5}, a_{7}) $ of the rogue wave solution $q^{[4]}(x,t)$ are given in Eq. \eqref{para2}, we choose three sets of values of $ (\kappa_{1,3}, \kappa_{2,5}, \kappa_{3,7})  $: 
\begin{equation}\label{kapa2}
	\begin{aligned}
		&(-0.32, -0.028, 0.063), \quad (-0.029 + 0.028\,\ii, -0.032 + 0.0077\,\ii, -0.0052 - 0.0016\,\ii), \\
		&(0.060 + 0.0098\,\ii,  0.0011 + 0.0045\,\ii, -0.0016 + 0.0013\,\ii).
	\end{aligned}
\end{equation}
The Adler--Moser polynomial $ \Theta_{4}(z) $ with free parameters $ (\kappa_{1}, \kappa_{2}, \kappa_{3}) $ defined by Eq. \eqref{kap1}-\eqref{kap3} has two distinct triple root $ (z_{0,1}, z_{0,2})$, which approximatively are one of $(-1.73,1)  $, $ (-0.86-0.51\,\ii, 1) $ and $ (-0.40-0.92\,\ii,1) $. If the parameters $ (\kappa_{1,1}, \kappa_{2,3},\kappa_{3,5}) =(0,0,0) $, we can gain the modified TTR-type patterns of the rogue wave solution $ q^{[4]}(x,t) $, as shown in Figs. \ref{Fig4} $(a)$-$(c)$. If $ (\kappa_{1,1}, \kappa_{2,3}, \kappa_{3,5}) $ are selected as one of 
    \begin{equation}\label{kapa3}
    \begin{aligned}
		&(-0.050, -0.17, 0), \quad (0.13 + 0.043\,\ii, 0.14 - 0.029\,\ii, 0),\quad (0.044 + 0.27\,\ii, 0.040 + 0.040\,\ii, 0),
    \end{aligned}
    \end{equation}
we can obtain the TTR-type patterns of the rogue wave solution $ q^{[4]}(x,t) $, as shown in Figs. \ref{Fig4} $(d)$-$(f)$. If  $ (\kappa_{1,1}, \kappa_{2,3}, \kappa_{3,5}) $ are selected as one of 
    \begin{equation}\label{kapa3a}
	\begin{aligned}
		&(-\frac{1}{6}, 0, 0), \quad (0.29, 0, 0), \quad (0.14 +0.34\,\ii,0, 0), \quad (0.067 + 0.61\,\ii, 0, 0),
	\end{aligned}
    \end{equation}
then the semi-modified TTR-type patterns of the rogue wave solution $ q^{[4]}(x,t) $ are generated, as shown in Fig. \ref{Fig5}.

\begin{figure*}[!htbp]
	\centering
	\includegraphics[width=0.8\textwidth]{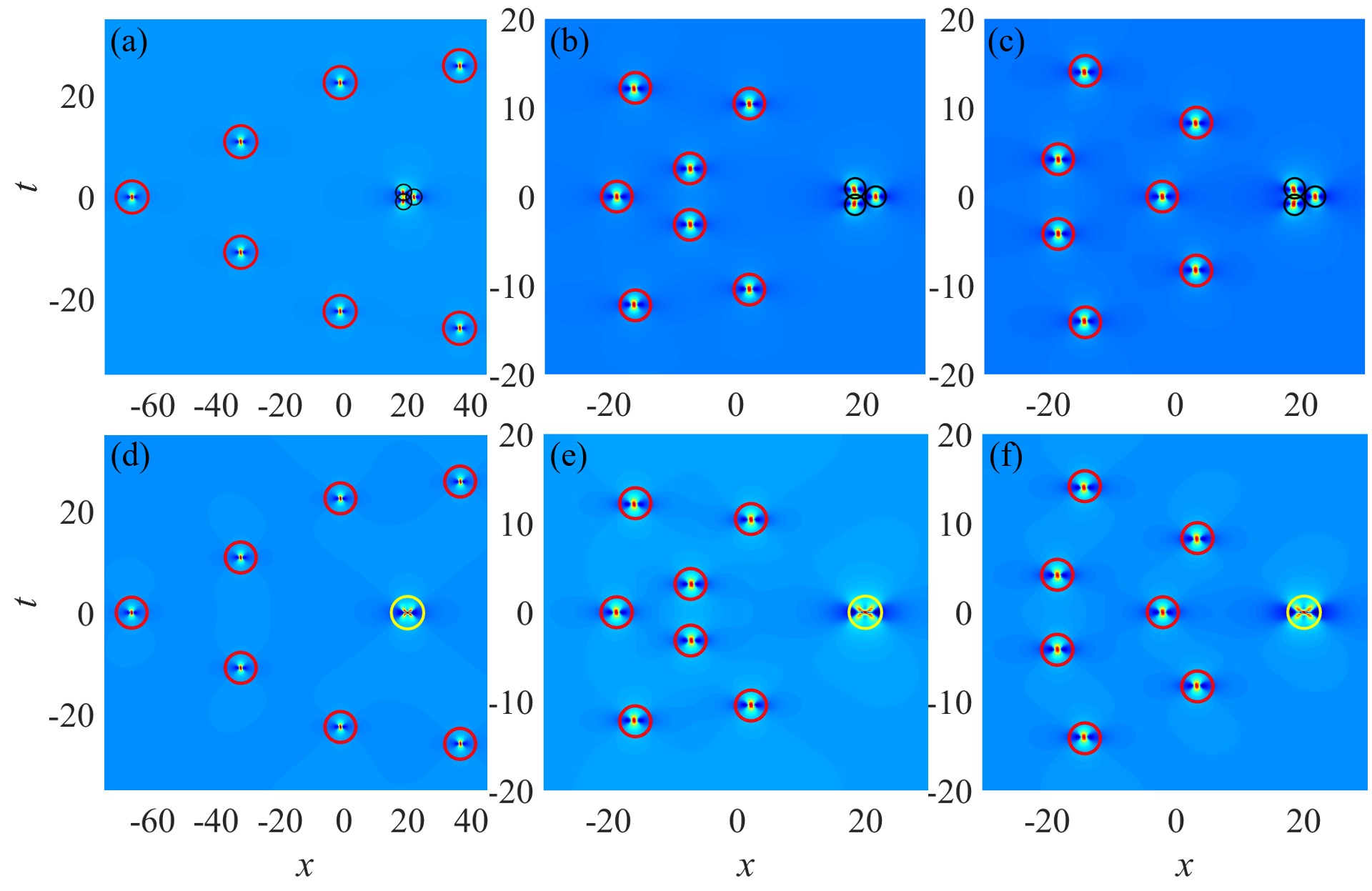}
	\caption{The OTR-type and modified OTR-type patterns of the rogue wave solution $ q^{[4]}(x,t) $ for NLS equation \eqref{NLS} with internal large parameters $ (a_{3}, a_{5}, a_{7}) $ given in Eq. \eqref{para2} and $ A=20 $. From left to right columns, the parameters $ (\kappa_{1,3}, \kappa_{2,5}, \kappa_{3,7}) $ in internal large parameters \eqref{para2} are sequentially equal to the values in Eq. \eqref{kapa1}. In the first row, these patterns are the modified OTR-type with $ (\kappa_{1,1}, \kappa_{2,3}, \kappa_{3,5})=(0,0,0) $. In the second row, these patterns are the OTR-type with $ (\kappa_{1,1}, \kappa_{2,3}, \kappa_{3,5})=(-\frac{1}{6},0,0) $. These red, yellow, and black circles represent the predicted positions of first-order rogue waves in the single-root region, lower-order rogue waves in the multiple-root region, and first-order rogue waves in the multiple-root region, respectively.}
	\label{Fig3}
\end{figure*}

It is evident that the TTR-type, semi-modified, and modified TTR-type patterns of the rogue wave solution $ q^{[4]}(x,t) $ all have four first-rogue waves in the single-root region. Based on Proposition \ref{prop4}, we find that positions of these first-rogue waves in the single-root region all correspond to the single roots of the corresponding Adler--Moser polynomial $ \Theta_{4}(z) $. However, these three types of rogue wave patterns have different structures in the multiple-root region. The TTR-type rogue wave patterns exhibit two dispersed second-order fundamental rogue waves in the multiple-root region. For the modified TTR-type rogue wave patterns, two dispersed triangular models are in the multiple-root region, where each triangular model is composed of three first-order rogue waves. For the semi-modified TTR-type rogue wave patterns, the structures of the multiple-root region comprise one second-order fundamental rogue wave and one triangular model composed of three first-order rogue waves. In addition, in the multiple-root region of these three types of patterns, the positions of the second-order fundamental rogue waves all correspond to the nonzero triple roots of the corresponding polynomial $ \Theta_{4}(z) $. Meanwhile, for the triangular models in the multi-root region, the positions of the first-order rogue waves are related not only to the triple roots of the corresponding polynomial $ \Theta_{4}(z) $ but also to three single roots of the polynomial $ Q_{2, i}(\bar{z}) $ $ (i=1,2) $ defined by Proposition \ref{prop4}.

\end{enumerate}

Note that the argument of factor $A$ in the large parameter $ a_{j} $ of the rogue wave solutions is equal to zero for these examples. According to Theorems \ref{Theo1}-\ref{Theo2} and Proposition \ref{prop3}-\ref{prop4}, it is known that when $\arg A \ne 0$, the structures of the rogue wave patterns will undergo a certain degree of rotation. For more details on the condition with $\arg A \ne 0$, refer to Refs. \cite{yang2021, LINGsu2024}.

\begin{figure*}[!htbp]
	\centering
	\includegraphics[width=0.8\textwidth]{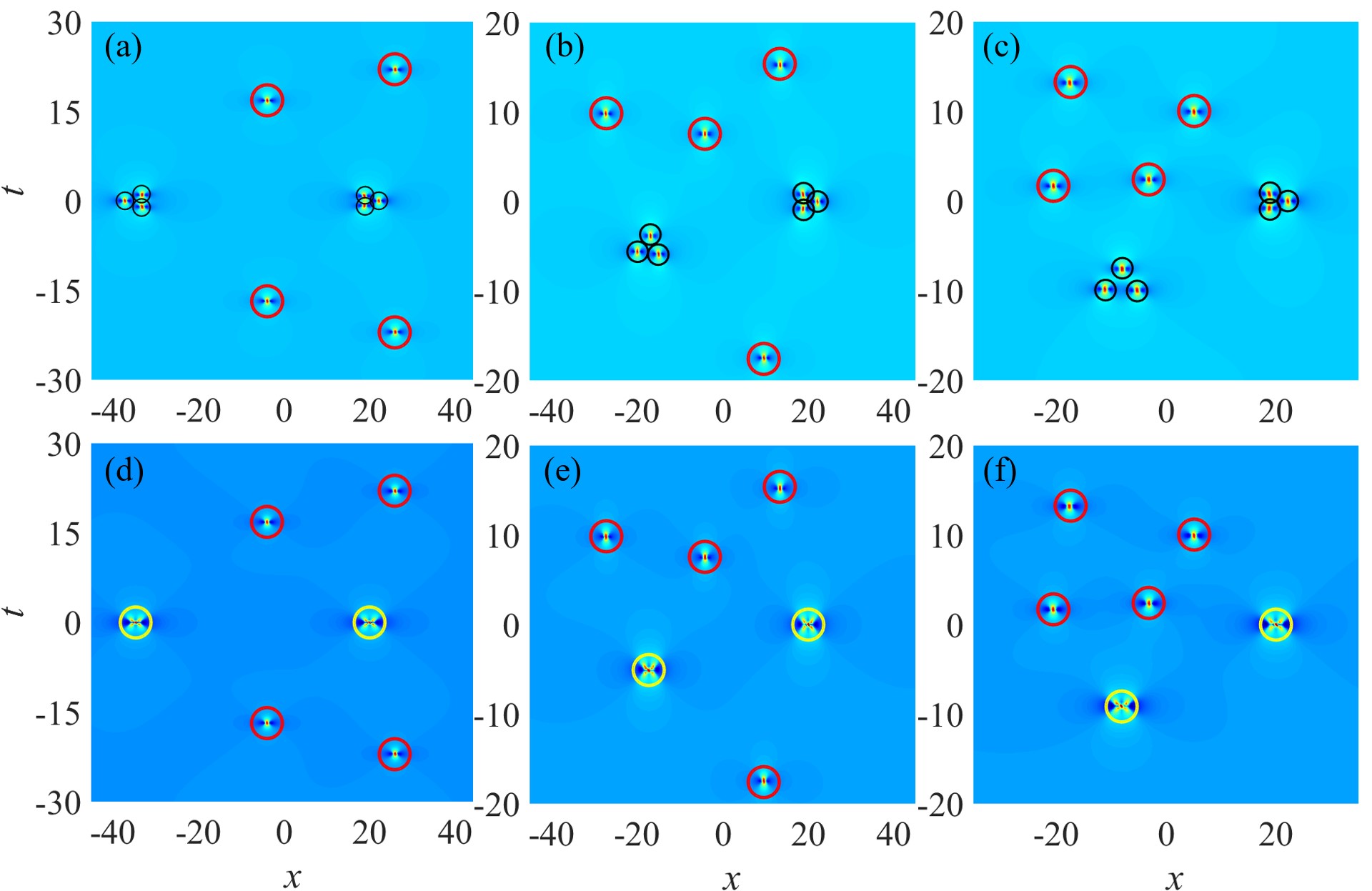}
	\caption{The TTR-type and modified TTR-type patterns of the rogue wave solution $ q^{[4]}(x,t) $ for NLS equation \eqref{NLS} with internal large parameters $ (a_{3}, a_{5}, a_{7}) $ given in Eq. \eqref{para2} and $ A=20 $. From left to right columns, the parameters $ (\kappa_{1,3}, \kappa_{2,5}, \kappa_{3,7})  $ are sequentially equal to the values in Eq. \eqref{kapa2}. In the first row, these patterns are the modified TTR-type with $ (\kappa_{1,1}, \kappa_{2,3}, \kappa_{3,5})=(0,0,0)$. In the second row, these patterns are the TTR-type with $ (\kappa_{1,1}, \kappa_{2,3}, \kappa_{3,5}) $ sequentially equal to the values in Eq. \eqref{kapa3}. These red, yellow, and black circles represent the predicted positions of first-order rogue waves in the single-root region, lower-order rogue waves in the multiple-root region, and first-order rogue waves in the multiple-root region, respectively.}
	\label{Fig4}
\end{figure*}

\section{Proofs of the main results}\label{Sec5}

\subsection{Proof of Proposition \ref{pro1a}}\label{ssec1a}

In this subsection, we provide the proof of Proposition \ref{pro1a}. For convenience, when $ \kappa_{j}=\frac{z_{0}^{2j+1}}{2j+1}$ $ (j\geq 1) $ and $ z_{0}\in \mathbb{C}\setminus \{0\} $, we define
\begin{equation}\label{funf1}
	g(\epsilon)=\exp\left( z_{0}\epsilon+\sum_{j=1}^{\infty}\frac{z_{0}^{2j+1}}{2j+1}\epsilon^{2j+1} \right) =\sum_{k=0}^{\infty} \theta_{k}\epsilon^{k}.
\end{equation}
Since $ g(-\epsilon)=\frac{1}{g(\epsilon)} $ and $ z_{0}\epsilon +\sum_{j=1}^{\infty}\frac{z_{0}^{2j+1}}{2j+1}\epsilon^{2j+1}=\frac{1}{2} \ln \left( \frac{1+z_{0}\epsilon}{1-z_{0}\epsilon}\right) $, we derive
\begin{equation}\label{funf5}
	\frac{g(\epsilon)-g(-\epsilon)}{g(\epsilon)+g(-\epsilon)}=\frac{g^{2}(\epsilon)-1}{g^{2}(\epsilon)+1} = \frac{\left( \exp\left( \frac{1}{2} \ln \left( \frac{1+z_{0}\epsilon}{1-z_{0}\epsilon}\right)\right)\right) ^{2} -1 }{\left( \exp\left( \frac{1}{2} \ln \left( \frac{1+z_{0}\epsilon}{1-z_{0}\epsilon}\right)\right)\right) ^{2} +1} =z_{0}\epsilon,
\end{equation}
i.e.,
\begin{equation}\label{funf3}
	g(\epsilon)-g(-\epsilon)= (g(\epsilon)+g(-\epsilon)) z_{0}\epsilon.
\end{equation}

Moreover, as
\begin{equation}\label{funf2}
\begin{aligned}
	&g(\epsilon)-g(-\epsilon)= 2\sum_{k=0}^{\infty}\theta_{2k+1}\epsilon^{2k+1},\\
	&(g(\epsilon)+g(-\epsilon)) z_{0}\epsilon= 2\sum_{k=0}^{\infty} \theta_{2k}z_{0}\epsilon^{2k+1},
\end{aligned}
\end{equation}
we have
\begin{equation}\label{funf4}
 \sum_{k=0}^{\infty}\theta_{2k+1}\epsilon^{2k+1} = \sum_{k=0}^{\infty} \theta_{2k}z_{0}\epsilon^{2k+1}.
\end{equation}
By grouping the terms according to the power of $\varepsilon$ in Eq. \eqref{funf4}, we obtain
\begin{equation}\label{funf6}
\theta_{2k}\theta_{1}-\theta_{2k+1} \theta_{0}=0, \quad k\leq 1,
\end{equation}
with $\theta_{0}=1$ and $\theta_{1}=z_{0}$.

Hence, we complete the proof of Proposition \ref{pro1a}.

\begin{figure*}[!htbp]
	\centering
	\includegraphics[width=0.8\textwidth]{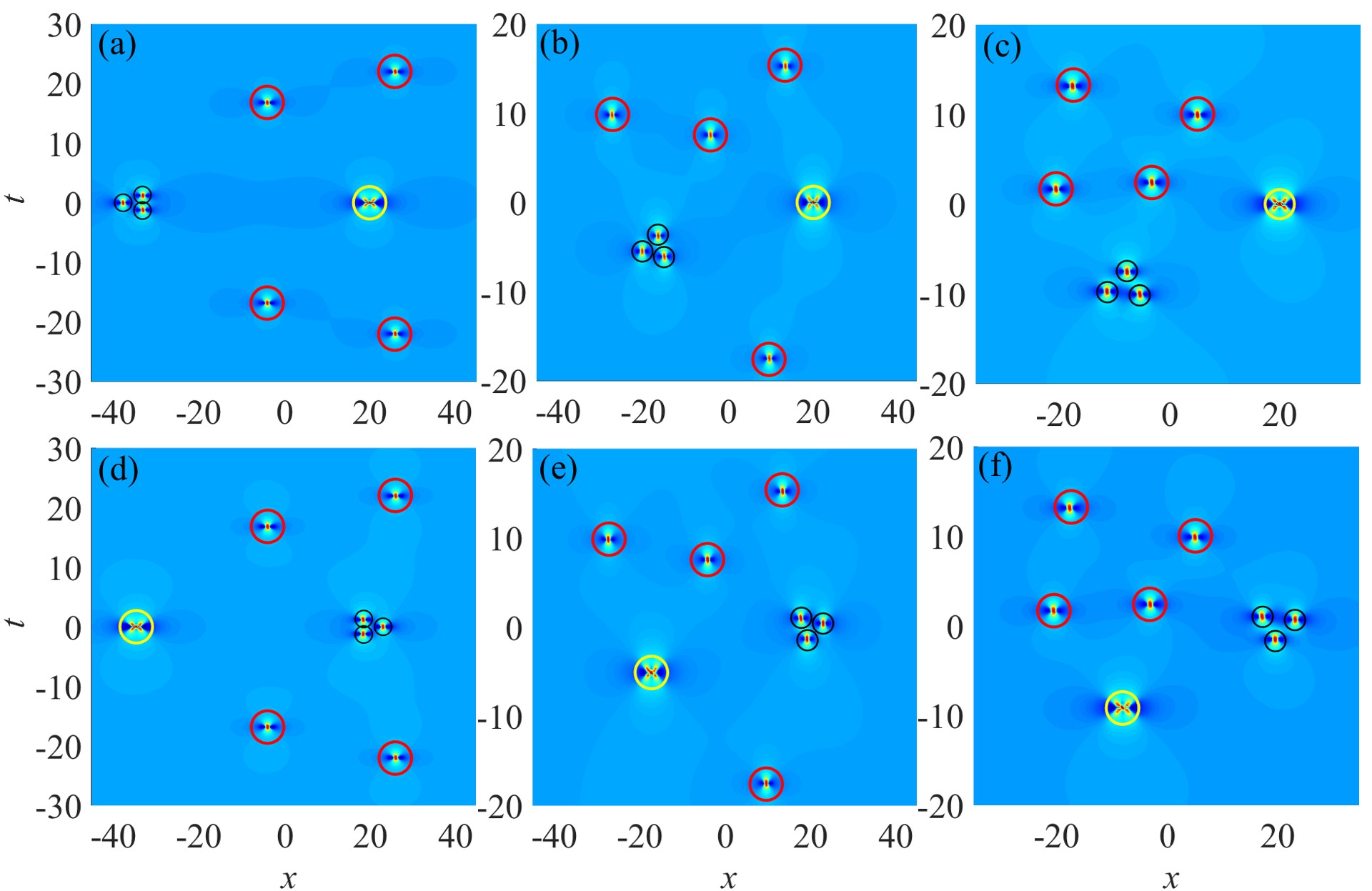}
	\caption{The semi-modified TTR-type patterns of the rogue wave solution $ q^{[4]}(x,t) $ for NLS equation \eqref{NLS} with internal large parameters $ (a_{3}, a_{5}, a_{7}) $ given in Eq. \eqref{para2}, $\kappa_{2,3}=\kappa_{3,5}=0$, and $ A=20 $. From left to right columns, the parameters $ (\kappa_{1,3}, \kappa_{2,5}, \kappa_{3,7}) $ in internal large parameters \eqref{para2} are sequentially equal to the values in Eq. \eqref{kapa2}. In $(a)$-$(c)$, the parameters $(\kappa_{1,1}, \kappa_{2,3}, \kappa_{3,5})=(-\frac{1}{6}, 0,0)$. From $(d)$ to $(f)$, the parameters $(\kappa_{1,1}, \kappa_{2,3}, \kappa_{3,5})$ $ = $ $  (0.29, 0, 0)$, $(0.14 +0.34\,\ii,0, 0)$, $(0.067 + 0.61\,\ii, 0, 0) $. These red, yellow, and black circles represent the predicted positions of first-order rogue waves in the single-root region, lower-order rogue waves in the multiple-root region, and first-order rogue waves in the multiple-root region, respectively.}
	\label{Fig5}
\end{figure*}

\subsection{Proof of Proposition \ref{pro1}}\label{ssec1}

In this subsection, we will present the proof of Proposition \ref{pro1}. For convenience, we denote the $ i $th derivative of the Adler--Moser polynomial $ \Theta_{4}(z) $ at $ z=z_{0} $ as $ (\Theta_{4}(z_{0}))^{(i)} $ with $ i\geq3 $.

%\begin{proof}
When $ (\kappa_{1}, \kappa_{2}, \kappa_{3}) $ are given in Eq. \eqref{kap1}, we can find that $ z_{0} $ is the triple root of $ \Theta_{4}(z) $. Then, substituting the value of $ \kappa_{2} $ into the polynomial $ \Theta_{3}(z_{0}) $, we can directly obtain $ \Theta_{3}(z_{0})=0 $. Moreover, assume $ (\Theta_{3}(z_{0}))'=0 $, then we find that $ z_{0} $ is a multiple root of $ \Theta_{3}(z) $. However, since $ \Theta_{3}(z) $ only has an nonzero multiple root at $ z=z_{0} $ if and only if $ \kappa_{1}=\frac{z_{0}^{3}}{3} $ and $ \kappa_{2}=\frac{z_{0}^{5}}{5} $ with $z_{0}\in \mathbb{C}\setminus \{0\} $, it contradicts the condition of $ \kappa_{1}\ne \frac{z_{0}^{3}}{3} $. Thus, $ (\Theta_{3}(z_{0}))'\ne0 $ holds.

As $ z_{0} $ is the triple root of $ \Theta_{4}(z) $, it is evident that $ \Theta_{4}(z_{0}) = (\Theta_{4}(z_{0}))' =(\Theta_{4}(z_{0}))''=0$ and $ (\Theta_{4}(z_{0}))^{(3)}\ne0 $. Expanding the determinants of $ \Theta_{4}(z_{0}) $ and $ (\Theta_{4}(z_{0}))' $ along the fourth column, we obtain
\begin{equation}\label{the3}
\begin{aligned}
	&\Theta_{4}(z_{0})=c_{4}\left(\frac{\theta_{4}}{c_{3}}\Theta_{3}(z_{0})-\theta_{2}\Theta_{3,1}(z_{0})+
	\Theta_{3,2}(z_{0}) \right) =0,\\
	&(\Theta_{4}(z_{0}))'=c_{4}\left(\frac{\theta_{3}}{c_{3}}\Theta_{3}(z_{0}) - \theta_{1}\Theta_{3,1}(z_{0})\right)=0,
\end{aligned}
\end{equation}
where $c_{i}$ $(i=3,4)$ and $ \Theta_{3,j}(z_{0}) $ $ (j=1,2) $ are defined by Eqs. \eqref{amp} and \eqref{nthe}, respectively.
Since $ \theta_{j}\ne 0 $ for $j\geq 0$, $c_{i}\ne0$ $(i=3,4)$, and $ \Theta_{3}(z_{0})=0 $, we derive $ \Theta_{3,1}(z_{0})=\Theta_{3,2}(z_{0})=0 $. Then, we calculate $ (\Theta_{4}(z_{0}))'' $ and $ (\Theta_{4}(z_{0}))^{(3)} $, as follows:
\begin{equation}\label{the4}
\begin{aligned}
	&(\Theta_{4}(z_{0}))'' =c_{4}\left(\Theta_{4,1}(z_{0}) + \frac{\theta_{2}}{c_{3}}\Theta_{3}(z_{0}) -\Theta_{3,1}(z_{0})\right)=0,\\
	&(\Theta_{4}(z_{0}))^{(3)} =c_{4}\left( 2\Theta_{4,2}(z_{0})+ \frac{2\theta_{1}}{c_{3}}\Theta_{3}(z_{0}) \right)
	\ne0,
\end{aligned}
\end{equation}
where $ \Theta_{4,i}(z_{0}) $ $ (i=1,2) $ are defined by Eq. \eqref{nthe}. Thus, we prove $ \Theta_{4,1}(z_{0})=0 $ and $ \Theta_{4,2}(z_{0})\ne 0 $.

Next, we will prove the third formula in Eq. \eqref{the2}. Suppose
\begin{equation}\label{the5}
\begin{vmatrix}
	\theta_{1} & \theta_{0} \\
	\theta_{3} & \theta_{2}
\end{vmatrix}=0,
\end{equation}
then we apply the facts of $ \Theta_{3}(z_{0})=\Theta_{3,1}(z_{0})=0 $ to derive that
\begin{equation}\label{the6}
\begin{vmatrix}
	\theta_{1} & \theta_{0} \\
	\theta_{5} & \theta_{4}
\end{vmatrix}=0, \quad
\begin{vmatrix}
	\theta_{1} & \theta_{0} \\
	\theta_{7} & \theta_{6}
\end{vmatrix}=0.
\end{equation}
Further, we can calculate that $ (\Theta_{4}(z_{0}))^{(3)}=0 $, which contradicts the fact that $ z_{0} $ is a triple root of $ \Theta_{4}(z) $. Hence, Eq. \eqref{the5} does not hold. Similarly, we also prove that Eq. \eqref{the6} is not valid. 

Thus, we complete the proof of Proposition \ref{pro1}.  

%\end{proof}

\subsection{Proof of Theorem \ref{Theo2}}\label{proof.Theo2}

In this subsection, we provide the proof of the asymptotics for the claw-like and modified claw-like patterns of the high-order rogue wave solution $ q^{[N]}(x,t) $ $ (N\geq 3) $ in the multiple-root region as stated in Theorem \ref{Theo2}.

First, we rewrite the determinant $ \tau^{(n)} $ in Eq. \eqref{qn1} as
\begin{equation}\label{tau}
	\begin{aligned}
		\tau^{(n)}=\begin{vmatrix}
			\mathbf{0}_{N\times N} & M^{(n,+)} \\
			-M^{(n,-)} & \mathbb{I}_{2N\times 2N}
		\end{vmatrix}, \quad n=0,1,
	\end{aligned}
\end{equation}
where
\begin{equation}\label{mpm}
	\begin{aligned}
		&M^{(n,+)}=\left[ 2^{1-j}S_{2i-j}(\mathbf{x}^{+}(n) +(j-1)\mathbf{s}) \right]_{1\leq i\leq N, 1\leq j\leq 2N},\\
		&M^{(n,-)}=\left[2^{1-i}S_{2j-i}(\mathbf{x}^{-}(n) +(i-1)\mathbf{s}) \right]_{1\leq i\leq 2N, 1\leq j\leq N}, 
	\end{aligned}
\end{equation}
$ \mathbb{I}_{2N\times 2N} $ is a $ 2N $th-order identity matrix, and $ \mathbf{s} $, $ \mathbf{x}^{\pm}(n) $, and $ S_{k}( \mathbf{x}) $ are defined by Eqs. \eqref{sj1}, \eqref{xpm}, and \eqref{schur}, respectively.

Next, we will simplify the matrices $ M^{(n,\pm)} $ in Eq. \eqref{mpm}. Here, we assume that $ (x_{0}, t_{0}) $ is the $ \frac{N(N-1)}{2} $-multiple root of the Adler--Moser polynomial $ \Theta_{N}(A^{-1}(x+2\ii t)) $ with free parameters $(\kappa_{1}, \kappa_{2}, \ldots, \kappa_{N-1})$ given in Eq. \eqref{kap0}. Then, we perform a coordinate transformation:
\begin{equation}\label{ct1}
	\bar{x}=x-\hat{x}_{0}|A|, \quad \bar{t}=t-\hat{t}_{0}|A|,
\end{equation}
where $( \hat{x}_{0},\hat{t}_{0})=(x_{0}|A|^{-1}, t_{0}|A|^{-1}) $. If the internal large parameters $ (a_{3}, a_{5}, \ldots, a_{N-1}) $ of the solution $ q^{[N]}(x,t) $ are defined by Eq. \eqref{para0}, we can obtain
\begin{equation}\label{xpm2}
	\begin{aligned}
		& x_{1}^{\pm}(n)=y_{1}^{\pm}(n) + |A|(\hat{x}_{0} \pm2\ii \hat{t}_{0}),  \quad y_{1}^{\pm}(n)=\bar{x} \pm 2\ii \bar{t} \pm n,
		\\ 
		& x_{2j+1}^{+}=y_{2j+1}^{+} + \kappa_{j,2j+1}A^{2j+1}, \quad x_{2j+1}^{-}= y_{2j+1}^{-}+(\kappa_{j,2j+1}A^{2j+1})^{*}, \\
		& y_{2j+1}^{+}=\frac{\bar{x} +2^{2j+1}\ii\bar{t}}{(2j+1)!}+ |A|\frac{\hat{x}_{0} +2^{2j+1}\ii\hat{t}_{0}}{(2j+1)!} + \kappa_{j,1}A, \quad y_{2j+1}^{-}=( y_{2j+1}^{+})^{*}, \quad j\geq 1,
	\end{aligned}
\end{equation}
where $ x_{1}^{\pm}(n) $ and $ x_{2j+1}^{\pm} $ are given in Eq. \eqref{xpm}.

When $ |A|\gg 1 $, we have
\begin{equation}\label{sch1}
	\begin{aligned}
		\sum_{k=0}^{\infty} S_{k}(\mathbf{x}^{+}(n) + v\mathbf{s}) (A^{-1}\epsilon)^{k} =& \exp\left( (\hat{x}_{0} +2\ii \hat{t}_{0})\ee^{-\ii \arg A}\epsilon + \kappa_{1,3}\epsilon^{3} +\kappa_{2,5}\epsilon^{5} +\cdots \right) \\
		&\times \exp\left( y_{1}^{+}(n)A^{-1}\epsilon +vs_{2}(A^{-1}\epsilon)^{2} +y_{3}^{+}(A^{-1}\epsilon)^{3}+vs_{4}(A^{-1}\epsilon)^{4}+\cdots \right) \\
		=&\left( \sum_{i=0}^{\infty}\theta_{i}(z_{0})\epsilon^{i}\right) 
		\left( \sum_{k=0}^{\infty}S_{k}(\mathbf{y}^{+}(n) + v\mathbf{s})(A^{-1}\epsilon)^{k} \right) \\
		=& \sum_{k=0}^{\infty} \left(\sum_{i=0}^{k} S_{k-i}(\mathbf{y}^{+}(n) + v\mathbf{s}) \theta_{i}(z_{0}) A^{i-k}  \right) \epsilon^{k},
	\end{aligned}
\end{equation}
and then derive
\begin{equation}\label{schk}
	S_{k}(\mathbf{x}^{+}(n) + v\mathbf{s}) =\sum_{i=0}^{k} S_{k-i}(\mathbf{y}^{+}(n) + v\mathbf{s}) \theta_{i}(z_{0}) A^{i},
\end{equation}
where $ \theta_{i}(z_{0}) $ is defined by Eq. \eqref{schur2}, $ v $ is a non-negative integer, and
\begin{equation}\label{yz}
	\mathbf{y}^{+}(n) =(y_{1}^{+}(n), 0, y_{3}^{+}, 0,\cdots),\quad z_{0}= (\hat{x}_{0} +2\ii \hat{t}_{0})\ee^{-\ii \arg A}= ({x}_{0} +2\ii {t}_{0})A^{-1}.
\end{equation}
According to the definition of $ (x_{0}, t_{0}) $ in Theorem \ref{Theo2}, it is evident that $ z_{0} $ is the $ \frac{N(N-1)}{2} $-multiple root of the polynomial $ \Theta_{N}(z) $.

By employing the expansion \eqref{schk}, we expand all elements of the matrix $ M^{(n,+)} $ \eqref{mpm}, as follows:
\begin{equation}\label{mpe1}
	M^{(n,+)}=\left[ 2^{1-j} \sum_{l=0}^{2i-j}S_{2i-j-l}(\mathbf{y}^{+}(n) + (j-1)\mathbf{s})\theta_{l}(z_{0}) A^{l}\right]_{1\leq i\leq N, 1\leq j\leq 2N}.
\end{equation}
Then, we apply the row transformations for the above matrix \eqref{mpe1} 
\begin{equation}\label{rtr0}
	r_{i+1}-\frac{\theta_{2i+1}}{\theta_{1}}A^{2i}r_{1}, \quad 1\leq i \leq N-1,
\end{equation}
to eliminate the highest-order terms of $ A $ in the first column elements of the $ i $th $ (2\leq i\leq N) $ rows, as follows:
%\begin{equation}\label{mpe2}
%	\begin{bmatrix}
%		\theta_{1}A+S_{1} &  S_{1}(\theta_{2}-\frac{\theta_{3}}{\theta_{1}})A^{2}+ \cO(A) & S_{1}(\theta_{4}-\frac{\theta_{5}}{\theta_{1}})A^{4} +\cO(A^{3}) & \cdots &  S_{1}(\theta_{6}-\frac{\theta_{6}}{\theta_{1}})A^{6} +\cO(A^{5}) \\
%		2^{-1} & 2^{-1}\left( (\theta_{2}-\frac{\theta_{3}}{\theta_{1}})A^{2} +\cO(A)\right) & 2^{-1}\left(  (\theta_{4}-\frac{\theta_{5}}{\theta_{1}})A^{4} +\cO(A^{3})\right)   & \cdots & 2^{-1}\left( (\theta_{6}-\frac{\theta_{6}}{\theta_{1}})A^{6} +\cO(A^{5})\right) \\
%		0 & 2^{-2}(\theta_{1}A+S_{1}) & 2^{-2}\sum_{l=0}^{3} S_{3-l}\theta_{l}A^{l} & \cdots & 
%		2^{-2} \sum_{l=0}^{2N-3} S_{2N-3-l}\theta_{l}A^{l} \\
%		\vdots & \vdots & \vdots & \ddots &  \vdots &\\
%		0 & 0 & 0 & \cdots & 2^{2-2N} (\theta_{1}A+S_{1})\\
%		0 & 0 & 0 & \cdots & 2^{1-2N} 
%	\end{bmatrix}^{T},
%\end{equation}
\begin{equation}\label{mpe2}
	\begin{bmatrix}
		\theta_{1}A+S_{1} & 2^{-1} &  \cdots  & 0 \\
		S_{1}(\theta_{2}-\frac{\theta_{3}}{\theta_{1}})A^{2}+ \cO(A) & 2^{-1} \left( (\theta_{2}-\frac{\theta_{3}}{\theta_{1}})A^{2} +\cO(A)\right)  & \cdots & 0 \\
		S_{1}(\theta_{4}-\frac{\theta_{5}}{\theta_{1}})A^{4} +\cO(A^{3}) &  2^{-1}\left(  (\theta_{4}-\frac{\theta_{5}}{\theta_{1}})A^{4} +\cO(A^{3})\right) & \cdots & 0 \\
		\vdots & \vdots  & \ddots &  \vdots \\
		S_{1}(\theta_{2N-2}-\frac{\theta_{2N-1}}{\theta_{1}})A^{2N-2} +\cO(A^{2N-3}) & 2^{-1}\left((\theta_{2N-2}-\frac{\theta_{2N-1}}{\theta_{1}})A^{2N-2} +\cO(A^{2N-3})\right) &  \cdots  & 2^{1-2N}
	\end{bmatrix},
\end{equation}
where $ r_{i} $ is denoted as the $ i $th row of the matrix, and $ S_{k} $ and $ \theta_{k} $ represent $ S_{k}(\mathbf{y}^{+}(n) + (j-1)\mathbf{s}) $ and $ \theta_{k}(z_{0}) $, respectively. For convenience, we will take the same notations $ S_{k} $ and $ \theta_{k} $ in the proof below.

From Proposition \ref{pro1a}, it is found that when $ \kappa_{j} $ $ (j\geq 1) $ are defined as Eq. \eqref{kap0}, the equations
\begin{equation}\label{the1}
	\theta_{2j}-\frac{\theta_{2j+1}}{\theta_{1}}=0 , \quad 1\leq j\leq N,
\end{equation}
are satisfied. Thus, the matrix \eqref{mpe2} is {reduced to}
\begin{equation}\label{mpe3}
\begin{bmatrix}
	\theta_{1}A+S_{1} &  \sum_{l=0}^{1}S_{3-l}\theta_{l}A^{l} & \sum^{3}_{l=0}S_{5-l}\theta_{l}A^{l} & \cdots &  \sum^{2N-3}_{l=0}S_{2N-1-l}\theta_{l}A^{l} \\
	2^{-1} & 2^{-1}\left(S_{1}\theta_{1}A +S_{2} \right)  & 2^{-1}\sum^{3}_{l=0}S_{4-l}\theta_{l}A^{l}   & \cdots & 2^{-1}\sum^{2N-3}_{l=0}S_{2N-2-l}\theta_{l}A^{l} \\
	0 & 2^{-2}(\theta_{1}A+S_{1}) & 2^{-2}\sum_{l=0}^{3} S_{3-l}\theta_{l}A^{l} & \cdots & 
	2^{-2} \sum_{l=0}^{2N-3} S_{2N-3-l}\theta_{l}A^{l} \\
	0 & 2^{-3} & 2^{-3}\sum_{l=0}^{2} S_{2-l}\theta_{l}A^{l} & \cdots & 2^{-3} \sum_{l=0}^{2N-4} S_{2N-4-l}\theta_{l}A^{l} \\
	0 & 0 & 2^{-4} \theta_{1}A+S_{1} & \cdots & 2^{-4} \sum_{l=0}^{2N-5} S_{2N-5-l}\theta_{l}A^{l} \\
	\vdots & \vdots & \vdots & \ddots &  \vdots &\\
	0 & 0 & 0 & \cdots & 2^{2-2N} (\theta_{1}A+S_{1})\\
	0 & 0 & 0 & \cdots & 2^{1-2N} 
\end{bmatrix}^{T}.
\end{equation}
%\begin{equation}\label{mpe3}
%	\begin{bmatrix}
%		\theta_{1}A+S_{1} &  \sum_{l=0}^{1}S_{3-l}\theta_{l}A^{l} & \sum^{3}_{l=0}S_{5-l}\theta_{l}A^{l} & \cdots &  \sum^{2N-3}_{l=0}S_{2N-1-l}\theta_{l}A^{l} \\
%		2^{-1} & 2^{-1}\left(S_{1}\theta_{1}A +S_{2} \right)  & 2^{-1}\sum^{3}_{l=0}S_{4-l}\theta_{l}A^{l}   & \cdots & 2^{-1}\sum^{2N-3}_{l=0}S_{2N-2-l}\theta_{l}A^{l} \\
%		0 & 2^{-2}(\theta_{1}A+S_{1}) & 2^{-2}\sum_{l=0}^{3} S_{3-l}\theta_{l}A^{l} & \cdots & 
%		2^{-2} \sum_{l=0}^{2N-3} S_{2N-3-l}\theta_{l}A^{l} \\
%		0 & 2^{-3} & 2^{-3}\sum_{l=0}^{2} S_{2-l}\theta_{l}A^{l} & \cdots & 2^{-3} \sum_{l=0}^{2N-4} S_{2N-4-l}\theta_{l}A^{l} \\
%		0 & 0 & 2^{-4} \theta_{1}A+S_{1} & \cdots & 2^{-4} \sum_{l=0}^{2N-5} S_{2N-5-l}\theta_{l}A^{l} \\
%		\vdots & \vdots & \vdots & \ddots &  \vdots &\\
%		0 & 0 & 0 & \cdots & 2^{2-2N} (\theta_{1}A+S_{1})\\
%		0 & 0 & 0 & \cdots & 2^{1-2N} 
%	\end{bmatrix}^{T}.
%\end{equation}

For the above matrix \eqref{mpe3}, we continue with row transformations sequentially:
\begin{equation}\label{rtr1}
r_{i+j}-\frac{\theta_{2j+1}}{\theta_{1}}A^{2j}r_{i}, \quad 2\leq i\leq N-1, \quad 1\leq j \leq N-i, \quad i,j \in \mathbb{N}.
\end{equation}
Now, the matrix $ M^{(n,+)} $ in Eq. \eqref{mpm} is reduced to the following matrix:
\begin{equation}\label{mpe4}
	\begin{bmatrix}
		\theta_{1}A+S_{1} &   S_{2}\theta_{1}A +S_{3} & S_{4}\theta_{1}A +S_{5} & \cdots &  S_{2N-2}\theta_{1}A +S_{2N-1} \\
		2^{-1} & 2^{-1}\left(S_{1}\theta_{1}A +S_{2} \right)  & 2^{-1}\left( S_{3}\theta_{1}A +S_{4} \right)    & \cdots & 2^{-1}\left( S_{2N-3}\theta_{1}A +S_{2N-2}\right)  \\
		0 & 2^{-2}(\theta_{1}A+S_{1}) & 2^{-2}\left( S_{2}\theta_{1}A +S_{3}\right)  & \cdots & 
		2^{-2} \left( S_{2N-4}\theta_{1}A +S_{2N-3} \right)  \\
		0 & 2^{-3} & 2^{-3}\left(S_{1}\theta_{1}A +S_{2} \right) & \cdots & 2^{-3} \left( S_{2N-5}\theta_{1}A +S_{2N-4} \right) \\
		0 & 0 & 2^{-4} \left( \theta_{1}A+S_{1} \right)  & \cdots & 2^{-4}\left( S_{2N-6}\theta_{1}A +S_{2N-5} \right) \\
		\vdots & \vdots & \vdots & \ddots &  \vdots \\
		0 & 0 & 0 & \cdots & 2^{2-2N} (\theta_{1}A+S_{1})\\
		0 & 0 & 0 & \cdots & 2^{1-2N} 
	\end{bmatrix}^{T}.
\end{equation}

Furthermore, we can similarly reduce the matrix $ M^{(n,-)} $ in Eq. \eqref{mpm}. Here, we omit the computational process and denote $\mathbf{y}^{-}(n)$ as 
\begin{equation}\label{yn0}
    \mathbf{y}^{-}(n)=(y_{1}^{-}(n), 0, y_{3}^{-}, 0,\cdots),
\end{equation}
with $y_{1}^{-}(n)$ and $y_{2j+1}^{-}$ given in Eq. \eqref{xpm2}.

Thus, as $ \theta_{1}=z_{0} $, the determinant $ \tau^{(n)}$ in Eq. \eqref{tau} can be simplified to
\begin{equation}\label{tau2}
	\begin{aligned}
		&\tau^{(n)} = |Az_{0}|^{2N}\begin{vmatrix}
			\mathbf{0}_{(N-1)\times (N-1)} & \hat{M}^{(n,+)} \\
			-\hat{M}^{(n,-)} & \mathbb{I}_{2(N-1)\times 2(N-1)}
		\end{vmatrix}\left[ 1+\cO(|A|^{-1})\right], \\
		&\hat{M}^{(n,+)}=\left[ 2^{-j}S_{2i-j}(\mathbf{y}^{+}(n) + j \,\mathbf{s}) \right]_{1\leq i\leq (N-1), 1\leq j\leq 2(N-1)}, \\
		&\hat{M}^{(n,-)}=\left[ 2^{-i}S_{2j-i}(\mathbf{y}^{-}(n) + i \, \mathbf{s}) \right]_{1\leq i\leq 2(N-1), 1\leq j\leq (N-1)},
	\end{aligned}
\end{equation}
where $ S_{k}(\mathbf{y}^{\pm}(n) + v \, \mathbf{s})=0 $ for $ k<0 $. 

Based on the definition \eqref{schur} of the Schur polynomials, we have 
\begin{equation}\label{ss1}
	S_{k}(\mathbf{y}^{+}(n) + v \,\mathbf{s}) = \sum_{i=0}^{\lfloor k/2\rfloor} S_{i}(\mathbf{s}) S_{k-2i}(\mathbf{y}^{+}(n) + (v-1) \,\mathbf{s}),
\end{equation}
where $ \lfloor k\rfloor $ represents the maximum integer less than or equal to $ k $. Then, by applying the expansion \eqref{ss1}, we expand all elements of the matrix $ \hat{M}^{(n,+)} $ in Eq. \eqref{tau2}, as follows:
\begin{equation}\label{hatm1}
\begin{aligned}
	\hat{M}^{(n,+)}= \frac{1}{2} 
	\begin{bmatrix}
		S_{1} & 2^{-1} & 0 & \cdots & 0 \\
		\bar{S}_{1}S_{1} +S_{3} & 2^{-1}(\bar{S}_{1} +S_{2}) & 2^{-2}S_{1} & \cdots & 0 \\
		\sum_{i=0}^{2} \bar{S}_{i} S_{5-2i} &  2^{-1}\sum_{i=0}^{2} \bar{S}_{i} S_{4-2i} & 2^{-2}(\bar{S}_{1}S_{1} +S_{3}) & \cdots & 0 \\
		\vdots & \vdots & \vdots & \ddots &  \vdots \\
		\sum_{i=0}^{N-2} \bar{S}_{i} S_{2N-3-2i} & 2^{-1} \sum_{i=0}^{N-2} \bar{S}_{i} S_{2N-4-2i} &  2^{-2} \sum_{i=0}^{N-3} \bar{S}_{i} S_{2N-5-2i} &  \cdots & 2^{2-2N}
	\end{bmatrix},
\end{aligned}
\end{equation}
where $ \bar{S}_{i} $ represents $ S_{i}(\mathbf{s}) $. By employing some row transformations, we can simplify the above matrix \eqref{hatm1} to
\begin{equation}\label{hatm2}
	\frac{1}{2}
	\begin{bmatrix}
		2^{1-j} S_{2i-j}(\mathbf{y}^{+}(n) + (j-1) \,\mathbf{s})
	\end{bmatrix}_{1\leq i\leq (N-1), 1\leq j\leq 2(N-1)}.
\end{equation}
Similarly, the matrix $ \hat{M}^{(n,-)} $ given in Eq. \eqref{tau2} can be also {reduced} to 
\begin{equation}\label{hatm3}
	\frac{1}{2}
	\begin{bmatrix}
		2^{1-i} S_{2j-i}(\mathbf{y}^{+}(n) + (i-1) \,\mathbf{s})
	\end{bmatrix}_{1\leq i\leq 2(N-1), 1\leq j\leq (N-1)}.
\end{equation}
This implies that the expression \eqref{tau2} of $ \tau^{(n)} $ can be rewritten as
\begin{equation}\label{tau2b}
\begin{aligned}
	&\tau^{(n)} = 2^{-2(N-1)}|Az_{0}|^{2N}
	\begin{vmatrix}
		\mathbf{0}_{(N-1)\times (N-1)} & \tilde{M}^{(n,+)} \\
		-\tilde{M}^{(n,-)} & \mathbb{I}_{2(N-1)\times 2(N-1)}
	\end{vmatrix} \left[ 1+\cO(|A|^{-1})\right], \\
	&\tilde{M}^{(n,+)}=\left[ 2^{1-j}S_{2i-j}(\mathbf{y}^{+}(n) + (j-1) \,\mathbf{s}) \right]_{1\leq i\leq (N-1), 1\leq j\leq 2(N-1)}, \\
	&\tilde{M}^{(n,-)}=\left[ 2^{1-i}S_{2j-i}(\mathbf{y}^{-}(n) + (i-1) \, \mathbf{s}) \right]_{1\leq i\leq 2(N-1), 1\leq j\leq (N-1)}, \quad n=0,1.
\end{aligned}	
\end{equation}

Next, we will discuss the asymptotic behaviors of the high-order rogue wave solution $ q^{[N]}(x,t) $ in the multiple-root region based on two different values of the parameters $ \kappa_{j,1} $ $ (1\leq j \leq N-2) $ in the internal large parameters $ (a_{3}, a_{5}, \ldots, a_{2N-1}) $ in Eq. \eqref{para0}.

\begin{itemize}
\item[(1).] When the parameters $ \kappa_{j,1} $ $ (1\leq j \leq N-2) $ are given in Eq. \eqref{para01}, we have
\begin{equation}\label{ypmj}
	y^{\pm}_{2j+1}= \frac{\bar{x} {\pm}2^{2j+1}\ii\bar{t}}{(2j+1)!}, \quad 1\leq j\leq N-2,
\end{equation}
based on the definitions \eqref{xpm2} and \eqref{yz} of $ y^{\pm}_{2j+1} $ and $ z_{0} $.

Therefore, when $ \sqrt{(x-x_{0})^{2}+(t-t_{0})^{2}}= \cO(1) $ and $ |A|\gg 1 $, we substitute the expression \eqref{tau2b} of $ \tau^{(n)} $ into the formula \eqref{qn1} of the high-order rogue wave solution $ q^{[N]}(x,t) $. Then, the asymptotic expression \eqref{asym1} of the claw-like rogue wave pattern of $ q^{[N]}(x,t) $ can be demonstrated. In other words, the high-order rogue wave solution $ q^{[N]}(x,t) $ tends to an $ (N-1) $th-order fundamental rogue wave near the position $ (x_{0}, t_{0}) $ in the $ (x,t) $-plane, where $ (x_{0}, t_{0}) $ is the $ \frac{N(N-1)}{2} $-multiple root of the Adler--Moser polynomial $ \Theta_{N}(A^{-1}(x+2\ii t)) $ with free parameters $(\kappa_{1}, \kappa_{2}, \ldots, \kappa_{N-1})$ given in Eq. \eqref{kap0}.

\item[(2).] When the parameters $ \kappa_{j,1} $ $ (1\leq j \leq m-1, 1\leq m \leq N-2) $ satisfy Eq. \eqref{para01} but $\kappa_{m,1}$ does not, we firstly rewrite the determinant in Eq. \eqref{tau2b} as
\begin{equation}\label{tau3}
\begin{aligned}
	\begin{vmatrix}
		\mathbf{0}_{(N-1)\times (N-1)} & \tilde{M}^{(n,+)} \\
		-\tilde{M}^{(n,-)} & \mathbb{I}_{2(N-1)\times 2(N-1)}
	\end{vmatrix}
	=& \sum_{0\leq v_{1}<v_{2}<\cdots<v_{N-1}\leq 2(N-1)-1} \det_{1\leq i,j\leq N-1} \left[2^{-v_{j}}S_{2i-v_{j}-1}(\mathbf{y}^{+}(n) + v_{j} \,\mathbf{s}) \right]  \\
 & \qquad \times\det_{1\leq i,j\leq N-1} \left[ 2^{-v_{i}}S_{2j-v_{i}-1}(\mathbf{y}^{+}(n) + v_{i} \,\mathbf{s}) \right], \quad n=0,1,
\end{aligned}
\end{equation}
by using the Cauchy--Binet formula. Then, we denote
\begin{equation}\label{eta}
	 \bar{y}_{2k+1}^{\pm}=\frac{\bar{x} \pm2^{2k+1}\ii\bar{t}}{(2k+1)!},  \quad \eta_{k}=|A|\frac{\hat{x}_{0} +2^{2k+1}\ii\hat{t}_{0}}{(2k+1)!} +\kappa_{k,1}A, \quad k\geq 1,
\end{equation}
and then derive
\begin{equation}\label{yk}
\begin{aligned}
    y_{2k+1}^{+}= \left\{
 \begin{array}{ll}
     \bar{y}_{2k+1}^{+}, & 1\leq k< m, \\
     \bar{y}_{2k+1}^{+} +\eta_{k}, & k\geq m,
 \end{array}\right. \quad
    y_{2k+1}^{-}= \left\{
 \begin{array}{ll}
     \bar{y}_{2k+1}^{-}, & 1\leq k< m, \\
     \bar{y}_{2k+1}^{-} +(\eta_{k})^{*}, & k\geq m.
 \end{array}\right.
\end{aligned}
\end{equation}
Since
\begin{equation}\label{sch2}
	\begin{aligned}
		\sum_{k=0}^{\infty}S_{k}(\mathbf{y}^{+}(n) + v \,\mathbf{s})\epsilon^{k} 
		&=\exp\left( y_{1}^{+}(n)\epsilon +vs_{2}\epsilon^{2} + y_{3}^{+}\epsilon^{3} +vs_{4}\epsilon^{4}+\cdots \right)\\
		&= \exp\left( y_{1}^{+}(n)\epsilon-\frac{2^{2m}}{2m+1} \left( \bar{\eta}_{m}^{1/(2m+1)} \epsilon\right)^{2m+1} \right) \exp\left( vs_{2}\epsilon^{2} +\bar{y}^{+}_{3}\epsilon^{3} + vs_{4}\epsilon^{4} +\cdots\right) \\ 
		&= \left( \sum_{k=0}^{\infty} p_{k}^{[m]}(\bar{z})\left( \bar{\eta}_{m}^{1/(2m+1)} \epsilon\right)^{k} \right) \left( 1+vs_{2}\epsilon^{2}+\cdots \right),
	\end{aligned}
\end{equation}
we yield
\begin{equation}\label{schk2}
	S_{k}(\mathbf{y}^{+}(n) + v \,\mathbf{s}) = \bar{\eta}_{m}^{k/(2m+1)}p_{k}^{[m]}(\bar{z})\left[ 1+\cO(|A|^{-2/(2m+1)})\right], 
\end{equation}
where $ \bar{\eta}_{m}=-\frac{2m+1}{2^{2m}}\eta_{m} =\cO(|A|) $, and $ p_{k}^{[m]}(\bar{z}) $ is defined by Eq. \eqref{yvpm} with
\begin{equation}\label{zbar}
	\bar{z}=y_{1}^{+}(n) \bar{\eta}_{m}^{-1/(2m+1)}=(\bar{x} + 2\ii \bar{t} + n)\bar{\eta}_{m}^{-1/(2m+1)}.
\end{equation}

To obtain the first two highest-order terms of $ A $ in the determinant \eqref{tau3}, we only consider two index choices, in which one is $ (v_{1}, v_{2}, \ldots, v_{N-2}, v_{N-1})=(0,1,\cdots, N-3, N-2 ) $, and another is $ (v_{1}, v_{2}, \ldots, v_{N-2}, v_{N-1})=(0,1, \cdots, N-3, N-1 ) $.

For the first index choice of $ (v_{1}, v_{2}, \ldots, v_{N-2}, v_{N-1})=(0,1,\cdots, N-3, N-2 ) $, we can apply the formula \eqref{schk2} to generate
\begin{equation}\label{mhat1}
	\det_{1\leq i,j\leq N-1} \left[2^{1-j}S_{2i-j}(\mathbf{y}^{+}(n) + (j-1) \,\mathbf{s}) \right] = \alpha_{1} \bar{\eta}_{m}^{\frac{N(N-1)}{2(2m+1)}} Q_{N-1}^{[m]}(\bar{z}) \left[ 1+\cO(|A|^{-2/(2m+1)})\right],
\end{equation}
where $ \alpha_{1}=2^{-(N-1)(N-2)/2}c_{N-1}^{-1} $, and $ c_{N-1} $ and $ Q_{N-1}^{[m]}(z) $ are defined by Eqs. \eqref{schur2} and \eqref{yvpm}. 

Similarly, we can also obtain the part containing $S_{k}(\mathbf{y}^- (n))$ in the determinant \eqref{tau3} with the first index choice, and omit it here. Therefore, we find that when $ |A|\gg 1 $, the highest $ A $-power terms of the determinant \eqref{tau3} is approximately
\begin{equation}\label{mhat0}
	{\alpha_{1}}^{2} \left| \bar{\eta}_{m}^{\frac{N(N-1)}{2(2m+1)}}\right|^{2}  \left|Q_{N-1}^{[m]}(\bar{z})\right|^{2}.
\end{equation}

In this article, we assume that the nonzero roots of the polynomial $ Q_{N-1}^{[m]}(\bar{z}) $ are all simple roots, but its zero root may be a multiple root. Thus, we discuss the asymptotic behaviors of the solution $ q^{[N]}(x,t)$ corresponding to the positions of zero root and nonzero roots of the polynomial $ Q_{N-1}^{[m]}(\bar{z}) $, respectively.

\begin{itemize}

\item[(a).] Suppose that
\begin{equation}\label{bz0}
	\bar{z}_{0}=(\bar{x}_{0} + 2\ii \bar{t}_{0} )\bar{\eta}_{m}^{-1/(2m+1)},
\end{equation}
is a nonzero simple root of the polynomial $ Q_{N-1}^{[m]}(\bar{z}) $. Then, when $ \sqrt{(\bar{x}-\bar{x}_{0})^{2}+(\bar{t}-\bar{t}_{0})^{2}}=\cO(1) $, we expand $ Q_{N-1}^{[m]}(\bar{z}) $ around $ \bar{z}=\bar{z}_{0} $, as follows:
\begin{equation}\label{yv31}
	Q_{N-1}^{[m]}(\bar{z}) = \bar{\eta}_{m}^{-1/(2m+1)} \left( \bar{x}-\bar{x}_{0} +2\ii (\bar{t}-\bar{t}_{0}) +n \right) (Q_{N-1}^{[m]}(\bar{z}_{0}))' \left[ 1+\cO(|A|^{-1/(2m+1)})\right].
\end{equation}
Then, Eq. \eqref{mhat1} is {reduced to}
\begin{equation}\label{mhat2}
	\alpha_{1}\bar{\eta}_{m}^{\frac{N(N-1)-2}{2(2m+1)}} (Q_{N-1}^{[1]}(\bar{z}_{0}))' \left( \bar{x}-\bar{x}_{0} +2\ii (\bar{t}-\bar{t}_{0}) +n \right) \left[ 1+\cO(|A|^{-1/(2m+1)})\right].
\end{equation}

On the other hand, for the second index choice of $ (v_{1}, v_{2}, \ldots, v_{N-2}, v_{N-1})=(0,1, \cdots, N-3, N-1 ) $, we generate 
\begin{equation}\label{mhat4}
\begin{aligned}
    &\det_{1\leq i\leq N-1} \left[S_{2i-1}(\mathbf{y}^{+}(n)), 2^{-1}S_{2i-2}(\mathbf{y}^{+}(n) + \,\mathbf{s}), \ldots, 2^{4-N}S_{2i-N+2}(\mathbf{y}^{+}(n) + (N-4) \,\mathbf{s}), \right.\\ 
 &\left. \qquad \qquad 2^{2-N}S_{2i-N}(\mathbf{y}^{+}(n) + (N-2) \,\mathbf{s}) \right]\\ 
&= \frac{1}{2}\alpha_{1}\bar{\eta}_{m}^{\frac{N(N-1)-2}{2(2m+1)}} (Q_{N-1}^{[m]}(\bar{z}))' \left[ 1+\cO(|A|^{-2/(2m+1)})\right].
\end{aligned} 
\end{equation}
When $ \sqrt{(\bar{x}-\bar{x}_{0})^{2}+(\bar{t}-\bar{t}_{0})^{2}}=\cO(1) $, Eq. \eqref{mhat4} is {simplified to}
\begin{equation}\label{mhat5}
	\frac{1}{2}\alpha_{1}\bar{\eta}_{m}^{\frac{N(N-1)-2}{2(2m+1)}} (Q_{N-1}^{[m]}(\bar{z}_{0}))' \left[ 1+\cO(|A|^{-1/(2m+1)})\right].
\end{equation}

Similarly, for the conjugate parts involving $S_{k}(\mathbf{y}^{-}(n))$ in the determinant \eqref{tau3} with these two index choice, we can get their asymptotic expressions near the position $ (\bar{x}_{0}, \bar{t}_{0}) $ in the $ (\bar{x}, \bar{t}) $ plane, which are omitted them here.

Combining the two leading terms \eqref{mhat2}, \eqref{mhat5}, and their conjugate parts for the determinant \eqref{tau3}, we obtain
\begin{equation}\label{tau4}
\begin{aligned}
\tau^{(n)}= \, &2^{-2(N-1)}|Az_{0}|^{2N} \alpha_{1}^{2} |\bar{\eta}_{m}|^{\frac{N(N-1)-2}{2m+1}} |(Q_{N-1}^{[m]}(\bar{z}_{0}))'|^{2} \\
&\times \left( (\bar{x}-\bar{x}_{0})^{2} +4(\bar{t}-\bar{t}_{0})^{2} -4\ii n(\bar{t}-\bar{t}_{0}) -n^{2} +\frac{1}{4} \right) \left[ 1+\cO(|A|^{-1/(2m+1)})\right], \quad n=0,1,
\end{aligned}
\end{equation}
by the formula \eqref{tau2b}. Then, we utilize Eq. \eqref{ct1} to rewrite the above expression \eqref{tau4} of $ \tau^{(n)} $ as 
\begin{equation}\label{tau5}
\begin{aligned}
	\tau^{(n)}= \, &2^{-2(N-1)}|Az_{0}|^{2N} \alpha_{1}^{2} |\bar{\eta}_{m}|^{\frac{N(N-1)-2}{2m+1}} |(Q_{N-1}^{[m]}(\bar{z}_{0}))'|^{2} \\
	&\times \left( ({x}-\tilde{x}_{0})^{2} +4({t}-\tilde{t}_{0})^{2} -4\ii n({t}-\tilde{t}_{0}) -n^{2} +\frac{1}{4} \right) \left[ 1+\cO(|A|^{-1/(2m+1)})\right], \quad n=0,1,
\end{aligned}	
\end{equation}
where $ \tilde{x}_{0}=\bar{x}_{0}+\hat{x}_{0}|A| $, $ \tilde{t}_{0}=\bar{t}_{0}+\hat{t}_{0}|A| $. According to the definitions of $ (\hat{x}_{0}, \hat{t}_{0}) $ and $ (\bar{x}_{0}, \bar{t}_{0}) $ in \eqref{ct1} and \eqref{bz0}, we find that $(\tilde{x}_{0}, \tilde{t}_{0}) $ satisfy Eq. \eqref{xt2} in Theorem \ref{Theo2}.

Therefore, when $ |A|\gg 1 $ and $ \sqrt{({x}-\tilde{x}_{0})^{2} +({t}-\tilde{t}_{0})^{2}}=\cO(1) $, we substitute the expression \eqref{tau5} into the formula \eqref{qn1} of the high-order rogue wave solution $ q^{[N]}(x,t) $ $ (N\geq 3) $. Then, we can obtain the following asymptotic expression:
\begin{equation}\label{qm1}
	q^{[N]}(x,t)=\left( 1- \dfrac{4(4\ii ({t}-\tilde{t}_{0})+1)}{4({x}-\tilde{x}_{0})^{2} +16({t}-\tilde{t}_{0})^{2} +1} \right)\ee^{2\ii t} +\cO(|A|^{-1/(2m+1)}),
\end{equation}
which approximately {tends to} a first-order rogue wave $ \hat{q}^{[1]}(x-\tilde{x}_{0}, t-\tilde{t}_{0})\ee^{2\ii t} $ with the approximate error $ \cO(|A|^{-1/(2m+1)}) $ and $ \hat{q}^{[1]}(x, t) $ given in Eq. \eqref{q1}.

\item[(b).] Next, we will discuss the asymptotics of the solution $ q^{[N]}(x,t) $ near the point $(\bar{x}, \bar{t})=(0,0) $, i.e., $(x,t)=(x_{0}, t_{0})$. By similar calculations to Eq. \eqref{sch2}, we obtain
\begin{equation}\label{tmpm}
    \begin{aligned}
        &S_{k}(\mathbf{y}^{+}(n) + v \,\mathbf{s})= \sum_{l=0}^{\lfloor k/(2m+1)\rfloor} \frac{{\bar{\eta}_{m}}^l}{l!} S_{k-l(2m+1)}(\mathbf{y}_{m}^{+}(n) + v \,\mathbf{s}),\\
        &\mathbf{y}_{m}^{+}(n)=\mathbf{y}^{+}(n)-(0,\ldots,0,\bar{\eta}_{m},0,\ldots)
        =(y_{1}^{+}(n), 0, \bar{y}_{3}^{+},0,\bar{y}_{5}^{+},0,\ldots,\bar{y}_{2m+1}^{+},0,y_{2m+3}^{+},0, \ldots ),
    \end{aligned}
\end{equation}
where ${y}_{2j+1}^{+}$, $\bar{y}_{2j+1}^{+}$, and $\bar{\eta}_{m}$ are given in Eqs. \eqref{eta}, \eqref{yk} and \eqref{schk2}, individually.
We apply the above formula \eqref{tmpm} to expand all elements of the matrix $\tilde{M}^{(n,+)}$ in the formula \eqref{tau2b} of $\tau^{(n)}$, as follows:
\begin{equation}\label{mpe7}
    \tilde{M}^{(n,+)}=\left[ 2^{1-j}\sum_{l=0}^{\lfloor \frac{2i-j}{2m+1}\rfloor}\frac{{\bar{\eta}_{m}}^l}{l!}S_{2i-j-l(2m+1)}(\mathbf{y}_{m}^{+}(n) + (j-1) \,\mathbf{s}) \right]_{1\leq i\leq (N-1), 1\leq j\leq 2(N-1)}.
\end{equation}

Then, we define $N_{0}$ by Eq. \eqref{gamma} with $N$ replaced by $N-1$. By similar row transformations to Eqs. \eqref{mpe1}-\eqref{mpe4}, the matrix \eqref{mpe7} can be simplified to the block matrix form below:
\begin{equation}\label{mrb01}
\begin{aligned}
    \begin{pmatrix}
    \mathbf{{B}}_{1,1} & \mathbf{0}_{(N-N_{0}-1)\times(2(N-1)-N_{0}) }\\
    \mathbf{{B}}_{2,1}& \mathbf{{B}}_{2,2}
\end{pmatrix}_{(N-1))\times 2(N-1)},\\
\end{aligned}
\end{equation}
where
\begin{equation}\label{mrb02}
\begin{aligned}
    &\mathbf{{B}}_{1,1}=
    \begin{pmatrix}
        S_{0} & 0 & \ldots & 0\\
        S_{1} & S_{0} & \ldots & 0\\
        \vdots & \vdots & \ddots & \vdots\\
        S_{N-N_{0}-2} & S_{N-N_{0}-3} & \ldots & S_{0}
    \end{pmatrix}_{(N-N_{0}-1)\times (N-N_{0}-1)},\\
    &\mathbf{{B}}_{2,1}=
    \begin{pmatrix}
        S_{N-N_{0}} & S_{N-N_{0}-1} & \ldots & S_{2}\\
        S_{N-N_{0}+2} & S_{N-N_{0}+1} & \ldots & S_{4}\\
        \vdots & \vdots & \ddots & \vdots\\
        S_{N+N_{0}-2} & S_{N+N_{0}-3} & \ldots & S_{2N_{0}}
    \end{pmatrix}_{N_{0}\times (N-N_{0}-1)},\\
    &\mathbf{{B}}_{2,2}=
    \begin{pmatrix}
        S_{1} & S_{0} & \ldots \\
        S_{3} & S_{2} & \ldots \\
        \vdots & \vdots & \ddots \\
        S_{2N_{0}-1} & S_{2N_{0}-2} & \ldots 
    \end{pmatrix}_{N_{0}\times (N+N_{0}-1)},
\end{aligned}
\end{equation}
and $S_{k}$ represents $S_{k}(\mathbf{y}_{m}^{+}(n) + v \,\mathbf{s}) $. Note that for the reduced matrix \eqref{mrb01}, we only provide the highest-order terms in each row with respect to $\bar{\eta}_{m}$, and omit their constant coefficients and the $\bar{\eta}_{m}$-power terms.

Moreover, the matrix $\tilde{M}^{(n,-)}$ in the formula \eqref{tau2b} of $\tau^{(n)}$ can be similarly reduced. Here, we omit these trivial calculations and define $\mathbf{y}_{m}^{-}(n)$ by
\begin{equation}\label{yn02}
    \mathbf{y}_{m}^{-}(n)=\mathbf{y}^{-}(n)-(0,\ldots,0,{\bar{\eta}_{m}}^*,0,\ldots)
        =(y_{1}^{-}(n), 0, \bar{y}_{3}^{-},0,\bar{y}_{5}^{-},0,\ldots,\bar{y}_{2m+1}^{-},0,y_{2m+3}^{-},0, \ldots ),
\end{equation}
where $\mathbf{y}^{-}(n)$, $y_{2j+1}^{-}$ and $\bar{y}_{2j+1}^{-}$ are given in Eqs. \eqref{yn0} and \eqref{yk}.

Therefore, by similar calculations to Eqs. \eqref{tau2}-\eqref{tau2b}, we can rewrite the formula \eqref{tau2b} of $\tau^{(n)}$ into the following expression again:
\begin{equation}\label{taut1}
\begin{aligned}
	&\tau^{(n)} = 2^{-2(N-1)}|Az_{0}|^{2N}\alpha_{2} |{\bar{\eta}}_{m}|^{\alpha_{3}}
	\begin{vmatrix}
		\mathbf{0}_{(N_{0})\times (N_{0})} & \tilde{M}_{1}^{(n,+)} \\
		-\tilde{M}_{1}^{(n,-)} & \mathbb{I}_{2N_{0}\times 2N_{0}}
	\end{vmatrix} \left[ 1+\cO(|A|^{-1})\right], \\
	&\tilde{M}_{1}^{(n,+)}=\left[ 2^{1-j}S_{2i-j}(\mathbf{y}_{m}^{+}(n) + (j-1) \,\mathbf{s}) \right]_{1\leq i\leq N_{0}, 1\leq j\leq 2N_{0}}, \\
	&\tilde{M}_{1}^{(n,-)}=\left[ 2^{1-i}S_{2j-i}(\mathbf{y}_{m}^{-}(n) + (i-1) \, \mathbf{s}) \right]_{1\leq i\leq 2N_{0}, 1\leq j\leq N_{0}}, \quad n=0,1,
\end{aligned}	
\end{equation}
where $\alpha_{j}$ $(j=2,3)$ are positive constants and the vectors $\mathbf{y}_{m}^{\pm}(n)$ are defined by Eqs. \eqref{tmpm} and \eqref{yn02}.

Since the subscripts of the elements $S_{k}(\mathbf{y}_{m}^{\pm}(n) + v \, \mathbf{s})$ in the matrices $\tilde{M}_{1}^{(n,\pm)}$ \eqref{taut1} are all less than $2N_{0}$ with $N_{0}\leq m$, we deduce from the definition \eqref{schur} of Schur polynomials that the matrices $\tilde{M}_{1}^{(n,\pm)}$ \eqref{taut1} does not contain the large parameter $A$. 

Now, when $ |A|\gg 1 $ and $ \sqrt{\bar{x}^{2}+\bar{t}^{2}}=\sqrt{(x-x_{0})^{2}+(t-t_{0})^{2}}= \cO(1) $ , we substitute the expression \eqref{taut1} of $ \tau^{(n)} $ into the formula \eqref{qn1} of the high-order rogue wave solution $ q^{[N]}(x,t) $. Then, the asymptotic expression \eqref{asym21} of $ q^{[N]}(x,t) $ can be demonstrated. In other words, the high-order rogue wave solution $ q^{[N]}(x,t) $ tends to an $ N_{0} $th-order fundamental rogue wave near the position $ (x_{0}, t_{0}) $ of the $ (x,t) $-plane with approximation error $ \cO(|A|^{-1}) $, where $ (x_{0}, t_{0}) $ is the $ \frac{N(N-1)}{2} $-multiple root of the Adler--Moser polynomial $ \Theta_{N}(A^{-1}(x+2\ii t)) $. 
    
\end{itemize}

\end{itemize}

Therefore, we complete the proof of Theorem \ref{Theo2}.

\subsection{Proof of Proposition \ref{prop3}}\label{prop3.proof}

In this subsection, we mainly prove the asymptotics of the OTR-type and modified OTR-type patterns for the $4$th-order rogue wave solution $ q^{[4]}(x,t) $ \eqref{qn1} in the multiple-root region. In the single-root region, the asymptotic behaviors of these patterns resemble those given in Theorem \ref{Theo1}, and detailed proofs refer to Ref. \cite{yang2024}.

We can easily obtain the expression of the solution $ q^{[4]}(x,t) $ from the formula \eqref{qn1}. Then, we will separately demonstrate the asymptotic behaviors of the OTR-type and modified OTR-type patterns of the rogue wave solution $ q^{[4]}(x,t) $ given in Proposition \ref{prop3} in the multiple-root region with $ \sqrt{(x-x_{0})^{2}+(t-t_{0})^{2}}\leq \cO(|A|) $, where the internal large parameters $ a_{2j+1} $, $ \kappa_{j,2j+1} $, and $ \kappa_{j,2j-1} $ $ (j=1,2,3) $ are given in Proposition \ref{prop3}. 

\begin{itemize}

\item[(1). ] Asymptotics of the OTR-type patterns for the solution $ q^{[4]}(x,t) $ in the multiple-root region.

First, we perform the coordinate transformation \eqref{ct1} with the triple root $ (x_{0}, t_{0}) $ of the Adler--Moser polynomial $ \Theta_{N}(A^{-1}(x+2\ii t)) $. Here, the polynomial $ \Theta_{N}(z) $ contain the free parameters $ (\kappa_{1}, \kappa_{2},\kappa_{3}) $ defined by Eq. \eqref{kap1} with $ \kappa_{1} $ not satisfying Eq. \eqref{kap2}. Then, we have
\begin{equation}\label{xpmk}
\begin{aligned}
 & x_{1}^{\pm}(n)=y_{1}^{\pm}(n) + |A|(\hat{x}_{0} \pm2\ii \hat{t}_{0}),  \quad y_{1}^{\pm}(n)=\bar{x} \pm 2\ii \bar{t} \pm n, \quad n=0,1,
 	\\ 
 & x_{2k+1}^{+}=\bar{y}_{2k+1}^{+} +\hat{\eta}_{k}A + \kappa_{k,2k+1}A^{2k+1} +\kappa_{k,2k-1}A^{2k-1}, \quad \hat{\eta}_{k}=\frac{\hat{x}_{0}+2^{2k+1}\ii \hat{t}_{0}}{(2k+1)!} \ee^{-\ii \arg A},\\ 
 & x_{2k+1}^{-}= \bar{y}_{2k+1}^{-} +(\hat{\eta}_{k} A +\kappa_{k,2k+1}A^{2k+1} +\kappa_{k,2k-1}A^{2k-1})^{*}, \quad 1\leq k\leq 3, \\
\end{aligned}
\end{equation}
where $ x_{1}^{\pm}(n) $, $ x_{2k+1}^{\pm} $, $ (\bar{x},\bar{t}) $, $ (\hat{x}_{0}, \hat{t}_{0}) $, and  $ \bar{y}_{2k+1}^{\pm} $ are defined by Eqs. \eqref{xpm}, \eqref{ct1} and  \eqref{eta}, respectively. 

Moreover, based on the definition \eqref{schur} of Schur polynomials and the similar calculation in Eq. \eqref{sch1}, we can obtain the following expansion
\begin{equation}\label{sch3}
	\begin{aligned}
		&S_{k}(\mathbf{x}^{+}(n)+v\mathbf{s}) = \sum_{j=0}^{k} S_{k-j}(\bar{\mathbf{y}}^{+}+v\mathbf{s}) \theta_{j}(z_{0})A^{i} +\sum_{j=0}^{k-3} S_{k-3-j}(\bar{\mathbf{y}}^{+}+v\mathbf{s}) \theta_{j}(z_{0}) (\hat{\eta}_{1}+\kappa_{1,1})A^{j+1} \\ 
		&\quad+\sum_{j=0}^{k-5}S_{k-5-j}(\bar{\mathbf{y}}^{+}+v\mathbf{s}) \theta_{j}(z_{0})\left( \kappa_{2,3}A^{j+3} + \hat{\eta}_{2}A^{j+1} \right)  +\sum_{j=0}^{k-6} S_{k-6-j}(\bar{\mathbf{y}}^{+}+v\mathbf{s}) \theta_{j}(z_{0})\frac{(\hat{\eta}_{1}+\kappa_{1,1})^{2}}{2}A^{j+2}\\
		&\quad+ S_{k-7}(\bar{\mathbf{y}}^{+}+v\mathbf{s})\left( \kappa_{3,5}A^{5} +\hat{\eta}_{3}A \right),  \quad 0\leq k\leq7,
	\end{aligned}
\end{equation}
where 
\begin{equation}\label{vecxy1}
\begin{aligned}
	\bar{\mathbf{y}}^{\pm}(n) =({y}_{1}^{\pm}(n), 0,\bar{y}_{3}^{\pm}, 0, \bar{y}_{5}^{\pm}, 0, \ldots),
\end{aligned}
\end{equation}
and $ \theta_{j}(z_{0}) $, $ \mathbf{s} $, and $ \mathbf{x}^{\pm}(n)$ are defined by Eqs. \eqref{schur2}, \eqref{sj1} and \eqref{xpm}, separately.

Next, we utilize the above expansion \eqref{sch3} to expand all elements of the matrix $ M^{(n,+)} $ in the determinant \eqref{tau} of $ \tau^{(n)} $ with $ N=4 $ for the rogue wave solution $ q^{[4]}(x,t) $. Further, we perform some row transformations on the expanded matrix $ M^{(n,+)} $ and simplify it by using the properties of Proposition \ref{pro1}, similar to those employed in the proof of the claw-like rogue wave patterns in the multiple-root region in Sec. \ref{proof.Theo2}. This leads to $ M^{(n,+)} $ being reduced to the following matrix:
\begin{equation}\label{mep6}
	\begin{bmatrix}
		A\theta_{1} + S_{1} & A^{2}S_{1}\beta_{1}+\cO(A) & A^{2}S_{3}\beta_{2} +\cO(A) & A^{3}{S_{2}}{\beta_{3}} \rho_{1} +A^{2}S_{5}\beta_{3}  +\cO(A) \\
		\frac{1}{2} & \frac{1}{2}A^{2}\beta_{1}+\cO(A) & \frac{1}{2}A^{2}S_{2}\beta_{2} +\cO(A) & \frac{1}{2}A^{3} {S_{1}} {\beta_{3}} \rho_{1} +\frac{1}{2}A^{2}S_{4}\beta_{3}  +\cO(A) \\
		0 & \frac{1}{2^{2}}(A\theta_{1}+S_{1}) & \frac{1}{2^{2}}A^{2}S_{1}\beta_{2} +\cO(A) & \frac{1}{2^{2}}A^{3}{\beta_{3}} \rho_{1} +\frac{1}{2^{2}}A^{2}S_{3}\beta_{3} +\cO(A) \\
		0 & \frac{1}{2^{3}} & \frac{1}{2^{3}}A^{2} \beta_{2} +\cO(A) &  \frac{1}{2^{3}}A^{2}S_{2}\beta_{3} +\cO(A) \\
		0 & 0 & \frac{1}{2^{4}}(A\theta_{1}+S_{1}) &  \frac{1}{2^{4}}A^{2}S_{1}\beta_{3}  +\cO(A) \\
		0 & 0 & \frac{1}{2^{5}} &  \frac{1}{2^{5}} A^{2} \beta_{3} +\cO(A) \\
		0 & 0 & 0 & \frac{1}{2^{6}} (A\theta_{1}+S_{1})\\
		0 & 0 & 0 & \frac{1}{2^{7}} 
	\end{bmatrix}^{T},
\end{equation}
where 
\begin{equation}\label{beta}
	\begin{aligned}
		&\beta_{1}=\theta_{2}-\frac{\theta_{3}}{\theta_{1}} \ne 0, \quad \beta_{2}=\theta_{2}-\frac{\theta_{1}\theta_{4}-\theta_{5}}{\theta_{1}\theta_{2}-\theta_{3}}\ne 0, \quad \beta_{3}=\theta_{2}-\frac{\theta_{4}(\theta_{1}\theta_{2}-\theta_{3}) -(\theta_{1}\theta_{6}-\theta_{7})}{\theta_{2}(\theta_{1}\theta_{2}-\theta_{3}) -(\theta_{1}\theta_{4}-\theta_{5})}\ne 0,\\
        &\rho_{1}= \frac{\hat{x}_{0} +8\ii \hat{t}_{0}}{6}\ee^{-\ii \arg A}+\kappa_{1,1} + \dfrac{\kappa_{2,3}(\beta_{1}+{\beta_{3}-{\theta_{1}}^{2}})}{\beta_{1}\beta_{3}} +\frac{\kappa_{3,5}}{\beta_{1}\beta_{3}},
	\end{aligned}	
\end{equation}
and the symbols $ S_{k} $ and $ \theta_{k} $ represent $ S_{k}(\bar{\mathbf{y}}^{+}(n)+v\mathbf{s})$ and $ \theta_{k}(z_{0}) $, respectively. Since the facts of $\Theta_{3}(z_{0})=\Theta_{3,1}(z_{0}) = \Theta_{4,1}(z_{0})=0$ are given in Proposition \ref{pro1}, we can find that $\beta_{1}=\beta_{2}=\beta_{3}= \frac{z_{0}^{3}-3\kappa_{1}}{3z_{0}}\ne 0$ and $ \rho_{1}=\rho(z_{0}, \kappa_{1,1}, \kappa_{1,3}, \kappa_{2,3}, \kappa_{3,5}) $ with $ \rho(z_{0}, \kappa_{1,1}, \kappa_{1,3}, \kappa_{2,3}, \kappa_{3,5}) $ defined by Eq. \eqref{para4}.
Similarly, another matrix $ M^{(n,-)} $ in the formula \eqref{tau} can be also simplified, which is omitted here.

Therefore, if the parameters $ (\kappa_{1,1}, \kappa_{2,3}, \kappa_{3,5}) $ satisfy the parameter equation $ \rho(z_{0}, \kappa_{1,1}, \kappa_{1,3}, \kappa_{2,3}, \kappa_{3,5})=0 $ in Eq. \eqref{para4}, then the determinant $ \tau^{(n)} $ \eqref{tau} of the solution $ q^{[4]}(x,t) $ can be rewritten as
\begin{equation}\label{tau8}
	\begin{aligned}
		&\tau^{(n)} = 2^{-10}|A|^{14}|z_{0}{\beta_{1}}^{3}|^{2}\begin{vmatrix}
			\mathbf{0}_{2\times 2} & \bar{M}^{(n,+)} \\
			-\bar{M}^{(n,-)} & \mathbb{I}_{4\times 4}
		\end{vmatrix}\left[ 1+\cO(|A|^{-1})\right], \\
		&\bar{M}^{(n,+)}=\left[ 2^{1-j}S_{2i-j}(\bar{\mathbf{y}}^{+}(n) + (j+1) \,\mathbf{s}) \right]_{1\leq i\leq 2, 1\leq j\leq 4}, \\ &\bar{M}^{(n,-)}=\left[ 2^{1-i}S_{2j-i}(\bar{\mathbf{y}}^{-}(n) + (i+1) \, \mathbf{s}) \right]_{1\leq i\leq 4, 1\leq j\leq 2}, \quad n=0,1.
	\end{aligned}
\end{equation}
Furthermore, through a similar approach as in Eqs. \eqref{tau2}-\eqref{tau2b} for the proof of Theorem \ref{Theo2}, we can simplify the matrix $ \bar{M}^{(n,\pm)} $ \eqref{tau8} to the expressions blow:
\begin{equation}\label{tmnpm1}
	\bar{M}^{(n,+)}=\left[ 2^{1-j}S_{2i-j}(\bar{\mathbf{y}}^{+}(n) + (j-1) \,\mathbf{s}) \right]_{1\leq i\leq 2, 1\leq j\leq 4}, \quad  \bar{M}^{(n,-)}=\left[ 2^{1-i}S_{2j-i}(\bar{\mathbf{y}}^{-}(n) + (i-1) \, \mathbf{s}) \right]_{1\leq i\leq 4, 1\leq j\leq 2}.
\end{equation}
 
Now, by substituting the simplified $ \tau^{(n)} $ \eqref{tau8} into the solution formula \eqref{qn1}, we find that when $ |A|\gg 1$, the OTR-type patterns of the rogue wave solution $ q^{[4]}(x,t) $ asymptotically approach a second-order fundamental rogue wave in the multiple-root region, and the asymptotical expression \eqref{asym3} of $ q^{[4]}(x,t) $ holds.

\item[(2).] Asymptotics of the modified OTR-type patterns for the solution $ q^{[4]}(x,t) $ in the multiple-root region.

We perform the coordinate transformation similar to \eqref{ct1}, where $ z_{0}=(x_{0}+2\ii t_{0})A^{-1} $ is the triple root of the Adler--Moser polynomial $ \Theta_{N}(z) $. By the definition \eqref{schur} of Schur polynomials and the similar calculation in Eq. \eqref{sch1}, we can obtain
\begin{equation}\label{sch4}
	S_{k}(\mathbf{x}^{+}(n) + v\mathbf{s}) =\sum_{i=0}^{k} S_{k-i}(\hat{\mathbf{y}}^{+}(n) + v\mathbf{s}) \theta_{i}(z_{0}) A^{i}, \quad 1\leq k\leq 7,
\end{equation}
where
\begin{equation}\label{hyz0}
	\hat{\mathbf{y}}^{+}(n) = (y_{1}^{+}(n), 0, \hat{y}_{3}^{+},0, \hat{y}_{5}^{+}, 0, \hat{y}_{7}^{+},0,\ldots) , \quad \hat{y}_{2j+1}^{+}= \bar{y}_{2j+1}^{+} + {\hat{\eta}_{j}A +\kappa_{j,2j-1}A^{2j-1}}, \quad {1 \leq j\leq 3},
\end{equation}
and $ \mathbf{s} $, $\mathbf{x}^{+}(n)$, $ \theta_{i}(z_{0}) $, $\bar{y}_{2j+1}^{+}$, and $\hat{\eta}_{k}$ are defined by Eqs. \eqref{sj1}, \eqref{xpm}, \eqref{schur2} and \eqref{xpmk}, respectively.

To simplify the determinant $ \tau^{(n)} $ \eqref{tau} of the solution $q^{[4]}(x,t)$, we employ the expansion \eqref{sch4} to expand all elements of the matrix $ M^{(n,+)} $ \eqref{mpm} with $ N=4 $, as follows:
\begin{equation}\label{mpe6}
	M^{(n,+)}=\left[ 2^{1-j} \sum_{l=0}^{2i-j}S_{2i-j-l}(\hat{\mathbf{y}}^{+}(n) + (j-1)\mathbf{s})\theta_{l}(z_{0}) A^{l}\right]_{1\leq i\leq 4, 1\leq j\leq 8}.
\end{equation}

Then, we can perform some row transformations on the matrix \eqref{mpe6}, similar to those employed in the proof of the claw-like rogue wave patterns in the multiple-region in Sec. \ref{proof.Theo2}. Therefore, we can simplify $M^{(n,+)}$ \eqref{mpe6} to the following form:
\begin{equation}\label{mpe5}
	\begin{bmatrix}
		S_{0}\theta_{1}A+S_{1} &  S_{1}\beta_{1}A^{2}+S_{2}\theta_{1}A +S_{3} & S_{3}\beta_{1}A^{2}+S_{4}\theta_{1}A +S_{5} & S_{5}\beta_{1}A^{2} +S_{6}\theta_{1}A +S_{7} \\
		2^{-1} & 2^{-1}(\beta_{1}A^{2}+S_{1}\theta_{1}A+S_{2})  & 2^{-1}(S_{2}\beta_{1}A^{2}+S_{3}\theta_{1}A+S_{4})  & 2^{-1}(S_{4}\beta_{1}A^{2}+S_{5}\theta_{1}A+S_{6}) \\
		0 & 2^{-2}(\theta_{1}A+S_{1}) & 2^{-2}(S_{1}\beta_{1}A^{2}+S_{2}\theta_{1}A +S_{3}) & 
		2^{-2}(S_{3}\beta_{1}A^{2}+S_{4}\theta_{1}A +S_{5}) \\
		0 & 2^{-3} & 2^{-3}(\beta_{1}A^{2}+S_{1}\theta_{1}A+S_{2})  & 2^{-3}(S_{2}\beta_{1}A^{2}+S_{3}\theta_{1}A+S_{4}) \\
		0 & 0 & 2^{-4}(\theta_{1}A +S_{1}) & 2^{-4}(S_{1}\beta_{1}A^{2}+S_{2}\theta_{1}A +S_{3}) \\
		0 & 0 & 2^{-5} & 2^{-5}(\beta_{1}A^{2}+S_{1}\theta_{1}A+S_{2}) \\
		0 & 0 & 0 & 2^{-6}(\theta_{1}A +S_{1})\\
		0 & 0 & 0 & 2^{-7}\\
	\end{bmatrix}^{T},
\end{equation}
where the parameter $ \beta_{1} $ is defined by Eq. \eqref{beta}, and the symbols $ S_{k} $ and $ \theta_{k} $ represent $ S_{k}(\hat{\mathbf{y}}^{+}+v\mathbf{s})$ and $ \theta_{k}(z_{0}) $, respectively. 

When the parameters $ (\kappa_{1,1}, \kappa_{2,3}, \kappa_{3,5}) $ do not satisfy the parameter equation $ \rho(z_{0}, \kappa_{1,1}, \kappa_{1,3}, \kappa_{2,3}, \kappa_{3,5})=0 $ in Eq. \eqref{para4}, we yield the expansion:
\begin{equation}\label{sch5}
\begin{aligned}
    S_{k}(\hat{\mathbf{y}}^{+}+v\mathbf{s}) =& \left( p_{k}(\bar{z}) A^{k/3}+ p_{k-5}(\bar{z})\kappa_{2,3} A^{(k+4)/3} + p_{k-7}(\bar{z})A^{k/3}(vs_{2}\kappa_{2,3}A^{2/3}+ \kappa_{3,5}A^{8/3}) \right) \\
    &\times \left[ 1+\cO(|A|^{-2/3})\right], \quad 1\leq k\leq 7,
\end{aligned}
\end{equation}
by the similar calculation in Eq. \eqref{sch1}, where $ p_{k}(\bar{z}) $ are defined by
\begin{equation}\label{ampk1}
\begin{aligned}
	\exp(\bar{z} \varepsilon + (\hat{\eta_{1}}+\kappa_{1,1})A\varepsilon^3)=\sum_{k=0}^{\infty} p_{k}(\bar{z}) A^{k/3}\varepsilon^{k}, \quad \bar{z}= (\bar{x}+2\ii \bar{t})A^{-1/3},
\end{aligned}
\end{equation}
with $(\bar{x},\bar{t})$, $(\hat{x}_{0}, \hat{t}_{0})$ and $ {\hat{\eta}}_{1} $ given in Eqs. \eqref{ct1} and \eqref{xpmk}. It is evident that $ p_{k}(\bar{z}) $ represents $ \theta({\bar{z}}) $ defined by Eq. \eqref{schur2} with only one free parameter $ \kappa_{1}=\hat{\eta_{1}}+\kappa_{1,1} $.

Then, we substitute the expansion \eqref{sch5} into the matrix \eqref{mpe5} and perform some similar row transformations as done previously, thus simplifying the matrix \eqref{mpe5} to the following form:
\begin{equation}\label{mpe8}
	\begin{bmatrix}
\theta_{1}A +\cO(A^{\frac{1}{3}}) & \beta_{1} p_{1}(\bar{z})A^{\frac{7}{3}} +\cO(A^{\frac{5}{3}}) & \beta_{1}p_{3}(\bar{z})A^{3} +\cO(A^{\frac{7}{3}}) & ( p_{5}(\bar{z})\beta_{1} + \rho_{2}p_{2}(\bar{z}) ) A^{\frac{11}{3}} +\cO({A}^{3}) \\
	2^{-1} & 2^{-1}\beta_{1}A^{2} +\cO(A^{\frac{4}{3}})  & 2^{-1}\beta_{1}p_{2}(\bar{z}) A^{\frac{8}{3}} +\cO(A^{2})  & 2^{-1}( p_{4}(\bar{z})\beta_{1} + \rho_{2}p_{1}(\bar{z}) ) A^{\frac{10}{3}} +\cO({A}^{\frac{8}{3}}) \\
	0 & 2^{-2}\theta_{1}A +\cO(A^{\frac{1}{3}}) & 2^{-2}\beta_{1}p_{1}(\bar{z}) A^{\frac{7}{3}} +\cO(A^{\frac{5}{3}}) & 
	2^{-2} (p_{3}(\bar{z})\beta_{1} + \rho_{2}) A^{3} +\cO({A}^{\frac{7}{3}}) \\
	0 &	2^{-3} & 2^{-3}\beta_{1}A^{2}+\cO({A}^{\frac{4}{3}})  & 2^{-3}\beta_{1}p_{2}(\bar{z})A^{\frac{8}{3}} +\cO(A^{2}) \\
	0 & 0 & 2^{-4}\theta_{1}A +\cO(A^{\frac{1}{3}}) & 2^{-4}p_{1}(\bar{z})A^{\frac{7}{3}} +\cO(A^{\frac{5}{3}}) \\
	0 & 0 & 2^{-5} & 2^{-5}\beta_{1}A^{2}+\cO({A}^{\frac{4}{3}}) \\
	0 & 0 & 0 & 2^{-6}\theta_{1}A +\cO(A^{\frac{1}{3}})\\
	0 & 0 & 0 & 2^{-7}\\
\end{bmatrix}^{T},
\end{equation}
with $ \rho_{2}=(\kappa_{2,3}(2\beta_{1}-\theta_{1}^{2})+\kappa_{3,5}){\beta_{1}}^{-1} $ with $ \beta_{1} $ defined by Eq. \eqref{beta}.

Similarly, the matrix $ M^{(n,-)} $ in Eq. \eqref{mpm} is also simplified in the same ways, which is omitted here.

Now, for the expression \eqref{tau} of $\tau^{(n)}$ with $ N=4 $ and the simplified matrices $ M^{(n,\pm)} $ \eqref{mpe8}, we apply the same method as the proof of the asymptotics for the modified claw-like rogue wave pattern in the multiple-root region presented in Sec. \ref{proof.Theo2}. Here, we omit the detailed calculation process of $\tau^{(n)}$. 

Then, we can obtain the following expression:
\begin{equation}\label{tau6}
\begin{aligned}
	\tau^{(n)}= & |A|^{46/3} |\theta_{1}\beta_{1}^{3}|^{2} \alpha_{4}^{2} \left|(Q_{2}(\bar{z}_{0}))'\right|^{2} \left( ({x}-\tilde{x}_{0})^{2} +4({t}-\tilde{t}_{0})^{2} -4\ii n({t}-\tilde{t}_{0}) -n^{2} +\frac{1}{4} \right)\\
	&\times  \left[ 1+\cO(|A|^{-1/3})\right], \quad n=0,1,
\end{aligned}
\end{equation}
where $ \alpha_{4}=(2^{6}c_{2})^{-1} $ with $ c_{2} $ given in Eqs. \eqref{schur2}, and $ (\tilde{x}_{0}, \tilde{t}_{0}) $ are defined by 
\begin{equation}\label{tixt0}
	\tilde{x}_{0}+2\ii \tilde{t}_{0} = \bar{z}_{0}A^{-1/3}+z_{0}A,
\end{equation}
with the unique nonzero triple root $z_{0}$ of the Adler--Moser polynomial $\Theta_{4}(z)$ and the single root $ \bar{z}_{0} $ of the polynomial $ Q_{2}(\bar{z}) $. Here, the polynomial $Q_{2}(\bar{z})$ represents the special Adler--Moser polynomial $\Theta_{2}(\bar{z})$ defined by Eq. \eqref{amp} with only one free parameter $ \kappa_{1}=\rho(z_{0}, \kappa_{1,1}, \kappa_{1,3}, \kappa_{2,3}, \kappa_{3,5})$ given in Eq. \eqref{para4}.

Therefore, by substituting the expression \eqref{tau6} of $ \tau^{(n)} $ into the formula \eqref{qn1} of the rogue wave solution $ q^{[4]}(x,t) $, we can conclude that when $ |A|\gg 1$, the solution $ q^{[4]}(x,t) $ asymptotically approaches to a first-order rogue wave near the position of $ (\tilde{x}_{0}, \tilde{t}_{0}) $ and the asymptotic expression \eqref{asym4} holds. Moreover, when the parameters $ (\kappa_{1,1}, \kappa_{2,3}, \kappa_{3,5}) $ do not satisfy the parameter equation $ \rho=0 $ in Eq. \eqref{para4}, we can easily find that the three roots of $ Q_{2}(\bar{z}) $ are all single. This implies that when $ |A|\gg 1$, the modified OTR-type patterns of the rogue wave solution $ q^{[4]}(x,t) $ asymptotically split into three first-order rogue waves in the multiple-root region, and the asymptotics \eqref{asym4} of $ q^{[4]}(x,t) $ holds.

\end{itemize}

Furthermore, according to the asymptotic behavior of the single-root region and the asymptotics of the multiple-root region for the OTR-type and modified OTR-type patterns of the rogue wave solution $ q^{[4]}(x,t) $, it is easily found that when $ |A|\rightarrow \infty $ and $ (x,t) $ is not near the locations of any of the above first-order rogue waves and the lower-order rogue wave, the solution $ q^{[4]}(x,t) $ asymptotically approaches to the plane wave background $ \ee^{2\ii t} $.

Finally, we complete the proof of Proposition \ref{prop3}.

%\end{proof}

\section{Conclusions and Discussions}\label{Sec6}

In this paper, we generate the determinant formula for high-order rogue wave solutions of the NLS equation by applying the DT method. We then conduct a detailed analysis of the asymptotic behavior for the rogue wave solutions with multiple internal large parameters and obtain some new rogue wave patterns: the claw-like, the OTR-type, the TTR-type, the semi-modified TTR-type, and their modified patterns. We mainly study the claw-like and modified claw-like patterns of the $ N $th-order rogue wave solution $ q^{[N]}(x,t) $, as well as the OTR-type, the TTR-type, semi-modified TTR-type, and their modified patterns of fourth-order rogue wave solution $ q^{[4]}(x,t) $. Moreover, we demonstrate the correlation between these rogue wave patterns and root structures of the Adler--Moser polynomials with multiple roots. The structures of these patterns are divided into two regions on the $ (x, t) $ plane: the single-root region and the multiple-root region, which correspond to the single roots and multiple roots of the Adler--Moser polynomials, respectively. It is found that these patterns of the rogue wave solution $ q^{[N]}(x,t) $ obtained in this paper asymptotically approach scattered first-order rogue waves in the single-root region, where their positions on the $(x, t)$-plane correspond to single roots of the Adler--Moser polynomials $ \Theta_{N}(z) $. In the multiple-root region, these rogue wave patterns asymptotically approach lower-order fundamental rogue waves, dispersed first-order rogue waves, or mixed structures of these rogue waves. The number and structures of the peaks in the multiple-root region are related to the root structures of special Adler--Moser polynomials with new free parameters, such as the Yablonskii--Vorob'ev polynomial hierarchy, among others. Notably, for the multiple-root region of the rogue wave patterns, we can control their positions based on the value of multiple roots of the Adler--Moser polynomials.

Furthermore, we also provide examples of the claw-like patterns of the rogue wave solutions $ q^{[N]}(x,t) $ $ (N=3,4,5,6) $ and the modified claw-like patterns of $ q^{[6]}(x,t) $, as well as the OTR-type, TTR-type, semi-modified TTR-type and their modified patterns of $ q^{[4]}(x,t) $. In general, we can similarly construct analogous patterns for arbitrary high-order rogue wave solutions of integrable equations and explore even more diverse patterns, where these patterns correspond to the root structures of the Adler--Moser polynomials $ \Theta_{N}(z) $ with some multiple roots. {Our results answer the open question posed by Yang \textit{et al}. in Ref. \cite{yang2024}: what will happen if some roots of the Adler–Moser polynomials are not simple? This serves as a complement and extension to the study presented in Ref. \cite{yang2024}, thereby enriching the theoretical research on rogue wave patterns.}

\section*{Conflict of interests}
The authors have no conflicts to disclose.

\section*{DATA AVAILABILITY}
Data sharing is not applicable to this article as no new data were created or analyzed in this study.

\section*{Acknowledgments}
%The authors appreciate the editor and referees for their valuable suggestions for this work. 
Liming Ling is supported by the National Natural Science Foundation of China (No. 12122105) and the Guangzhou Municipal Science and Technology Project (Guangzhou Science and Technology Plan) (No. 2024A04J6245).


\bibliographystyle{elsarticle-num-names}
\bibliography{Ref_RWofNLS}


%\begin{thebibliography}{00}


%\bibitem[()]{}

%\end{thebibliography}

\end{document}